\PassOptionsToPackage{unicode}{hyperref}
\PassOptionsToPackage{hyphens}{url}
\PassOptionsToPackage{dvipsnames,svgnames,x11names}{xcolor}
\documentclass[
]{article}
\usepackage{xcolor}
\usepackage[margin=1in]{geometry}
\usepackage{amsmath,amssymb}
\usepackage{iftex}
\ifPDFTeX
  \usepackage[T1]{fontenc}
  \usepackage[utf8]{inputenc}
  \usepackage{textcomp} 
\else 
  \usepackage{unicode-math} 
  \defaultfontfeatures{Scale=MatchLowercase}
  \defaultfontfeatures[\rmfamily]{Ligatures=TeX,Scale=1}
\fi
\usepackage{lmodern}
\ifPDFTeX\else
\fi
\IfFileExists{upquote.sty}{\usepackage{upquote}}{}
\IfFileExists{microtype.sty}{
  \usepackage[]{microtype}
  \UseMicrotypeSet[protrusion]{basicmath} 
}{}
\makeatletter
\@ifundefined{KOMAClassName}{
  \IfFileExists{parskip.sty}{%
    \usepackage{parskip}
  }{
    \setlength{\parindent}{0pt}
    \setlength{\parskip}{6pt plus 2pt minus 1pt}}
}{
  \KOMAoptions{parskip=half}}
\makeatother
\usepackage{color}
\usepackage{fancyvrb}

\DefineVerbatimEnvironment{Highlighting}{Verbatim}{commandchars=\\\{\}}
\usepackage{framed}
\definecolor{shadecolor}{RGB}{248,248,248}
\newenvironment{Shaded}{\begin{snugshade}}{\end{snugshade}}

\newcommand{\AttributeTok}[1]{\textcolor[rgb]{0.13,0.29,0.53}{#1}}

\newcommand{\CommentTok}[1]{\textcolor[rgb]{0.56,0.35,0.01}{\textit{#1}}}

\newcommand{\ConstantTok}[1]{\textcolor[rgb]{0.56,0.35,0.01}{#1}}
\newcommand{\ControlFlowTok}[1]{\textcolor[rgb]{0.13,0.29,0.53}{\textbf{#1}}}

\newcommand{\DecValTok}[1]{\textcolor[rgb]{0.00,0.00,0.81}{#1}}

\newcommand{\FloatTok}[1]{\textcolor[rgb]{0.00,0.00,0.81}{#1}}
\newcommand{\FunctionTok}[1]{\textcolor[rgb]{0.13,0.29,0.53}{\textbf{#1}}}

\newcommand{\NormalTok}[1]{#1}

\newcommand{\OtherTok}[1]{\textcolor[rgb]{0.56,0.35,0.01}{#1}}

\newcommand{\SpecialCharTok}[1]{\textcolor[rgb]{0.81,0.36,0.00}{\textbf{#1}}}

\newcommand{\StringTok}[1]{\textcolor[rgb]{0.31,0.60,0.02}{#1}}

\usepackage{longtable,booktabs,array}
\usepackage{calc} 
\usepackage{etoolbox}
\makeatletter
\patchcmd\longtable{\par}{\if@noskipsec\mbox{}\fi\par}{}{}
\makeatother
\IfFileExists{footnotehyper.sty}{\usepackage{footnotehyper}}{\usepackage{footnote}}
\makesavenoteenv{longtable}
\usepackage{graphicx}
\makeatletter
\newsavebox\pandoc@box
\newcommand*\pandocbounded[1]{
  \sbox\pandoc@box{#1}%
  \Gscale@div\@tempa{\textheight}{\dimexpr\ht\pandoc@box+\dp\pandoc@box\relax}%
  \Gscale@div\@tempb{\linewidth}{\wd\pandoc@box}%
  \ifdim\@tempb\p@<\@tempa\p@\let\@tempa\@tempb\fi
  \ifdim\@tempa\p@<\p@\scalebox{\@tempa}{\usebox\pandoc@box}%
  \else\usebox{\pandoc@box}%
  \fi%
}
\def\fps@figure{htbp}
\makeatother
\providecommand{\tightlist}{%
  \setlength{\itemsep}{0pt}\setlength{\parskip}{0pt}}
\usepackage[]{natbib}
\usepackage{amsmath, amssymb}
\usepackage{booktabs, array}
\usepackage{hyperref}
\usepackage{graphicx}
\usepackage{natbib}
\usepackage{multirow}
\usepackage{float}

\makeatletter\@namedef{ver@colortbl.sty}{9999/01/01}\makeatother
\DeclareUnicodeCharacter{03BB}{\ensuremath{\lambda}}
\DeclareUnicodeCharacter{03C3}{\ensuremath{\sigma}}
\DeclareUnicodeCharacter{03B8}{\ensuremath{\theta}}
\DeclareUnicodeCharacter{03BC}{\ensuremath{\mu}}
\usepackage{booktabs}
\usepackage{longtable}
\usepackage{array}
\usepackage{multirow}
\usepackage{wrapfig}
\usepackage{float}
\usepackage{colortbl}
\usepackage{pdflscape}
\usepackage{tabularx}
\usepackage{xltabular}
\usepackage{threeparttable}
\usepackage{threeparttablex}
\usepackage[normalem]{ulem}
\usepackage{makecell}
\usepackage{xcolor}
\usepackage{bookmark}
\IfFileExists{xurl.sty}{\usepackage{xurl}}{} 
\hypersetup{
  pdftitle={Logistic Credibility with Temporal Decay: Extending Bühlmann--Straub for Commercial Lines},
  pdfauthor={Jake Morris FIA CSPA},
  colorlinks=true,
  linkcolor={black},
  filecolor={Maroon},
  citecolor={black},
  urlcolor={blue},
  pdfcreator={LaTeX via pandoc}}

\title{Logistic Credibility with Temporal Decay: Extending Bühlmann--Straub for Commercial Lines}
\author{Jake Morris FIA CSPA}
\date{November 2026}

\begin{document}
\maketitle

\begin{abstract}
Bühlmann--Straub (B-S) credibility blends an account's own experience with a complement (the benchmark rate), using the weight $Z_i = E_i/(E_i+K)$, where $E_i$ is the account's exposure and $K$ is a constant estimated from the portfolio. The weight assumes the same $K$ for every account, equal influence for recent and older years, and a fixed complement. On held-out US Commercial Auto data the constant-$K$ assumption fails for small accounts: standard B-S gives them a calibration slope of \textbf{4.0} (actual on predicted; ideal 1.0; 90\% bootstrap CI 2.5--4.9), meaning they are under-credited.

\medskip
We propose a joint framework that relaxes all three assumptions while keeping Bühlmann--Straub's interpretable credibility weight. The credibility weight $Z_i$ is a logistic function of observable account characteristics; older experience is down-weighted by an exponentially weighted moving average whose decay rate $\lambda$ is estimated; and $Z$, $\lambda$ and the complement are fitted in one likelihood. Rolling B-S (B-S over the model’s lookback window) is a special case, so extensions can be tested by likelihood-ratio test, and the complement can be estimated jointly or taken from an existing pricing model.

\medskip
On 96 companies with the last two accident years held out, the proposed model brings the small-account calibration slope from 4.0 to 1.0 and reduces exposure-weighted prediction error by \textbf{25\%} relative to standard B-S (90\% bootstrap interval 7\%--42\%), both using the prior-year portfolio mean as the complement. The fitted credibility weight is nearly flat in size (0.61--0.67), above B-S's weight for small accounts and below it for large ones. A step-by-step patched B-S comes within 2.5\% of the proposed model overall but deteriorates for small accounts when recency weighting is added; joint estimation avoids this. A mixed-effects benchmark, which shrinks almost exactly as B-S does, is outperformed overall; it does better for mid-sized accounts in the test period (not statistically significant) but miscalibrates small ones.

\medskip
In Commercial Auto the decay rate varies with account size ($\hat\lambda \approx 0.60$, $0.84$ and $0.13$ for Small, Mid and Large): small and mid accounts benefit from multi-year averaging, large accounts mainly from recent experience. In Other Liability, where all models perform similarly, the decay rate again falls with size. A simulation study (drift and heterogeneous $K$) supports the mechanisms.

\medskip
The model needs only account-year summaries and delivers transparent numbers for the underwriter: the credibility weight, the complement, the account's recency-weighted experience and the renewal rate they produce. A Bayesian implementation adds a posterior distribution for each account's credibility weight.

\end{abstract}

\pagebreak

\begin{center}
{\large\bfseries Practitioner Summary}
\end{center}

\noindent\textbf{What this paper delivers.}
A unified framework for experience rating that estimates credibility weight \(Z\), how quickly older years lose influence (\(\lambda\)) and the complement jointly in one model fit. It replaces the fixes practitioners apply one at a time (for example stratifying or judgementally selecting \(K\), adding recency weighting or adjusting the complement).

On a 10-year CAS Commercial Auto dataset (96 companies, AY 1998--2007, two-year held-out test set), it reduces exposure-weighted prediction error (wMSE) by 25\% over standard B-S with a deployable prior-year complement (proposed tercile-\(\lambda\) model; 27\% in the main tables, which use the realised portfolio mean as the complement). Fitted in a Bayesian framework, the model also provides predictive intervals. A team that has already patched B-S on all three fronts (the choice of \(K\), recency weighting and the complement; in the case study each varies with account size) has a wMSE within about 2.5\% of the proposed model's. The further gains are a single likelihood, which gives formal tests of which assumptions to relax, and, in the case study, avoiding the loss of small-account accuracy that arises when such patches are added one at a time.

\medskip

\noindent\textbf{Three numbers build each renewal rate.}
\[
  \hat{r}_i = (1 - \hat{Z}_i)\,\hat{\mu}_i + \hat{Z}_i\,\hat{f}_i
\]

\begin{itemize}
  \item $\hat{Z}_i \in (0,1)$: how much to trust this account's own history
  \item $\hat{\mu}_i$: what the portfolio or segment benchmark says the risk should cost
  \item $\hat{f}_i$: what the account's own history says, with recent years counting more (an exponentially weighted moving average, EWMA)
\end{itemize}

\medskip

\noindent\textbf{Go/no-go checklist (before any fitting).}
The minimum portfolio size (\(N\), the number of accounts) depends on what an ``account'' means and how noisy its loss statistic is.

\begin{itemize}
  \item Enough distinct rated accounts with meaningful lag history (accounts with only one lag year contribute negligible $\lambda$ signal; five or more lag years is preferable)
  \item Target loss statistic comparable to the renewal being priced (restated for changes in rate level, attachment or trend) and, where loss skewness is high, attritional (large losses capped or excluded, depending on how much volume each account represents; Section \ref{sec:empirical} discusses this for the case study)
  \item Account-level exposure by year available (to construct log lookback exposure, the structural $Z$ input)
  \item Pre-adoption signal check: next-year loss statistic vs EWMA $\hat{f}_i$; a positive slope overall or in at least one exposure band confirms the data support credibility weighting
\end{itemize}
\medskip

\noindent\textbf{Post-deployment check.}
Regress actual on predicted (exposure-weighted), overall and by exposure band. A slope whose bootstrap interval excludes 1.0 in more than one consecutive cycle warrants investigation: well below 1 indicates over-crediting (too much weight on the account's own history), well above 1 under-crediting (too much shrinkage toward the complement). Some deviation in the smallest band is expected, as exposure weighting favours larger accounts.

\medskip

\noindent\textbf{New accounts.} With no lookback exposure, \(\hat{Z}_i \to 0\) and the prediction is the complement \(\hat{\mu}_i\).

\medskip

\noindent\textbf{Where to go from here.}
Start with Section~\ref{sec:mvd} (Quick Start, p.~\pageref{sec:mvd}), whose Step 0 is the full checklist including a \(\lambda\) uncertainty diagnostic;
Sections~\ref{sec:applying}--\ref{sec:applying-renewal} give the full fitting and renewal guide and \hyperref[app:implementation]{Appendix~G} the R code.

\pagebreak

\section{The Experience Rating Problem}\label{the-experience-rating-problem}

Across commercial, specialty, and reinsurance lines, pricers face the same
fundamental question at every renewal.
There is a model or market benchmark: the rate implied by the account's
risk characteristics.
There is also the account's own claims history (three, five, perhaps
ten years of observed experience).
These two signals rarely agree perfectly.
The question is: how much weight should the history carry?

In changing market conditions this question carries direct pricing consequences: over-crediting a volatile small account may anchor the renewal to a past loss spike; under-crediting a stable large account may surrender margin unnecessarily.

The actuarial profession has a rigorous, theoretically grounded answer.
Bühlmann--Straub (B-S) credibility \citep{buhlmann1970} says the optimal blend is:

\begin{equation}
  \hat{r}_i = (1 - Z_i)\,\mu + Z_i\,\bar{f}_i,
  \qquad
  Z_i = \frac{E_i}{E_i + K},
  \label{eq:bs}
\end{equation}

where \(E_i = \sum_t E_{it}\) is the account's total exposure over its observed history (any volume measure, such as premium or vehicle-years), \(\bar{f}_i\) is the account's own observed
claim rate, \(\mu\) is the complement (the portfolio mean in textbook B-S), and \(K\) (in the same units as \(E_i\)) is estimated from the
portfolio by the method of moments.
The formula is the empirical-Bayes-optimal solution under a normal--normal hierarchical model \citep{jewell1974, morris1983},
has a clean exposure interpretation, and has been the textbook standard
for half a century.\footnote{The other classical approach is \emph{limited fluctuation credibility}
  \citep{mowbray1914, longley-cook1962}, in which
  \(Z_i = \min\!\bigl(\sqrt{n_i/N_{\mathrm{ref}}}, 1\bigr)\) with \(N_{\mathrm{ref}}\) set by a
  precision criterion rather than estimated from the portfolio.
  It answers a different question (when does an account have sufficient volume
  to be treated as self-rating?) and does not pool accounts against a portfolio
  prior in the B-S sense.
  We adopt B-S as the primary comparator throughout, as it is the theoretically
  grounded pooling estimator most directly comparable to the proposed framework.}

Wherever B-S is applied to a heterogeneous portfolio, however, it has four structural limitations that can matter in practice.

\textbf{1. \(K\) is assumed the same for all accounts.}
In a relatively homogeneous portfolio, this is reasonable.
But a volatile manufacturer and a stable logistics fleet
of the same exposure size are not equivalent risks: the manufacturer's loss ratio
may swing widely year on year (high within-company variance, warranting low credibility),
while the fleet's may be consistent (low within-company variance, warranting high credibility).
Standard Bühlmann--Straub applies the same pooled \(K\) to both, regardless of how predictive
each account's own history actually is of its future experience: the fleet's history is
underweighted and the manufacturer's is overweighted.
The moment estimator has no way to distinguish the two.

\textbf{2. \(K\) and the complement are estimated independently of each other.}
B-S can in principle accept any complement: a segment mean, a size-varying benchmark, or a GLM output.
In standard form, however, the complement defaults to the portfolio grand mean,
which may mis-state the prior for accounts whose risk level differs systematically
from the portfolio average.
The deeper constraint is that \(K\) is estimated independently of the complement:
substituting a better complement without re-estimating \(K\) renders the blend
inconsistent, and there is no internal mechanism to compensate for a wrong complement.
The proposed framework estimates \(Z_i\) jointly with the complement parameters.\footnote{A related assumption is that an account's own experience and the complement are statistically independent. In practice they often share sources of correlation (estimation error in common trend, development and on-levelling factors applied to both), which the standard derivation does not account for.}

\textbf{3. Historical experience is treated as homogeneous in time.}
A five-year claims history is blended uniformly even if the account's
underlying risk changed substantially two years ago.
Under risk drift, old experience can be actively harmful: it pulls the
indicated rate toward a level that may no longer exist.
Practitioners can apply manual year weights or exponentially weighted moving
average (EWMA) decay rates, but these are typically chosen by judgement
rather than estimated from the data, and different actuaries working
from the same portfolio may arrive at different choices with no principled
basis for selecting between them.

\textbf{4. The credibility weight itself carries no uncertainty.}
A weight of \(Z = 0.45\) might be well-identified from a stable, seven-year
history, or it might be estimated with wide uncertainty from three years of
a volatile account.
These are materially different risk positions, but the B-S point estimate
provides no way to distinguish them.

\medskip

Experienced practitioners often address these limitations
through manual adjustments: stratifying \(K\) by risk class, down-weighting older
years, or replacing the grand mean with a model, segment, or market-level benchmark.
The theoretical literature has also proposed extensions: hierarchical credibility
models \citep{jewell1975, sundt1980} allow \(K\) to vary across levels of a
portfolio tree, and \citet{mahler1997} provides a comprehensive treatment of
credibility estimation methods used in practice.
Section~\ref{sec:bs-failure} tests simpler versions of these ideas empirically, including
a size-varying complement and a stratified \(K\).
The finding is not that B-S is wrong in principle. It is that each manual patch
addresses one limitation while leaving the others intact, and the patches may
interact in ways that are hard to control without a joint model.

Bühlmann--Straub persists in practice for good reasons.
The output is a named credibility weight \(Z_i\) with a direct, auditable interpretation, which is operationally appealing in a commercial lines environment.
The formula holds as the minimum-variance linear estimator under any distribution with finite second moments \citep{buhlmann1970, buhlmann2005}, with \(K\) estimated by method of moments: no likelihood specification, no iterative optimisation.
It is therefore desirable for a replacement to preserve these properties rather than trade them away for statistical flexibility alone.

We propose a direct, data-driven approach to all four limitations.
The key structural feature is that all parameters (credibility weights,
temporal decay rates, and, optionally, complement parameters) are estimated jointly
in a single model fit.
There is no separate K estimation step, no method-of-moments iteration,
and no practitioner-specified prior weights unless desired.
Practitioners who already operate a complement model (a GLM, a market
benchmark, or an underwriter-adjusted rate) can supply it as a fixed
input and estimate only the \(Z\) and \(\lambda\) parameters, as described
in Section~\ref{sec:applying}.
A Bayesian implementation also delivers a posterior distribution over each account's
credibility weight \(Z_i\), giving the underwriter
a defensible range within which to exercise judgement rather than a single
number to accept or override.

The framework proposed here is applicable wherever a practitioner would reach for Bühlmann--Straub. The empirical validation uses the CAS Loss Reserve Database \citep{meyershi2011} (96 US Commercial Auto companies) as the closest publicly available structural analogue, but data requirements are not specific to Commercial Auto or the US market.

\textbf{Why this framework rather than a mixed-effects model?}
The natural statistical alternative is a generalised linear mixed model (GLMM) with random effects on account
\citep{nelder1997, frees1999, antonio2007}, and such models achieve similar shrinkage via variance-component estimation.
GLMMs are a legitimate and well-studied modelling choice: in contexts where an explicit credibility weight is not required, or where data volumes support richer random-effect specifications, they can be the right tool.
In commercial lines pricing, however, an explicit \(Z_i\) is both analytically useful and practically valuable for explaining and challenging renewal prices with underwriters: it answers ``how much should I trust this account's history?'' in plain language. A GLMM's credibility weight is implicit: for a random-intercept model it is approximately of the B-S form with \(K = \phi/\sigma^2_u\) (\hyperref[app:implied-z]{Appendix~E}), recoverable from the variance components but not reported by standard software. Most consequentially for renewal pricing, a GLMM estimates a random effect only for accounts that appeared in its training data, so history supplied for a new account (for example, a broker's loss record) cannot be credibility-weighted without refitting. The logistic model applies a credibility weight set by the account's own exposure directly to whatever history it is given, without refitting existing pricing models (Section~\ref{sec:applying-renewal}). GLMMs also inherit the same \(K\)-homogeneity limitation as B-S by default, and require the analyst to specify the functional form of any temporal trend. The empirical comparison in Section~\ref{sec:glmm-comparison} tests predictive accuracy against a GLMM directly.

\textbf{Why not a GLM with experience as a predictor?}
A common practitioner alternative is to include historical experience directly as a fixed predictor in a pricing GLM.
Observed variants include: (i) a single experience term (e.g.~a rolling average loss ratio); (ii) an experience--exposure interaction, allowing the impact of experience to vary with account size; and (iii) pre-credibility-weighting the experience statistic before it enters the GLM, approximating shrinkage outside the model.
Variants (i) and (ii) were tested empirically on the CAS dataset (variant (iii) was not) and underperform both standard B-S (by 5--6\% in wMSE) and the proposed framework. The results are in \hyperref[app:full-comparison]{Appendix~D}. The structural reason is shared across all three variants: the lookback window \(W\) and any size-stratification of that window are discrete hyperparameters that sit outside the likelihood.
A practitioner can compare candidate windows by log-likelihood or held-out error, but allowing the window to vary jointly by account size requires fitting a separate model for each combination.
The joint framework internalises this: \(\lambda\) (introduced in Section~\ref{sec:decay}) is a continuous parameter estimated within the same MLE pass as \(Z\) and the complement, so the optimal effective lookback and any size stratification are determined by the data, with no grid search required.

The logistic credibility framework therefore occupies a deliberate position:
it combines the \textbf{interpretability, transparency, and actuarial
tractability of Bühlmann--Straub} with the \textbf{statistical flexibility
of a mixed-effects approach}. These characteristics are well suited to the data
conditions of commercial insurance, where account histories are often short and
portfolio sizes can be measured in hundreds or thousands rather than millions. They also
support underwriter communication through explicit, auditable numbers
that can reflect underwriters' expertise and input.
Table~\ref{tab:comparison} summarises the comparison.

\begin{table}[H]
\centering
\small
\setlength{\tabcolsep}{8pt}
\caption{Comparison of credibility approaches for commercial insurance
  experience rating.
  \checkmark\ = satisfied;\enspace
  $\sim$ = possible but requires careful specification;\enspace
  $\times$ = not satisfied.
  $^*$A Normal--Normal GLMM with REML variance components nests B-S
  algebraically; Gamma, Poisson, and Tweedie GLMMs do not.
  The logistic model's nesting result (Section \ref{sec:bs-nesting}) holds
  regardless of the chosen likelihood, since the $Z_i$ formula depends
  only on exposure, not the distributional form.}
\label{tab:comparison}
\begin{tabular}{p{6.5cm}ccc}
\toprule
& \textbf{Bühlmann--Straub} & \textbf{GLMM} & \textbf{Logistic credibility} \\
\midrule
Explicit credibility weight $Z_i$          & \checkmark & $\sim$   & \checkmark \\
$K$ varies by size, sector, or any covariate       & $\times$   & $\sim$   & \checkmark \\
Temporal weighting, form-free              & $\times$   & $\sim$     & \checkmark \\
Nests rolling Bühlmann--Straub exactly     & ---        & $\sim$$^*$ & \checkmark \\
Three-number underwriter output            & \checkmark & $\times$   & \checkmark \\
Posterior uncertainty on $Z_i$ (Section \ref{sec:uq}) & $\times$   & $\sim$     & \checkmark \\
\bottomrule
\end{tabular}
\end{table}

Before presenting the framework, we document these limitations on real data:
because the empirical evidence is the most direct motivation.

\subsection{Four Limitations of Bühlmann-Straub: A Motivating Example}\label{sec:bs-failure}

To illustrate how the four limitations above manifest on real data, we applied both Bühlmann--Straub and the proposed framework to the Casualty Actuarial Society (CAS) Commercial Auto loss-ratio database \citep{meyershi2011}: 96 US insurers (the accounts) with a complete ten-year history of accident-year (AY) loss ratios (AY 1998--2007), evaluated against a common portfolio baseline.
We train on AY 2001--2005, with AY 1998--2000 supplying lookback history, and hold out AY 2006--2007 (192 company-year observations).
Net earned premium (NEP) is the credibility exposure throughout, as accident-year cohorts call for; gross written premium (GWP) would be the analogue for underwriting-year cohorts. Section~\ref{sec:empirical} gives the full description.

B-S produces miscalibrated predictions on this data in three separable, concrete ways, and a fourth limitation concerns the absence of uncertainty quantification on the credibility weight itself.

\begin{figure}[H]

{\centering \includegraphics[width=0.95\linewidth]{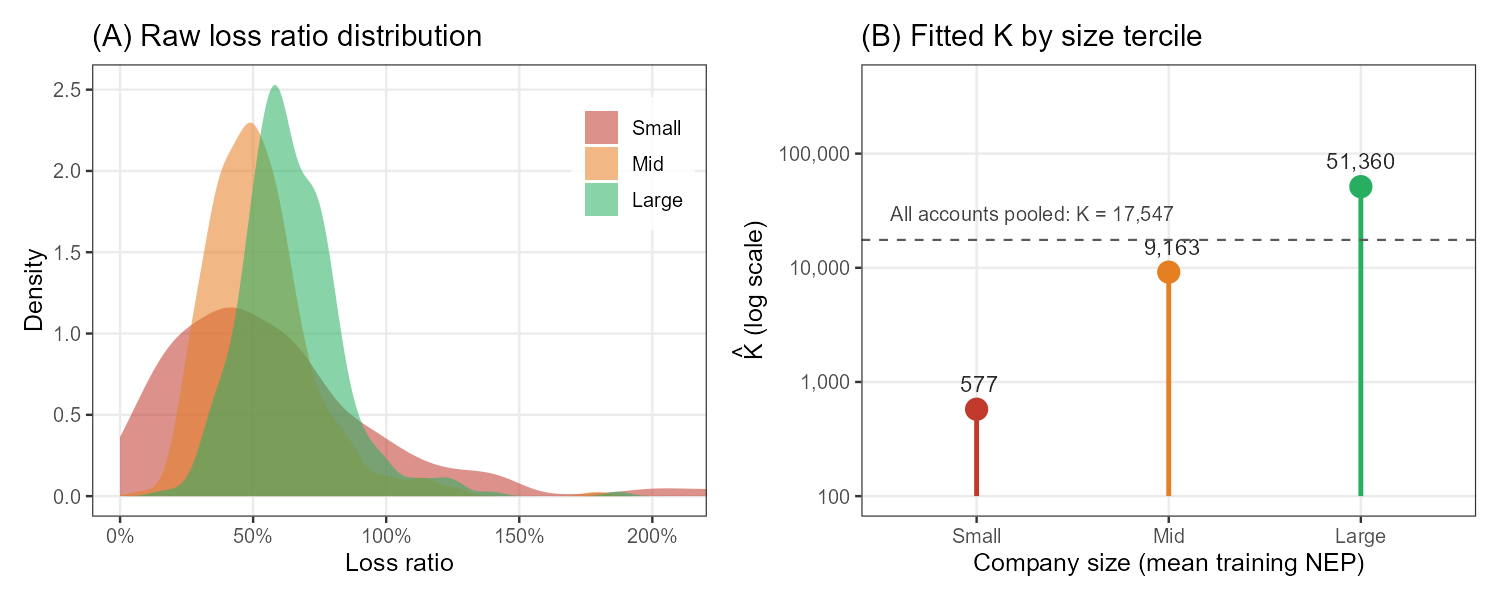} 

}

\caption{Loss ratio distributions and fitted credibility parameter by size tercile (Small: red, Mid: orange, Large: green), using the same tercile definition (mean training-year NEP) as the main results. (A) Raw loss ratio density across all accident years (x-axis truncated for readability; Small accounts' right tail extends further). (B) Fitted $\hat K$ (log scale) when estimated separately within each tercile. Small: 577, Mid: 9,163, Large: 51,360; dashed line: pooled $\hat{K}$ = 17,547 for all accounts.}\label{fig:fig-intro-lr-dist}
\end{figure}

\textbf{Limitation 1: a single pooled \(K\) does not fit all account sizes.}
B-S estimates one portfolio-wide \(K\), implicitly assuming the ratio of between-account variance to within-account (process) variance is constant across the portfolio.
Small and large accounts show visibly different raw loss-ratio dispersion (Figure~\ref{fig:fig-intro-lr-dist}, panel A), as expected if variance falls with exposure (\hyperref[app:variance]{Appendix~C} examines how quickly).
However, once we estimate \(K\) directly, we find that the constant-\(K\) assumption is violated: fitting \(K\) separately within each size tercile (companies grouped into three equal-sized bands by mean NEP) gives \(\hat K \approx 577\) for Small, \(9{,}163\) for Mid, and \(51{,}360\) for Large, against \(17{,}547\) for the portfolio pooled (Figure~\ref{fig:fig-intro-lr-dist}, panel B), a spread of roughly \(90\times\) across size groups.

\begin{figure}[H]

{\centering \includegraphics[width=0.85\linewidth]{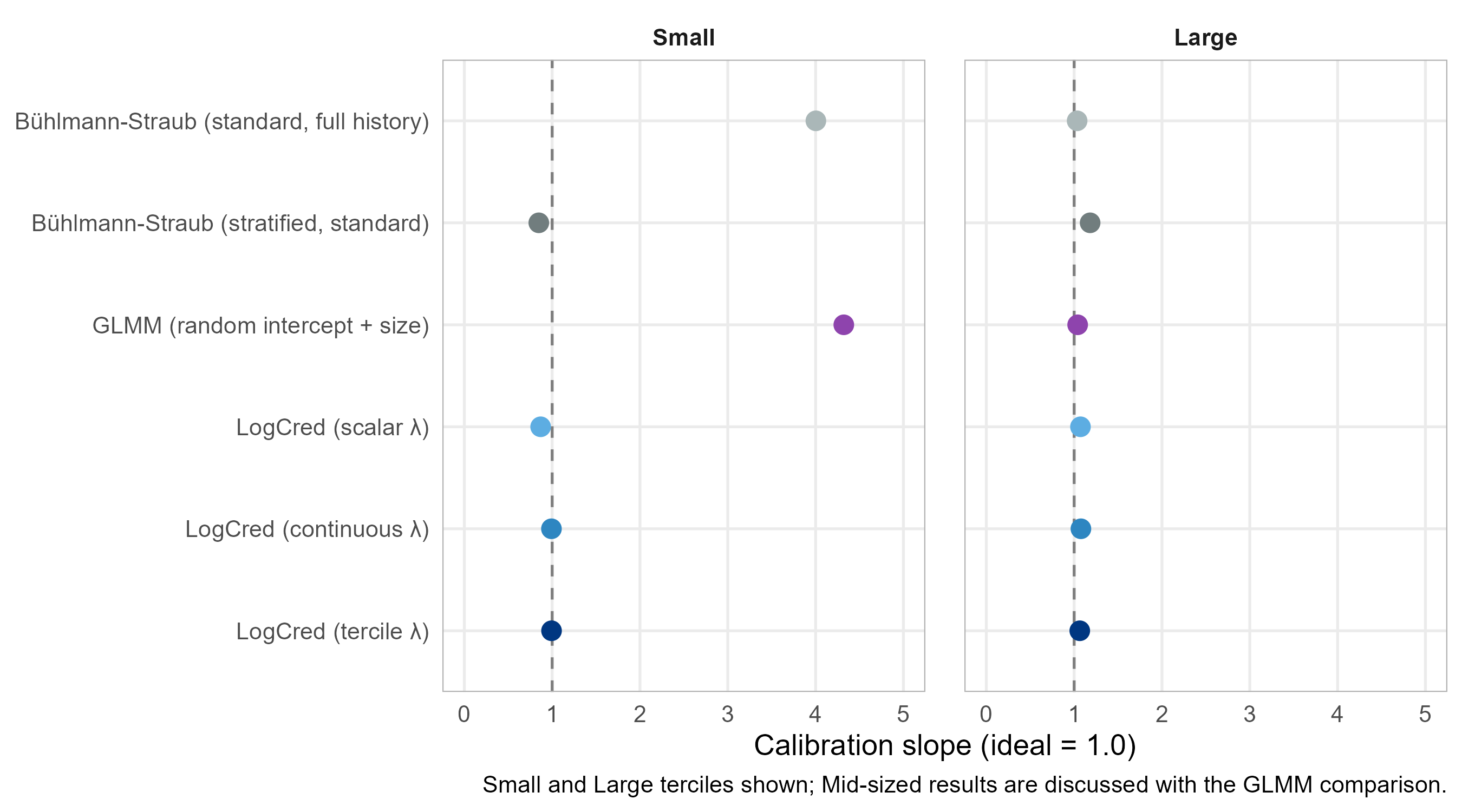} 

}

\caption{Calibration slope (ideal = 1.00) for small and large accounts, held-out test set (AY 2006--2007). Bühlmann--Straub (greys): 4 (90\% CI: 2.5--4.9) small, 1.03 large, so the under-crediting is concentrated in small accounts. Logistic credibility (blues): 0.99 small, 1.06 large. Mid-sized results in Section \ref{sec:empirical}.}\label{fig:fig-intro-slope}
\end{figure}

The pooled estimator (\(\hat K \approx 17{,}547\)) sits between the Mid and Large strata's own values, so it over-shrinks the smallest accounts and gives the largest more credibility than their own stratum's \(K\) would. Under \(Z_i = E_i/(E_i + \hat K)\), a Small account of typical size receives \(\hat{Z} \approx 0.16\) under the pooled \(\hat K\) but \(\hat{Z} \approx 0.85\) under its own stratum's \(\hat K\); a typical Large account moves from \(0.93\) to \(0.82\).
The consequence is visible in the held-out calibration slope\interfootnotelinepenalty=10000\footnote{The calibration slope is the exposure-weighted OLS slope of actual on predicted outcomes (ideal 1.0); above 1.0 indicates under-crediting (predictions compressed toward the complement).}\interfootnotelinepenalty=100{} (Figure~\ref{fig:fig-intro-slope}): for small accounts it is well above 1.0 (about 4), so predictions are compressed toward the complement and each account's own experience carries too little weight.

\textbf{Limitation 2: a grand mean complement introduces systematic bias.}
The standard moment estimator uses the portfolio grand mean as the complement
for every account.

\begin{figure}[H]

{\centering \includegraphics[width=0.65\linewidth]{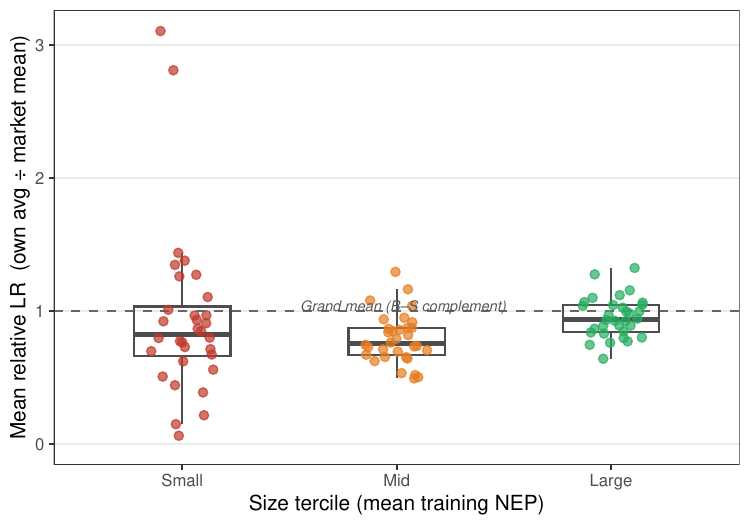} 

}

\caption{Distribution of company-level mean relative loss ratio (own multi-year mean ÷ market mean) by size tercile, training AY 2001--2005. The dashed line at 1.0 is the grand-mean complement B-S uses for low-credibility accounts. Small and Mid cluster below it (medians $\approx 0.83$, $0.76$); reverting to 1.0 biases the group means (the relevant comparison) upward by $\approx 6\%$ (Small) and $\approx 26\%$ (Mid).}\label{fig:fig-intro-complement}
\end{figure}

When B-S assigns a low credibility weight to a small account, it reverts
that account to the exposure-weighted portfolio mean, approximately 1.0
in relative (to market) loss-ratio space. This benchmark mostly reflects large
companies' risk profile, since large companies dominate the weighting.
In this dataset, small and mid-sized companies typically sit \emph{below} this mean, though the distributions have a long right tail from a few higher-outlier companies (Figure~\ref{fig:fig-intro-complement}).
Reverting a low-credibility small or mid-sized account to 1.0 therefore introduces a systematic
upward bias for most, though not all, such accounts.

\begin{figure}[H]

{\centering \includegraphics[width=0.82\linewidth]{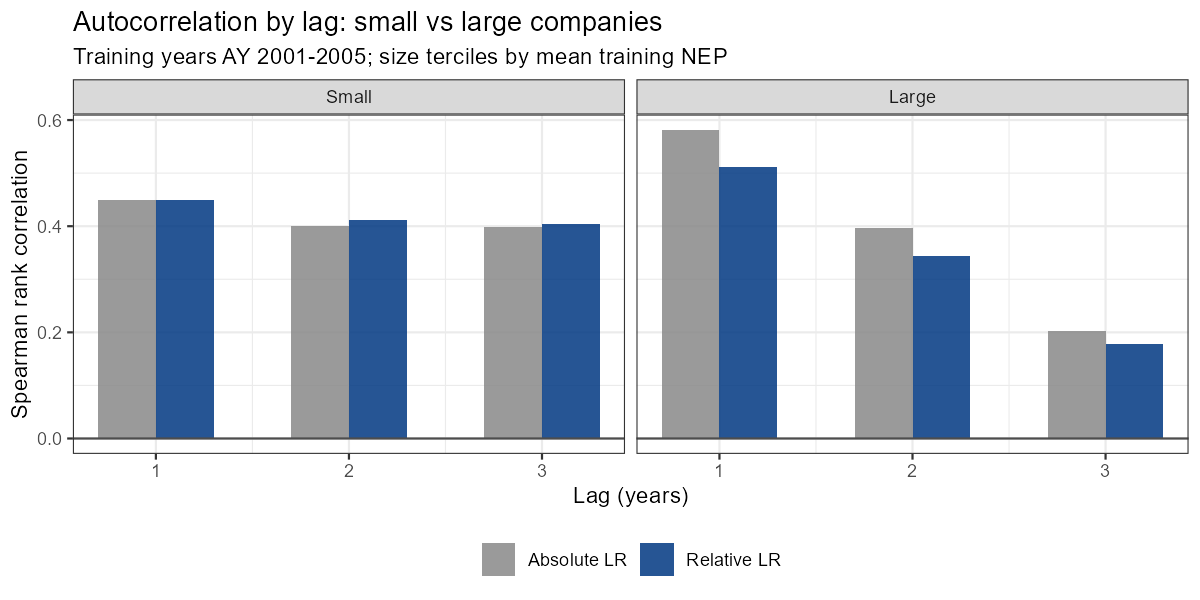} 

}

\caption{Year-on-year rank correlation of absolute (grey) and relative (blue) loss ratios by lag, split by company size (training years AY 2001--2005; size terciles by mean training NEP). Small companies (left): relative-LR autocorrelation stays high across lags ($\rho$ from 0.45 to 0.40); older years remain almost as informative as the most recent. Large companies (right): strong lag-1 signal ($\rho \approx 0.51$) that falls to $\rho \approx 0.18$ by lag 3; the most recent year dominates. B-S applies the same equal-weight average to both groups. Quoted $\rho$ values throughout refer to the relative-LR (blue) series.}\label{fig:fig-intro-autocorr}
\end{figure}

\textbf{Limitation 3: all historical experience years carry equal weight.}
B-S uses an exposure-weighted average of all available loss years with equal weights, treating all years as equally informative regardless of how old they are.
Yet Figure~\ref{fig:fig-intro-autocorr} shows the empirical autocorrelation
structure differs sharply by account size:
large companies show a strong lag-1 signal in the relative loss ratio (\(\rho \approx 0.51\)) that
decays rapidly by lag 3 (\(\rho \approx 0.18\)), so their most recent
year is by far the most informative.
Pooling all past years with equal weight dilutes this signal.
Small companies show relative-LR autocorrelation that stays high across lags (\(\rho \approx 0.45\) at lag 1 and 0.40 at lag 3),
indicating that equal weighting is a reasonable assumption.
The two size groups need fundamentally different lookback structures; a
single equal-weight rule fits neither as well.

\begin{figure}[H]

{\centering \includegraphics[width=0.95\linewidth]{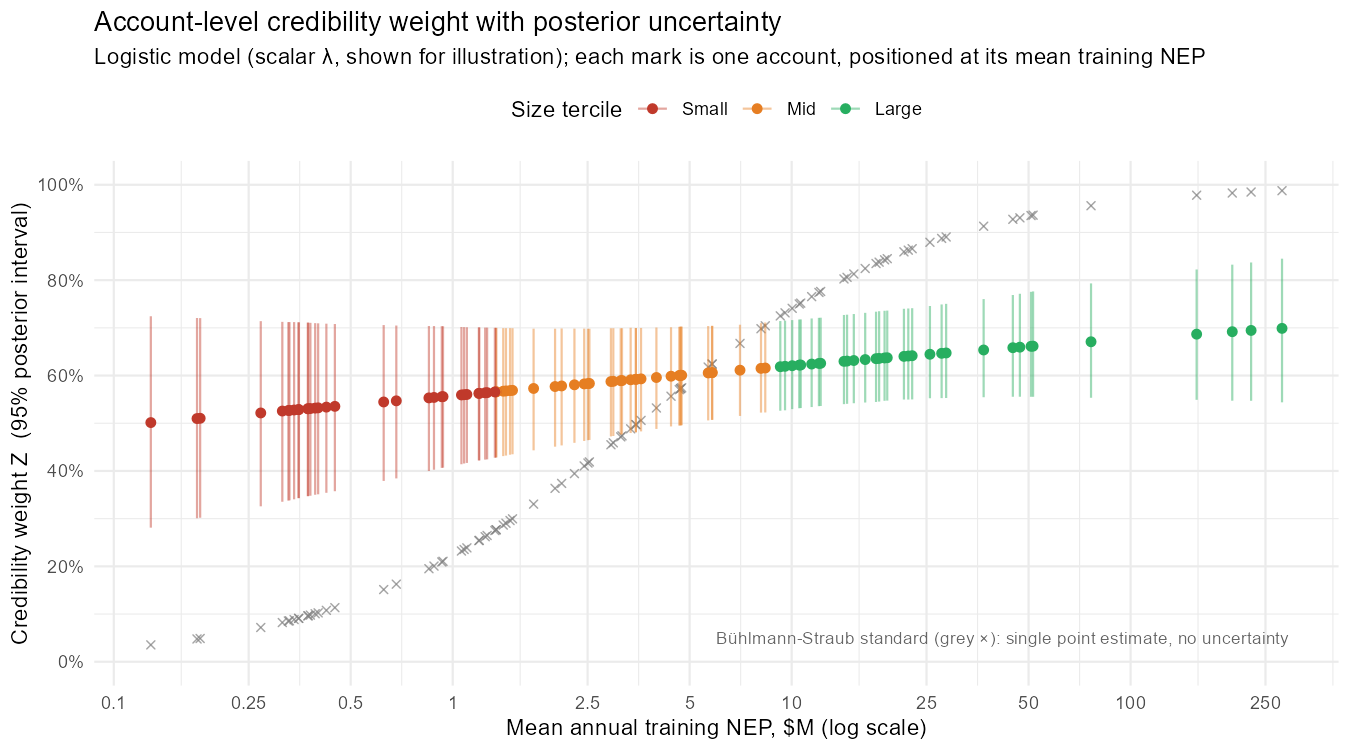} 

}

\caption{Account-level credibility weight $\hat{Z}_i$ against mean annual training NEP (log scale). Coloured points: logistic model (scalar $\lambda$, shown for illustration; $Z$ evaluated at mean training NEP as a stand-in for the lookback exposure it is fitted on) with 95\% posterior CI. Grey crosses: Bühlmann--Straub point estimate (no uncertainty). Small accounts (red, left): wide intervals, reflecting genuine uncertainty about how much experience to trust. Large accounts (green, right): narrow intervals, so departures from the model estimate carry a high evidential burden.}\label{fig:fig-emp-z-vs-nep}
\end{figure}

\textbf{Limitation 4: no quantification of estimation uncertainty.}
Even if B-S produced correct point estimates, it would still give no
indication of how reliable those estimates are.
Figure~\ref{fig:fig-emp-z-vs-nep} shows the posterior distribution of
credibility weights from the proposed framework: small accounts (red)
carry wide intervals (underwriter judgement has genuine scope)
while large accounts (green) are tightly identified.
B-S collapses this entire distribution to a single number.
The two curves also differ sharply in level, not just uncertainty: small accounts attract substantially higher credibility, and large accounts lower, under the logistic model than under pooled-\(K\) standard B-S, the \(b_Z < 1\) pattern derived in Section~\ref{sec:bs-nesting} and reported formally in Section~\ref{sec:results-main}.
Whether to treat this lack of uncertainty quantification as a limitation depends on the use case: distinguishing accounts where the weight is well-identified from those where it is uncertain can influence how much underwriter judgement is appropriate.

\textbf{Can the limitations be addressed by sequential patches?}
The empirical study in Section~\ref{sec:empirical} tests this directly.
On the CAS Commercial Auto panel we implemented one version of these patches, and they do improve overall wMSE: stratifying \(K\) by size reduces it by 6\% overall (a 41\% reduction for small accounts, which carry only 1.2\% of the exposure, and 9\% for Mid, with Large 1\% higher); adding a size-varying complement and EWMA with tercile \(\lambda\) (Section~\ref{sec:decay}) improves it a further 20 percentage points, reaching a 26\% reduction overall.

But the EWMA step creates a cross-size trade-off: small-account wMSE \emph{rises by 34\%} relative to the previous step while large accounts improve strongly, and mid accounts are essentially unchanged.
The underlying issue appears to be that stratified-\(K\) B-S assigns small accounts high credibility (\(\hat{Z} \approx 0.8\)), so a recency weighting that fits the training years but not the test years passes almost fully into the prediction; re-calibrating \(K\) or the complement to the EWMA experience does not remove this (\hyperref[app:full-comparison]{Appendix~D}).

\begin{figure}[H]

{\centering \includegraphics[width=1\linewidth]{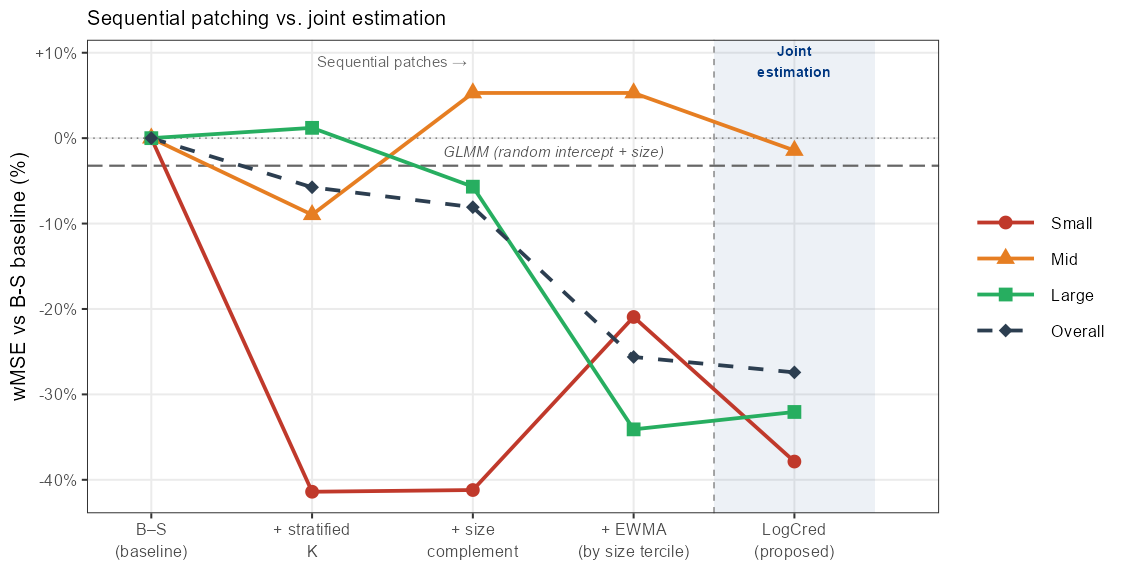} 

}

\caption{Prediction error (wMSE, \% change vs standard B-S) at each sequential patch step, by size tercile. "+ EWMA" is the fully patched B-S (tercile-stratified $K$, continuous size complement, tercile $\lambda$). Small accounts (red) worsen by 34\% at the "+ EWMA" step (20 percentage points on this scale), Mid accounts (orange) by 16\% at the "+ size complement" step, which also switches the experience term to the lookback window, while Large accounts (green) improve most at "+ EWMA". The logistic model with tercile $\lambda$ (right, shaded) estimates all components jointly. The GLMM+size benchmark (grey dashed) is shown for reference.}\label{fig:fig-intro-patching}
\end{figure}

The logistic framework, which estimates \(Z\), the complement and \(\lambda\) together, avoids this in the case study by assigning Small a lower credibility weight than the stratified-\(K\) patch (0.62 against 0.85), restoring small-account performance and achieving a 27\% wMSE reduction (with an oracle complement, i.e.~the realised future portfolio mean, unavailable in practice; 25\% with a deployable forecast complement). The fully patched B-S comes close overall (26\%), so the case for the logistic framework rests less on the headline gain than on avoiding the small-account reversal in wMSE and the chain of ad-hoc steps (Figure~\ref{fig:fig-intro-patching}).

Estimating the components under one likelihood has further advantages. Because the likelihood is explicit, the data show which B-S assumptions to relax, and the continuous slope \(b_Z\) on log-exposure gives the size adjustment to \(K\) directly, without arbitrary cut-offs. A patched B-S has neither a likelihood for testing such choices nor predictive intervals; the joint model supplies both.\footnote{Sequential patching is also methodologically inconsistent. Two actuaries working from the same data can reach different patched specifications with no principled basis for choosing between them. The three steps also use three different loss functions: method of moments for $K$ (variance matching), weighted least squares on the log complement (implicitly Gaussian on the log scale) and minimised wMSE for $\lambda$ (Gaussian on the raw scale). None matches a single likelihood for the response, and with Gamma loss ratios none is the Gamma MLE. This is independent of the small-account issue above and is a further reason to prefer joint estimation under a single likelihood.}

The logistic framework therefore addresses all four limitations in a single, estimable model.

\textbf{Relationship to prior literature.}
Prior extensions to Bühlmann--Straub have allowed \(K\) to vary across pre-specified strata \citep{jewell1975, sundt1980, mahler1997} or incorporated regression structure \citep{hachemeister1975}, but each requires the heterogeneity structure to be defined before estimation and estimated sequentially, precluding formal joint tests. In empirical Bayes terms, B-S is the normal--normal case with a fixed shrinkage form; a regression prior mean and a prior variance that depends on a size measure have precedent in \citet{fayherriot1979}, which we combine here with temporal decay and joint estimation.
The literature has further extended B-S via GLMM-based pooling \citep{nelder1997, antonio2007}, geometric- and exponential-smoothing-based weighting schemes \citep{sundt1988, kremer1990}, and autoregressive specifications on the random effects \mbox{\citep{bolance2003}}. A different use of logistic regression in credibility is the Logistic Regression Credibility model of \citet{jahanbani2024}, where a logistic regression gives the probability that a claim belongs to a component of a finite mixture and the Bayesian credibility mean follows from the mixture; here the logistic function parameterises the credibility weight itself.

This paper makes four contributions:

\begin{enumerate}
  \item \textbf{Joint estimation.} $Z_i$, the temporal decay rate $\lambda$, and the complement are estimated jointly under a single likelihood. To our knowledge this is the first such joint framework in the credibility literature, and in the case study it avoids the small-account deterioration that occurred when recency weighting was patched onto B-S sequentially.
  \item \textbf{Formal nesting result.} The logistic $Z$ generalises B-S's fixed exposure-slope relationship to one estimated from data, while containing rolling B-S (B-S over the same lookback window) exactly as a restricted special case (Section \ref{sec:bs-nesting}), providing a likelihood-ratio test for whether the heterogeneous-$K$ extension is actually warranted, rather than assumed.
  \item \textbf{Temporal decay gradient.} An empirical finding that the optimal temporal decay rate applied to past experience varies by account size, consistent with a law-of-large-numbers mechanism; Other Liability shows the same direction, with $\hat\lambda$ falling monotonically with size.
  \item \textbf{Bayesian uncertainty quantification.} A Bayesian implementation delivering a posterior distribution over each account's credibility weight, giving underwriters a defensible range for judgement rather than a single number to accept or override.
\end{enumerate}

The approach preserves the named, interpretable \(Z_i\) that makes B-S easy to explain and challenge in underwriting.

\textbf{Relationship to machine-learning approaches.}
Gradient boosting and neural networks could in principle capture non-linear effects in \(Z\), but even with post-hoc feature-attribution tools, neither yields an explicit, auditable credibility weight comparable to \(Z_i\), and renewal prices are often expected to be defensible in plain language. Data volume reinforces this: a 96-company panel with three to five years per account is already at the limit of stable identification for the logistic model's five to eight parameters, let alone a more flexible model's larger (effective) parameter count.
Neural architectures can embed similar credibility mechanisms directly (the Credibility Transformer of \citet{richman2025} integrates a credibility-weighted blend into a Transformer encoder) but at the cost of the named, auditable \(Z_i\).

\textbf{Paper structure.}

\begin{itemize}
\tightlist
\item
  \textbf{Section~\ref{sec:framework}} presents the logistic credibility framework: the core blend formula, the logistic \(Z\) function and its feature selection logic, the B-S nesting result (Section~\ref{sec:bs-nesting}), the temporal decay extension, and a three-tier uncertainty hierarchy.
\item
  \textbf{Section~\ref{sec:empirical}} validates the framework on 96 US Commercial Auto companies: prediction accuracy, calibration by size, the temporal decay gradient, and LOO-CV (leave-one-out cross-validation) model selection.
\item
  \textbf{Section~\ref{sec:simulation}} confirms the mechanisms under controlled conditions via three simulation scenarios (a standard B-S sanity check, temporal drift and industry-heterogeneous \(K\)).
\item
  \textbf{Section~\ref{sec:practical}} provides a self-contained implementation guide: a full fitting and renewal guide (Sections~\ref{sec:applying}--\ref{sec:applying-renewal}), a worked example on two real companies (Section~\ref{sec:worked-example}), and a Minimum Viable Deployment checklist (Section~\ref{sec:mvd}) for immediate deployment.
\item
  \textbf{Section~\ref{sec:conclusion}} concludes.
\end{itemize}

\section{The Framework}\label{sec:framework}

\subsection{The Core Formula}\label{sec:core-formula}

The predicted renewal rate for account \(i\) is a blend of the model price
and the account's own claims experience:

\begin{equation}
  \hat{r}_i = \bigl(1 - Z_i\bigr)\,\mu_i
             + Z_i\,\hat{f}_i,
  \label{eq:pred}
\end{equation}

where \(\mu_i\) is the complement (the prior estimate for account \(i\)
in the absence of its own experience) and \(\hat{f}_i\) is the
account's own claims history rate, exponentially weighted to favour recent years (Section~\ref{sec:decay}), both expressed on the same
scale (e.g.~frequency, pure premium, or loss ratio).
This blend formula is identical in form to Bühlmann--Straub, with the innovation lying entirely in how \(Z_i\) and \(\hat{f}_i\) are determined, as described in the following sections.\footnote{Hats mark estimators and fitted values; $Z_i$ and $\mu_i$ are unhatted where they define the model.}

The precision required of \(Z_i\) scales with how far apart the complement and experience already are: \(\partial\hat r_i/\partial Z_i = \hat f_i - \mu_i\), so a mis-specified \(Z_i\) moves the prediction only in proportion to this gap. Where \(\mu_i\) and \(\hat f_i\) are already close, even a crude credibility rule performs adequately; the framework's value is concentrated in the accounts where they diverge (see the gap check, Section~\ref{sec:mvd}), consistent with Clark's \citep{clark2025} observation that credibility weighting is forgiving of imprecision once a reasonable complement is chosen.

\textbf{Arithmetic vs geometric blend.}
The formulation above is the \textbf{arithmetic blend}.
An alternative \textbf{geometric blend} operates on the log scale:
\(\log\hat{r}_i = (1-Z_i)\log\mu_i + Z_i\log\hat{f}_i\),
equivalently \(\hat{r}_i = \mu_i^{1-Z_i}\cdot\hat{f}_i^{Z_i}\).
The arithmetic form is the natural B-S analogue and is preferred for underwriter interpretability. The geometric form is more natural when \(\mu_i\) comes from a multiplicative (log-link) GLM and treats proportional deviations above and below the complement symmetrically.
The choice must be fixed before fitting; the \(Z\)-function structure, nesting result, and estimation procedure are identical under either form.\footnote{The geometric form has one limitation: if an account has zero losses in all lookback years, \(\log\hat{f}_i = -\infty\); the arithmetic blend handles this gracefully.}

The complement \(\mu_i\) can be a base rate from a pricing GLM, a segment or portfolio mean, or any other actuarially appropriate prior.
When \(\mu_i\) is the output of an existing GLM, the credibility layer applies directly on top without re-estimating the base model: the blend simply adjusts each account's GLM rate toward its own experience in proportion to how much history it has accumulated.

Rather than computing \(Z_i = E_i/(E_i + K)\) with a single pooled \(K\),
we estimate the credibility weight as a logistic function of observable
account characteristics:

\begin{equation}
  Z_i = \Lambda\!\left(a_Z + \mathbf{b}_Z^\top \tilde{\mathbf{x}}_i\right),
  \label{eq:logistic}
\end{equation}

with \(\Lambda(x) = 1/(1+e^{-x})\) the logistic function and
\(\tilde{\mathbf{x}}_i\) a vector of standardised account-level features.
In the single-covariate case with \(\tilde{\mathbf{x}}_i = \log\tilde{E}_i\), the intercept and slope are the scalars \(a_Z\) and \(b_Z\).
Log exposure (specifically, log cumulative lookback exposure
\(\log\tilde{E}_i\), where \(\tilde{E}_i = \sum_{k=1}^{W} E_{i,t-k}\) and \(E_{it}\) is account \(i\)'s exposure in year \(t\)) is
a structural component of \(\tilde{\mathbf{x}}_i\).
It is the direct analogue of Bühlmann--Straub's exposure weight: larger
accounts are expected to accumulate more exposure, giving a lower-variance
estimate of their true underlying rate, so their experience is therefore more credible.

The logistic function is the natural choice for three reasons.
First, it maps any real-valued linear score to \((0,1)\), so \(Z_i\) is a probability by construction, requiring no clipping or post-hoc normalisation.
Second, with intercept \(a_Z = -\log K\) and slope \(b_Z = 1\) on \(\log\tilde{E}_i\), it reproduces the Bühlmann--Straub weight structure \(\tilde{E}_i/(\tilde{E}_i+K)\) exactly (Section~\ref{sec:bs-nesting}), enabling a formal likelihood-ratio test for whether freeing \(b_Z\) is warranted.
Third, with two parameters (\(a_Z\), \(b_Z\)), it is a minimal extension over B-S's single pooled \(K\). Unlike stratifying \(K\) by exposure band, it gives credibility weights that vary smoothly with exposure, and further covariates (industry, broker channel, etc.) can be added to the \(Z\) equation as the portfolio supports.

Note that \(\tilde{E}_i\) is the \emph{lookback-period} exposure, not the
full accumulated exposure used by Bühlmann--Straub; this distinction
matters for growing or shrinking accounts and is discussed further in
Section~\ref{sec:empirical}.

Account size is operationalised throughout by this exposure measure (net earned premium in the case study, but the framework applies equally to vehicle-years, turnover, TIV, headcount, or any volume measure appropriate to the portfolio).

\medskip

\noindent\textit{Notation summary for exposure variables used in this paper:}

\begin{itemize}\setlength\itemsep{0pt}
  \item $E_{it}$: current-year (or row-level) exposure; may weight the likelihood, depending on the target variable and likelihood family. B-S weights by exposure ($\mathrm{Var} \propto 1/E_{it}$); the case study's variance falls more slowly, so exposure-varying dispersion is an alternative (\hyperref[app:variance]{Appendix~C}). $E_{it}$ may also enter the complement, depending on its specification (Section \ref{sec:eda}).
  \item $\tilde{E}_{it} = \sum_{k=1}^{W} E_{i,t-k}$: cumulative lookback exposure over the $W$-lag window; enters the $Z$ logistic as the credibility exposure signal. (Suppressed to $\tilde{E}_i$, as are $Z_i$, $\mu_i$, $\hat{f}_i$ and $\hat{r}_i$, in subsequent notation for readability; strictly all are $t$-indexed since the lookback window shifts each prediction year.)
  \item $\bar{E}_i$: mean training exposure; may enter the temporal decay model (Section \ref{sec:decay}), as in this paper's case study, to capture a structural size--recency relationship.
\end{itemize}

Beyond log of lookback exposure, any characteristic informative about how much weight to
give an account's own history is a legitimate additional feature.
Natural enrichments include:

\begin{itemize}
\tightlist
\item
  \(\tilde{n}_i\): number of years of claims history available
\item
  \(\tilde{\kappa}_i\): coefficient of variation of past annual loss experience
  (a measure of stability, but noisy when history is short)
\item
  Industry segment indicators (e.g.~heavy industry vs.~light manufacturing
  vs.~chemical processing), illustrated by the industry-heterogeneity
  simulation in Section~\ref{sec:simulation}
\item
  Geographic territory or any other characteristic that plausibly affects
  how predictable the account's experience is
\end{itemize}

The framework learns the weight of each feature from the portfolio data.
The right set of enrichments depends on what data the practitioner has available
and what the portfolio is large enough to identify, as the empirical
study in Section~\ref{sec:empirical} demonstrates.

The linear predictor \(a_Z + \mathbf{b}_Z^\top \tilde{\mathbf{x}}_i\)
in equation \eqref{eq:logistic} is a modelling choice, not a theoretical
requirement.
Where the portfolio is large enough to support more flexibility, the linear
term \(\beta\log\tilde{E}_i\) can be replaced by a piecewise-linear or spline
function of \(\log\tilde{E}_i\):
\[
  Z_i = \Lambda\!\bigl(f(\log\tilde{E}_i)\bigr),
  \quad f \text{ a smooth function estimated from data.}
\]
The empirical case study uses the linear form throughout; Figure
\ref{fig:fig-z-shape} (Section~\ref{sec:ffvalidation}) confirms that free
exposure-tercile estimate ranges for \(\hat{Z}_i\) fall within the linear model's confidence
band on this dataset, providing limited evidence that a more flexible shape is needed.
For larger portfolios or settings with a pronounced non-linear exposure--credibility
relationship, the spline extension is a natural next step, albeit with the
potential cost of lower interpretability.

\begin{figure}[H]

{\centering \includegraphics[width=0.8\linewidth]{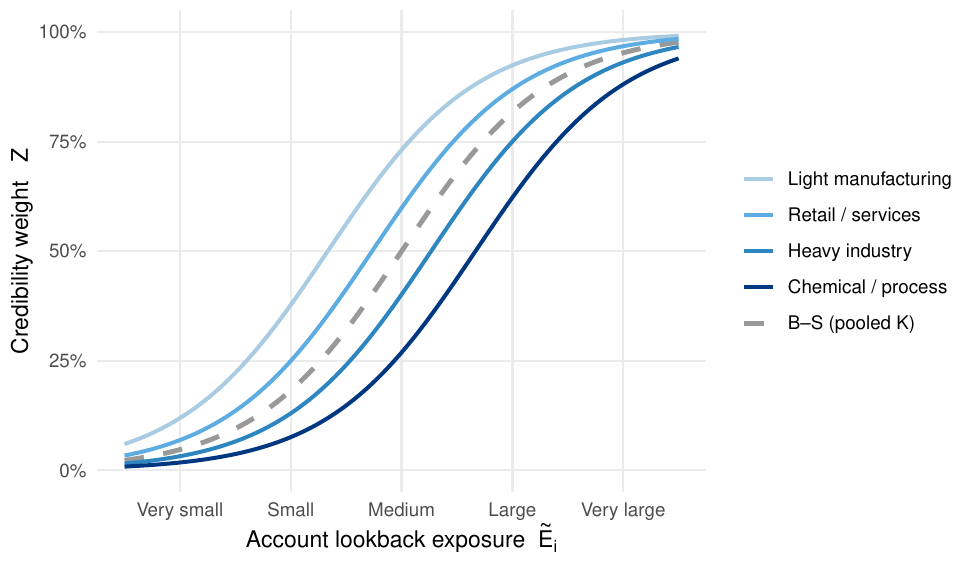} 

}

\caption{Illustrative credibility weight curves by industry segment. Each coloured curve shows $Z_i = \Lambda(a_{Z,\text{ind}} + b_Z\,\log\tilde{E}_i)$ for a different industry (light manufacturing, retail/services, heavy industry and chemical/process), with the intercept $a_{Z,\text{ind}}$ varying to reflect different inherent loss experience volatilities, and a slope shared with B-S, so each curve is a B-S curve with its own effective $K$. B-S by default applies a single pooled $K$ across all segments (dashed grey). Parameters do not correspond to the empirical case study.}\label{fig:fig-z-industry}
\end{figure}

Figure~\ref{fig:fig-z-industry} illustrates the effect of including an
industry indicator.
Each industry receives its own \(Z\) curve based on its own effective \(K\),
estimated jointly from the data.
Bühlmann--Straub can accommodate this with manual stratification, but by
default applies a single pooled \(K\) across all industries, represented by the
dashed reference line.

The model parameters are estimated from portfolio data by maximising a
log-likelihood, which may be weighted by exposure depending on the target
variable and likelihood family.
For loss-ratio targets (Gamma, Tweedie), explicit exposure
weights reflect the fact that larger accounts contribute more
information (in proportion to exposure under B-S; see \hyperref[app:variance]{Appendix~C}); for frequency targets (Poisson, negative binomial), the
exposure typically enters as an offset inside the likelihood rather than as an
external weight.
The appropriate likelihood depends on the target:
Gamma for aggregated loss ratios or individual severities (continuous,
strictly positive);
Tweedie for individual loss costs or loss ratios where zero-loss years
are possible (compound Poisson--Gamma, mass at zero plus positive tail);
Poisson or negative binomial for claim frequency (the latter accommodating
overdispersion that Poisson assumes away).

The simplest estimator is maximum likelihood (MLE):
\begin{equation}
  \hat\theta_{\mathrm{MLE}} = \arg\max_\theta
    \sum_i w_i \log p(y_i \mid \hat{r}_i(\theta)),
  \label{eq:mle}
\end{equation}
where \(w_i\) are exposure weights.
MLE is straightforward to implement (e.g.~via \texttt{nlminb} or
\texttt{L-BFGS-B} in statistical software \texttt{R} \citep{rcoreteam2024}) and delivers well-calibrated estimates whenever
the portfolio is large enough to identify all parameters.
When parameters are weakly identified (as \(\lambda\) can be for
small-account sub-portfolios), adding weakly informative priors
stabilises the fit without meaningfully biasing well-identified parameters.
The prior on \(b_Z\) (\(\mathrm{N}(+0.5,\,0.5)\), centred positive) encodes the structural constraint that credibility increases with exposure, consistent with B-S theory (larger accounts have lower between-year variance relative to their cumulative exposure, so their history is more informative and deserves higher weight):
\begin{equation}
  \hat\theta_{\mathrm{MAP}} = \arg\max_\theta
    \left[\sum_i w_i \log p(y_i \mid \hat{r}_i(\theta))
    + \log p(\theta)\right].
  \label{eq:map}
\end{equation}
Examples of both MLE and MAP estimation (via \texttt{nlminb}) are shown in the
listings of \hyperref[app:implementation]{Appendix~G}.
For uncertainty quantification (Section~\ref{sec:empirical-uq}), the same model
can be run via Hamiltonian Monte Carlo (HMC, implemented via \texttt{brms} \citep{burkner2017}/Stan \citep{stan2024}),
which delivers a full posterior distribution over \(\theta\);
posterior means can then serve as point estimates (in well-specified models
with informative data they typically lie close to the MLE, though this
is not guaranteed).

This is a one-step fit: no separate credibility estimation, no
method-of-moments iteration.
For the CAS case study the optimisation completes in seconds via \texttt{nlminb}.

In the limit, very small accounts tend toward \(Z_i \approx 0\): their
thin history carries little weight and the complement dominates.
Very large, stable accounts tend toward \(Z_i \approx 1\): their history
is sufficiently informative that the model assigns it near-full weight.
In practice, for a commercial portfolio of moderate-sized accounts,
most \(Z_i\) values will lie in an intermediate range rather than at the
extremes.

\subsection{Choosing Features}\label{sec:zfeatures}

\textbf{Log of lookback exposure (\(\log\tilde{E}_i\)) is the natural starting point for the \(Z\) equation}. It is the logistic analogue of the B-S exposure weight for which the log transformation delivers a nesting property (Section~\ref{sec:bs-nesting}).
Note that \(\tilde{E}_i\) is the account's total exposure over the lookback window, so it combines annual account size and years of history. For accounts with a full \(W\)-year history it varies only with size; for newer accounts it also reflects the shorter history, so a large account with only one or two years of data receives lower credibility than the same account with a full history.

However, there is no guarantee that a strong relationship between \(Z\) and exposure exists, as illustrated in Section~\ref{sec:empirical}. Including log of lookback exposure therefore enables testing of the extent to which the B-S credibility-exposure relationship holds, i.e.~whether \(b_Z = 1\). Whether \(b_Z > 0\) (credibility rises with exposure) is always testable, against the exposure-invariant null \(b_Z = 0\). The functional form of exposure can be varied: a spline, exposure-bucket indicator, or alternative exposure transformation may be substituted and statistically tested (specific tests are outlined below), though the log specification is recommended as the default.

Natural additional features to test include \textbf{years of history}, \textbf{industry segment} (Section~\ref{sec:simulation} quantifies the gain when genuine risk-class heterogeneity is present), and \textbf{account volatility}.

On account volatility: in Bühlmann--Straub, \(K = \sigma^2_\varepsilon/\sigma^2_u\) (within-account process variance over between-account structural variance). The logistic intercept \(a_Z \approx -\log K\) already encodes this pooled variance ratio (Section~\ref{sec:bs-nesting}).
B-S assumes homogeneous within-account variance \(\sigma^2_\varepsilon\) across all accounts, so including per-company CoV in \(Z_i\) directly tests that assumption. If CoV\(_i\) carries signal, it implies \(K\) should be company-specific (\(K_i = \sigma^2_{\varepsilon,i}/\sigma^2_u\)), with high-volatility accounts receiving lower credibility than the pooled \(K\) implies. In practice, per-company CoV requires sufficient history to estimate reliably: with short \(W\) it is dominated by sampling noise and will typically fail formal feature-selection testing (defined below). The empirical study confirms this: adding CoV yields no improvement in out-of-sample wMSE (Section~\ref{sec:empirical}).

Any other account characteristic that may plausibly impact experience predictability (geographic territory, coverage type) can also be tested and retained if they carry signal and make sense to the practitioner.

Calendar-year effects are not natural Z features: portfolio-wide temporal structure should be handled via existing actuarial/modelling approaches, and temporal structure within an account's history is the role of \(\lambda\) (Section~\ref{sec:decay}).

\textbf{Feature selection.}
With a likelihood-based fit, adding a feature costs one degree of freedom. The
primary selection criterion is a likelihood ratio test (LRT): compare maximised
log-likelihoods as an LRT statistic against \(\chi^2(1)\); this is asymptotic under
pure MLE, and approximate but directionally reliable when weakly informative
priors shift the optimum slightly.

For the Bayesian implementation, leave-one-out cross-validation (LOO-CV) on the training portfolio is the natural complement: fit the model with and without the candidate feature and compare ELPD (expected log predictive density) differences against twice their standard error. LOO-CV is performed at the observation level (one company-year at a time), which is optimistic for panel data where within-company observations are temporally linked \citep{vehtari2017practical} (leaving out a single year while retaining the company's other years allows the model to partially see the account's history). To address this weakness, selection findings can be directionally supported by an independent held-out period comparison.\footnote{ELPD values from observation-level LOO-CV should be read as indicative rather than as formally valid cross-validation estimates for panel data; company-level LOO-CV would avoid this optimism.}

\subsection{Formal Nesting: Bühlmann-Straub as a Restricted Case}\label{sec:bs-nesting}

The logistic model contains rolling Bühlmann--Straub (B-S over the same lookback window) as a restricted special case, and its maximum-likelihood estimates recover the B-S parameters when the data follow B-S with exact credibility (proof sketch in \hyperref[app:nesting-proof]{Appendix~A}).
The value of the nesting result is practical: any improvement over B-S comes from relaxing specific, identifiable constraints, and the proposition below states precisely which ones.

\textbf{Proposition.}
Assume logistic \(Z\) parameters \(a_Z\) (intercept) and \(b_Z\) (slope on log lookback exposure),
cumulative lookback exposure \(\tilde{E}_i = \sum_{k=1}^{W} E_{i,t-k}\) over the window \(W\),
and a decay parameter \(\lambda\) governing how quickly historical experience is discounted (Section~\ref{sec:decay}).
If the complement is flat, \(\lambda = 1\) (no decay, equal-weight experience average), \(b_Z = 1\) on the \emph{unstandardised} \(\log \tilde{E}_i\) scale,
and \(a_Z = -\log K\) (intercept set from the B-S moment estimator), then the
logistic weight \(Z_i = \Lambda(a_Z + \log \tilde{E}_i)\) reproduces the
Bühlmann--Straub weight \(\tilde{E}_i / (\tilde{E}_i + K)\) computed over the same \(W\)-year window (\emph{rolling} B-S, explained in the remark below) exactly for every account \(i\).
This \(Z\)-parameter constraint holds regardless of complement specification; recovering the full rolling B-S prediction additionally requires the complement to equal the portfolio grand mean (i.e.~no size or other effect on the base rate).

The proof is a single algebraic step: \(\Lambda(\log(\tilde{E}_i) - \log K) =
\tilde{E}_i / (\tilde{E}_i + K)\), which follows from the identity
\(\Lambda(\log(x/K)) = x/(x+K)\) for any \(x, K > 0\) \citep{buhlmann2005}.

In practice, the model is fit on standardised inputs; when fitting with \(\log\tilde{E}_i^{\mathrm{sc}} =
(\log \tilde{E}_i - \mu_{\tilde{\ell}}) / \sigma_{\tilde{\ell}}\), exact nesting
requires \(b_Z = \sigma_{\tilde{\ell}}\) (the training-set standard deviation
of \(\log \tilde{E}_i\)), not \(b_Z = 1\).

\textbf{Remark.} This corresponds to a \emph{rolling B-S} formulation, in which both the experience mean and the credibility weight use the same \(W\) years. Standard B-S instead typically estimates \(K\) by method of moments over the full panel and uses the total exposure across all years (\(E_i\) in Equation~\ref{eq:bs}). Its weight form can be nested by letting \(W\) cover all history before the prediction year, but this breaks the coherence between the credibility weight and the experience statistic (\(Z\) would be informed by years that do not contribute to \(\bar{f}_i\)), so the rolling formulation is the natural parameterisation.

MLE consistency under a true B-S data-generating process is given in \hyperref[app:nesting-proof]{Appendix~A}.

\textbf{Scope of the LRT.} The formal likelihood-ratio test (Section~\ref{sec:zfeatures}) tests whether the logistic extensions are warranted relative to rolling B-S, its direct nested special case.
It does \emph{not} test against \emph{standard} B-S (full training history), because the two use different data subsets and are not nested under the same observed likelihood. A practitioner currently running standard B-S should treat the LRT as evidence that the extension is warranted once a rolling-window baseline is adopted; a held-out error (e.g.~wMSE or other) comparison is the appropriate evidence base for the direct standard-B-S comparison, as outlined in Section~\ref{sec:empirical}.

\begin{figure}[H]

{\centering \includegraphics[width=1\linewidth]{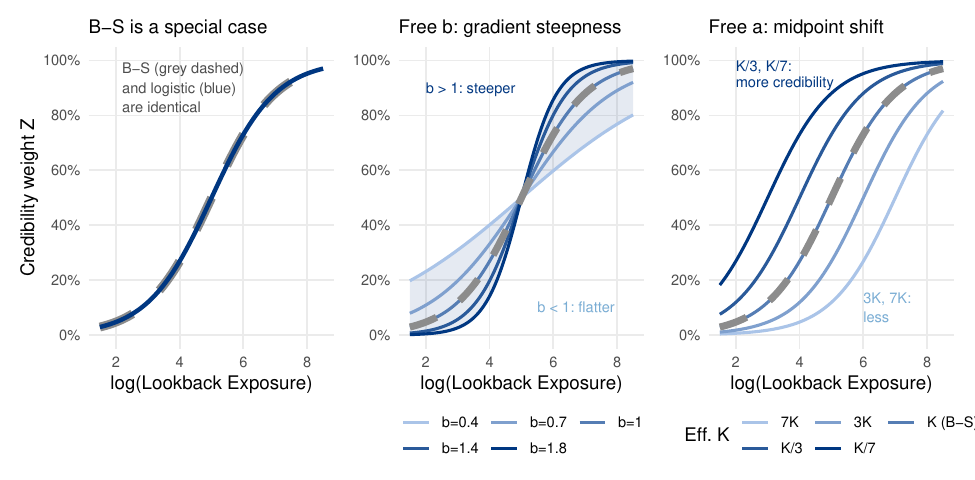} 

}

\caption{\textbf{Left}: Bühlmann--Straub (grey dashed) and logistic with $a_Z = -\log K$, $b_Z = 1$ (blue solid) are identical at every exposure level (the nesting identity). \textbf{Centre}: Freeing $b_Z$ controls gradient steepness; $b_Z > 1$ gives large accounts more credibility faster, $b_Z < 1$ flattens the curve. \textbf{Right}: Freeing $a_Z$ shifts the midpoint (effective $K$); lower effective $K$ means credibility is earned faster across all sizes. Both $a_Z$ and $b_Z$ are estimated jointly from data.}\label{fig:fig-bs-nesting}
\end{figure}

Figure~\ref{fig:fig-bs-nesting} illustrates the nesting relationship
geometrically.
By default, both models assign low \(Z\) to small accounts and high \(Z\) to large ones:
that much follows from any sensible credibility formula.
The difference is in the curve's shape.
B-S has a single free parameter \(K\), and once \(K\) is fixed, both the
position and steepness of the \(Z\)-curve are determined.
The logistic model has two free parameters, intercept \(a_Z\) and slope \(b_Z\),
so it can independently adjust where the \(Z\)-curve is centred and how
steeply it rises with account exposure.
If the data support a shallower gradient than the B-S curve implies (as they
do in the empirical study), the logistic model can
capture this; B-S cannot without compromising the fit for either small or large
accounts.

\textbf{Implied account-varying $K$.}
It is instructive to express the logistic model in B-S language.
Solving \(Z_i = \tilde{E}_i / (\tilde{E}_i + K_i)\) for the implied credibility parameter gives
\begin{equation}
  K_i = \tilde{E}_i\,\frac{1 - Z_i}{Z_i}
       = \tilde{E}_i\,e^{-(a_Z + b_Z\log \tilde{E}_i)}
       = e^{-a_Z}\,\tilde{E}_i^{1 - b_Z}.
  \label{eq:implied-K}
\end{equation}
When \(b_Z = 1\) this reduces to \(K_i = e^{-a_Z}\), a constant: the B-S
special case with a single pooled \(K\).
When \(b_Z < 1\), \(K_i\) grows with account exposure: large accounts face a higher
evidence threshold before their experience overrides the portfolio prior, a size-varying threshold that a single pooled \(K\) cannot produce.
One interpretation of this pattern is that large accounts retain residual within-account variance that does not diminish with exposure as quickly as the B-S variance assumption \(\mathrm{Var}(y_{it}) \propto 1/E_{it}\) implies, requiring more exposure to reach the same evidence threshold. The dispersion fit in \hyperref[app:variance]{Appendix~C} points the same way (the fitted variance falls roughly as \(E_{it}^{-0.57}\)).

When \(b_Z > 1\), \(K_i\) falls with size: large accounts earn credibility faster
than B-S predicts, the opposite pattern.
Two parameters \((a_Z, b_Z)\) generate a full continuum of account-specific
thresholds from a single estimable model.
Where an existing B-S implementation is to be retained, the implied \(K_i\) could be used to set exposure-band values of \(K\), with the fitted \(\hat{b}_Z\) indicating whether \(K\) needs to vary with exposure at all.

The model does not require the residual-variance explanation to hold: \(b_Z\) is estimated from the data, and its value is an empirical question for each portfolio. In the case study \(Z_i\) is nearly flat across account sizes under exposure weighting (\hyperref[sec:weights-sens]{Appendix~D}); a worked numerical illustration is given in Section~\ref{sec:worked-example}.

This equivalence also connects to penalised regression: Bühlmann--Straub with pooled \(K\) is algebraically equivalent to ridge regression, where the same \(K\) serves as a uniform penalty on deviations from the complement \citep{miller2015}, itself an instance of the general bias-variance tradeoff underlying regularised estimation \citep{brady2021}. A GLMM also corresponds to a penalised estimator (the BLUP penalty on random effects), but the penalty is implicit in the variance components and not directly parameterised by account characteristics. The logistic framework makes the penalty explicit and covariate-driven, preserving the interpretable \(Z_i\) decomposition while allowing the effective penalty to vary across accounts.

\subsection{Temporal Decay}\label{sec:decay}

Commercial lines accounts often change over time.
An account's risk profile from five years ago may be largely irrelevant
today: the business may have grown, contracted, or changed its operations;
the policy terms may have changed; the broader market
environment may have shifted.
Applying a uniform average across all historical years can actively harm
predictions when the account's underlying rate has drifted.

The framework addresses this through temporal decay: older years progressively lose weight, as the simple average \(\bar{f}_i\) used by B-S is replaced by an exponentially weighted moving average (EWMA), which becomes the experience rate \(\hat{f}_i\) in equation \eqref{eq:pred}:

\begin{equation}
  \hat{f}_i^{(\lambda)}
    = \frac{\sum_{k=1}^{W} \lambda^{k-1} C_{i,t-k}}
           {\sum_{k=1}^{W} \lambda^{k-1} E_{i,t-k}},
  \label{eq:ewma}
\end{equation}

where \(\lambda \in (0, 1]\) is estimated jointly with the \(Z\) parameters
and any complement parameters from the portfolio, rather than chosen by judgement.
We write \(\hat{f}_i^{(\lambda)}\) to show the dependence on the jointly estimated decay rate, and \(\hat{f}_i\) as shorthand where that dependence is understood.
Temporal weighting of credibility estimates has precedent in geometric- and exponential-smoothing-based approaches \citep{sundt1988, kremer1990}. Gluck's decay-factor extension of the Cape Cod method employs a structurally similar exponential decay weight \citep{gluck1997}, and Clark's recursive formulation can achieve a comparable decay pattern by a different route \citep{clark2025forgetting}.

When \(\lambda = 1\) all historical years receive equal weight, and the formula reduces to an exposure-weighted average of historical loss ratios, which is exactly the experience term in rolling Bühlmann--Straub (Section~\ref{sec:bs-nesting}).
As \(\lambda \to 0\) only the most recent year carries any weight.
Equivalently, \(\lambda\) determines an effective lookback window:
the number of years at which the cumulative EWMA weight first reaches a given threshold (\hyperref[app:effective-memory]{Appendix~B}).

The appropriate \(\lambda\) depends on the portfolio rather than being a universal constant, and can be estimated from the data where there is enough history.
A practitioner who applies \(\lambda = 1\) by default may be
over-weighting stale data for some accounts and under-using recent
data for others.
By estimating \(\lambda\) within the training likelihood, which fits the
EWMA-weighted experience average to observed outcomes, the framework learns
the degree of temporal persistence that characterises the portfolio.

\begin{figure}[H]

{\centering \includegraphics[width=0.75\linewidth]{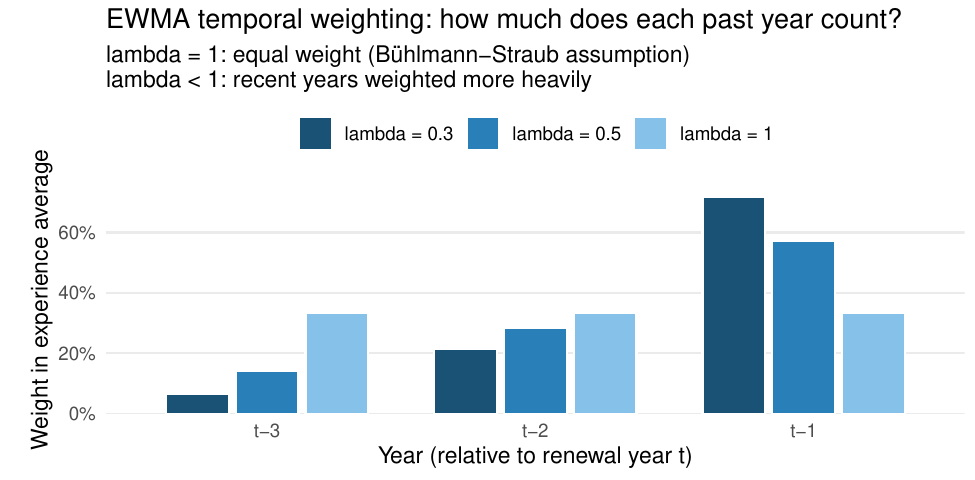} 

}

\caption{Relative weight assigned to each historical year as a function of the decay parameter \(\lambda\), for a three-year lookback window. At \(\lambda=1\) (Bühlmann-Straub assumption) all years receive equal weight. As \(\lambda\) decreases, the most recent year is upweighted and older years are discounted.}\label{fig:fig-ewma-weights}
\end{figure}

Figure~\ref{fig:fig-ewma-weights} shows how the relative weight of each
historical year changes with \(\lambda\).
At \(\lambda = 1\) (the B-S assumption), all three years contribute equally.
At \(\lambda = 0.5\) the most recent year receives twice the weight of
two years ago and four times the weight of three years ago, reflecting
the intuition that recent claims better predict next year's loss ratio
than claims from several years back.

Critically, the empirical study below reveals that the optimal \(\lambda\)
can vary substantially by company size.
The framework therefore allows \(\lambda\) to vary across accounts through a second logistic extension:
as with \(Z_i\), any account-level feature vector
\(\tilde{\mathbf{x}}_i^{(\lambda)}\) can drive the decay rate:

\begin{equation}
  \mathrm{logit}(\lambda_i) = \boldsymbol{\theta}_\lambda^\top \tilde{\mathbf{x}}_i^{(\lambda)},
  \label{eq:logit-lambda}
\end{equation}

where \(\boldsymbol{\theta}_\lambda\) is a parameter vector estimated jointly with
the \(Z\) and complement parameters.
Any characteristic that plausibly affects how quickly historical experience
becomes stale (account size, recent exposure growth, loss-ratio volatility,
or others) can enter \(\tilde{\mathbf{x}}_i^{(\lambda)}\).
The same principle applies as for \(Z\): let the data determine how temporal
relevance varies with account characteristics.
In the empirical study below, the primary specification uses size tercile
indicators (Small/Mid/Large by mean training exposure) to allow \(\lambda\) to
vary discretely by account size class. A continuous logistic form
\(\mathrm{logit}(\lambda_i) = a_\lambda + b_\lambda \log\bar{E}_i\) is tested as
a variant, with results reported in Section~\ref{sec:lambda-decay}.

The logistic link is a modelling choice, not a necessity. It constrains \(\lambda \in (0, 1)\), which encodes the assumption that recent years are at least as informative per unit exposure as older ones, which can be appropriate when data are fully developed and recent years carry more relevant information.

Where recent years are undeveloped, the recommended approach is to translate losses to an estimated ultimate basis before the credibility fit (Section~\ref{sec:applying}). Extending the model to estimate a development pattern jointly with \(Z\) and \(\lambda\) is an alternative to this pre-processing step. Open questions around both approaches are discussed under Further Work (Section~\ref{sec:conclusion}).

Independently of development, a softplus link allowing \(\lambda > 1\) remains a natural extension where the most recent year(s) legitimately warrant less weight for other reasons: residual estimation uncertainty after developing, or atypical adverse experience specific to the most recent year(s) that has since been corrected at the portfolio level (e.g.~a rate action or underwriting change), making it less representative of ongoing risk.

A natural concern with joint estimation is that \(\lambda\) and \(Z\) might trade off: a model with low \(\lambda\) (heavy discounting of old data) and one with low \(Z\) (little total weight on experience) can appear superficially similar.
The EWMA structure provides the identification: \(\lambda\) governs the \emph{shape} of decay across experience lags, while \(Z\) governs the \emph{total weight} given to experience. These are distinct data features.
A sufficient condition for local identification is that the portfolio exhibits variation in account sizes (informing \(a_Z\), \(b_Z\)); for a scalar \(\lambda\) this is sufficient, since the level of temporal autocorrelation identifies \(\lambda\) directly. For covariate-varying \(\lambda\) specifications, variation in temporal autocorrelation structure across account characteristics (e.g.~size) is additionally required; where this is absent, \(\lambda\) differences across accounts are weakly identified and a Bayesian implementation with informative priors may be preferable to MLE.

\subsection{Uncertainty Quantification}\label{sec:uq}

A point estimate of the renewal rate (MLE or posterior mean \(\hat{r}_i\)) gives
a single best-estimate prediction per account.
But two accounts with the same \(\hat{r}_i\) may carry very different
levels of uncertainty in that estimate.
An account with ten years of stable history and a credibility weight
well-identified from a rich portfolio may have a narrow posterior over \(Z_i\).
An account with two years of volatile history (where the logistic
model can be genuinely uncertain whether \(Z_i = 0.1\) or \(Z_i = 0.4\))
will have a wide posterior that the point estimate conceals.

Running the same model via HMC (implemented via \texttt{brms}/Stan)
delivers the full Bayesian posterior, from which three distinct uncertainty
objects can be extracted:

\medskip
\begin{center}
\begin{tabular}{@{}cp{3.2cm}p{5.5cm}p{5.5cm}@{}}
\hline
\textbf{Tier} & \textbf{Object} & \textbf{Practical output} & \textbf{Question answered} \\
\hline
1 & Posterior of $Z_i$ & Posterior SD or credible interval for $Z_i$ & How uncertain is the credibility weight for this account? \\[4pt]
2 & Posterior of $\hat{r}_i$ & Posterior mean and credible interval for $\hat{r}_i$; $R^*_i(q)$ (defined below) as conservative estimate & What blend rate can I not rule out at level $\alpha$? \\[4pt]
3 & Posterior predictive for $y_{i,T+1}$ & Predictive draws $\rightarrow$ prediction interval & What range of outcomes should I plan for next year? \\
\hline
\end{tabular}
\end{center}
\medskip

Note that Tier 2 uncertainty is not simply a rescaling of Tier 1: accounts far from the portfolio mean can carry material Tier 2 uncertainty even when \(Z_i\) is well-identified, because the complement \(\mu_i\) is itself estimated with error.

\subsubsection{\texorpdfstring{Conservative rate \(R^*_i\)}{Conservative rate R\^{}*\_i}}\label{sec:rstar}

Tier 2 (the posterior of the mean prediction \(\hat{r}_i\)) is
the natural basis for a decision-relevant uncertainty measure.
We define the \emph{conservative rate} \(R^*_i\) as the upper
\(q\)-quantile of that posterior:
\begin{equation}
  R^*_i(q) = F_{\hat{r}_i \mid \text{data}}^{-1}(q).
  \label{eq:rstar}
\end{equation}
The choice of \(q\) is a business decision. We use \(q = 0.95\) throughout as a conventional default.
The label ``conservative'' reflects that the rate is the blend the data cannot rule out at the chosen level.
The \emph{uncertainty load} \(R^*_i - \mathbb{E}[\hat{r}_i]\) is therefore a
\emph{blend-stability margin}: it measures how much the credibility blend
could shift under plausible parameter values, rather than how variable next year's
outcome will be. It is not an explicit premium loading.
Two sources contribute to this load: uncertainty in the \(Z\) parameters \((a_Z, b_Z, \ldots)\), where accounts in low-curvature regions of the logistic have poorly-identified \(Z_i\); and uncertainty in the complement parameters where these are estimated jointly, meaning accounts far from the portfolio mean carry material load even when \(Z_i\) is well identified.
A practical advantage of the Bayesian approach is that these two contributions can be isolated by fixing one parameter block and integrating over the other.

Operationally, a small uncertainty load means the credibility blend is well identified.
\(R^*_i\) is a parameter-uncertainty quantile answering ``how confident am I in this rate?''; a Tier 3 predictive interval for \(y_i\) answers the separate question ``what outcome range should I expect next year?'' and requires adding process noise on top.
The two quantities serve different purposes; full Tier 3 coverage comparisons across model specifications are in \hyperref[app:variance]{Appendix~C}.

Because \(Z_i\) is a nonlinear function of the full parameter vector \((\boldsymbol{\theta}_Z, \lambda)\), its posterior is not available analytically; the Bayesian HMC approach propagates parameter uncertainty through to \(Z_i\) directly, making \(R^*_i(q)\) exact rather than an approximation.
A worked demonstration of \(R^*_i\) for two real accounts is given in Section~\ref{sec:worked-example}.

\subsubsection{Variance structure and interval width}\label{sec:dispersion}

The Tier 3 interval width depends on the variance specification.
In the case study's Gamma model, constant \(\phi\) gives all accounts the same CV regardless of size, whereas letting \(\log\phi_{it} = \phi_0 + \phi_1\log E_{it}\) allows larger accounts narrower proportional spread.
The fitted \(\hat\phi_1\) is 0.57 (95\% credible interval {[}0.50, 0.63{]}) against 1 for the B-S assumption \(\mathrm{Var} \propto 1/E_{it}\), so the fitted variance falls more slowly with exposure (derivation and validation in \hyperref[app:variance]{Appendix~C}).
Because dispersion also sets the likelihood weighting, it can affect point predictions too (by about 1.8\% in wMSE here).
Families without a free dispersion, such as the Poisson, call for an added overdispersion term if their implied variance does not match the data.

The framework is now fully specified.
Section~\ref{sec:empirical} tests it on real company-level Commercial Auto data.

\section{Empirical Validation: 96 US Commercial Auto Companies}\label{sec:empirical}

\noindent\textit{The Worked Example (Section \ref{sec:worked-example}) applies the fitted model to two real companies drawn from this dataset; the portfolio-level evidence follows.}

\subsection{Data and Setup}\label{sec:data-setup}

The CAS loss reserve triangle database provides
accident-year loss triangles for US property-casualty insurers.
We restrict to Commercial Auto liability and apply two quality filters: companies below \$100k NEP are excluded as too small for meaningful credibility estimation, and companies without a full 10-year presence (AY 1998--2007) are excluded to ensure a balanced panel. The three years prior to the training window (AY 1998--2000) serve only to supply lookback history and to strengthen the balanced-panel filter itself; model fitting uses AY 2001--2005 and the held-out test is AY 2006--2007. The balanced panel gives a common holdout period across all accounts and avoids influential short-history accounts that can dominate complement calibration (\hyperref[app:deployment-diagnostics]{Appendix~H} quantifies both effects). This gives 96 companies across AY 1998--2007.
Section~\ref{sec:cross-lob} applies the same framework to a second line (Other Liability, 101 companies) to test whether the findings generalise beyond Commercial Auto.

Each insurance company plays the role of an \textbf{account}, where its annual loss
ratio (\(\text{incurred losses} \div \text{net earned premium}\)) is the experience
statistic, and its NEP is the credibility exposure.
Because each company loss ratio aggregates many underlying policies, volatility is lower than typical direct commercial lines accounts, and therefore practitioners applying the framework to direct accounts should expect noisier parameter estimates for equivalent sample sizes.

\textbf{Calendar-year normalisation.}
The CAS dataset does not include rate-change history, so on-levelling loss ratios to a common rate level in the usual way is not possible.
As a data-driven substitute, we instead work on a normalised scale: \(y^*_{it} = y_{it}/\mu_t\), where \(\mu_t\) is the NEP-weighted portfolio mean loss ratio for accident year \(t\).
The same normalisation is applied to the lag loss ratios used to construct the EWMA experience feature \(\hat{f}^*_i\), so the model operates entirely in relative space; the complement in relative space is \(\approx 1\) for a market-average account.
The absolute prediction for account \(i\) is \(\hat{r}^{\,\text{abs}}_i = \hat{r}^*_i \times \mu_t\).
Dividing by \(\mu_t\) removes the common calendar-year component of effects such as market pricing cycles and macroeconomic shocks. It does not correct for company-specific movements, whether trend or an uneven response to those effects (\hyperref[app:trajectories]{Appendix~H} illustrates the effect). To convert the relative prediction back to an absolute loss ratio requires \(\mu_t\), taken as the realised portfolio mean for each test year (values only available with hindsight).

\textbf{Balanced-panel design and survivorship bias.}
The balanced-panel filter (positive NEP across all 10 AYs 1998--2007) removes 61 of 157 companies: 20 fall below the \$100k NEP floor and 41 have incomplete panels. Of the 41 incomplete-panel exclusions, approximately 25 would classify as Small by the surviving tercile breaks, roughly twice the Small share in the retained set. Surviving Small accounts therefore skew toward companies that maintained a full 10-year presence, which may be more predictable than the excluded entrants and exiters; \(\hat{Z}\) for Small could be slightly overstated. This limitation applies equally to any credibility method estimated from this panel, and to any training set that omits lapsed accounts. \hyperref[app:deployment-diagnostics]{Appendix~H} reports a sensitivity check that relaxes the filter.

\textbf{Design choices.}
Training uses accident years 2001--2005; the test set is AY 2006--2007 combined.
All logistic credibility variants are fitted by Bayesian estimation (HMC) of the exposure-weighted Gamma likelihood (Section~\ref{sec:framework}), with the weakly informative priors in \hyperref[app:implementation]{Appendix~G} (Listing 2); maximum-likelihood refits are labelled as such.
All model parameters (including \(a_Z\), \(b_Z\), \(\lambda\) and the complement parameters) are estimated on the training years only; the test set is strictly held out.
We use a lookback window of up to \(W_{\max} = 7\) prior years; companies with shorter pre-2001 history use only their available lags.
Loss ratios are taken at 10-year development (the most mature evaluation available in this triangle) as a proxy for the unobserved ultimate, for both the prediction target and every lagged loss ratio feeding the EWMA experience feature \(\hat f_i^*\). In a live deployment, recent accident years will not yet be fully developed and must be brought to an estimated ultimate before entering the EWMA (Section~\ref{sec:applying}). Using these mature losses isolates the credibility and decay mechanism from development uncertainty, so the \(\lambda\) estimates reported here should not be assumed to transfer unchanged where recent years carry additional development noise (Section~\ref{sec:lambda-decay}).
All loss metrics are NEP-weighted throughout, so that larger accounts, where a given ratio error carries greater financial consequence, have correspondingly more influence (Section~\ref{sec:results-main}). Percentages quoted for wMSE are relative changes (for example, ``27\% lower than standard B-S''); differences between two such percentages are stated in percentage points.

Throughout this section, \textbf{size terciles} (Small / Mid / Large) are defined by each company's mean NEP over the training years (AY 2001--2005), giving a stable classification consistent across the training and test sets and avoiding any test-period information entering the tercile definition.
Three size groups balance simplicity against sample size: finer splits (quartiles or quintiles) would leave fewer companies per group in a 96-company panel, weakening per-group parameter estimation.
The tercile breaks are approximately
\$1M and
\$9M mean annual NEP.\footnote{The $Z$ equation uses cumulative lookback exposure $\tilde{E}_i = E_{i,t-1}+\cdots+E_{i,t-W}$ (the evidence base for credibility). The $\lambda$ equation uses the stable company-level mean training NEP $\bar{E}_i$: discretised into the size terciles in the proposed model, and used continuously in the continuous-$\lambda$ variant.}

Table~\ref{tab:tbl-model-map} maps each model group to the B-S assumption it relaxes, to help readers navigate the results.

\begin{table}[!h]
\centering
\caption{\label{tab:tbl-model-map}Models in Table~\ref{tab:tbl-emp-results} and the B-S assumptions each relaxes relative to standard B-S. Rows 5--9 are logistic credibility variants; the full model set (21 models including diagnostic variants) is in \hyperref[app:full-comparison]{Appendix~D}.}
\centering
\begin{tabular}[t]{ll}
\toprule
Model & B-S assumptions relaxed\\
\midrule
Baselines & ---\\
B-S (standard) & ---\\
B-S (best sequential patch) & Relaxes pooled \(K\), the flat complement and \(\lambda=1\)\\
GLMM (random intercept + size) & Relaxes the flat complement\\
LogCred (scalar \(\lambda\)) & Relaxes the fixed \(Z\)-slope (\(b_Z = 1\)), the flat complement and \(\lambda=1\)\\
\addlinespace
LogCred (continuous \(\lambda\)) & As scalar \(\lambda\); also relaxes common \(\lambda\) across sizes (parametric size form)\\
LogCred (tercile \(\lambda\)) [proposed] & As scalar \(\lambda\); also relaxes common \(\lambda\) across sizes (free tercile intercepts)\\
LogCred (B-S $Z$ (tercile $K$) + tercile $\lambda$) & As tercile \(\lambda\), but retains the B-S fixed \(Z\)-slope (\(b_Z = 1\))\\
LogCred (tercile \(\lambda\) + dispersion) & As tercile \(\lambda\); also relaxes constant dispersion\\
\bottomrule
\end{tabular}
\end{table}

Table~\ref{tab:tbl-emp-results} compares ten models across two baselines, two B-S competitors, one GLMM comparator, and five logistic credibility variants (a stratified-complement variant is in \hyperref[app:full-comparison]{Appendix~D}).

\emph{Baselines}

\begin{itemize}
\tightlist
\item
  \textbf{Market mean}: predict the portfolio loss ratio for every account (pure complement, no experience)
\item
  \textbf{Last year LR}: use last year's loss ratio as the prediction (pure recency, no shrinkage)
\end{itemize}

\emph{B-S competitors}

\begin{itemize}
\tightlist
\item
  \textbf{B-S (standard)}: pooled \(K\) by method of moments, implicit \(\lambda=1\), shrinking toward the exposure-weighted portfolio mean
\item
  \textbf{B-S (best sequential patch)}: stratified \(K\), size-varying complement, EWMA tercile \(\lambda\); the strongest sequential patch we tested
\item
  \textbf{GLMM (random intercept + size)}: random company intercept and log-exposure fixed effect (Gamma/log link)
\end{itemize}

\emph{Logistic credibility variants} (all share \(Z_i = \Lambda(a_Z + b_Z \cdot \log\tilde{E}_i)\) and complement \(\exp(\alpha + \beta \cdot \log E_{it})\), but differ in the EWMA \(\lambda\) specification)

\begin{itemize}
\tightlist
\item
  \textbf{LogCred (scalar \(\lambda\))}: single pooled \(\lambda\) (the parsimonious base specification)
\item
  \textbf{LogCred (continuous \(\lambda\))}: \(\lambda_i = \Lambda(a_\lambda + b_\lambda \cdot \log\bar{E}_i)\) varies continuously with mean training NEP
\item
  \textbf{LogCred (tercile \(\lambda\))} {[}proposed{]}: three free \(\lambda\) intercepts by size tercile; the primary specification\footnote{Designated on the basis of EDA motivation (Section \ref{sec:eda}) and LOO-CV model comparison (\hyperref[app:full-comparison]{Appendix~D}); the test set provides independent confirmation, not the selection criterion.}
\item
  \textbf{LogCred (B-S Z (tercile \(K\)) + tercile \(\lambda\))}: B-S weight \(Z = \tilde{E}/(\tilde{E} + K)\) with \(K\) estimated jointly for each size tercile, otherwise as the proposed model; tests whether a smooth logistic \(Z\) adds value over a stepwise B-S-style weight
\item
  \textbf{LogCred (tercile \(\lambda\) + dispersion)}: proposed model extended with size-varying dispersion \(\log\phi_{it} = \phi_0 + \phi_1\log E_{it}\); primarily affects interval width (\hyperref[app:variance]{Appendix~C})
\end{itemize}

The covariate for \(\lambda\) is a modelling choice: mean training NEP and lookback exposure are highly correlated unless history lengths differ widely or volumes change sharply (case-study sensitivity: \hyperref[app:full-comparison]{Appendix~D}).\footnote{Diagnostic variants (fixed-$\lambda$ comparisons, logistic $Z$-structure ladder with $b_Z$ fixed at 1) and additional B-S patching variants are in the full comparison in \hyperref[app:full-comparison]{Appendix~D}.}

\subsection{Exploratory Data Analysis}\label{sec:eda}

Three features of the data motivate the framework's key design choices.

\begin{figure}[H]

{\centering \includegraphics[width=0.95\linewidth]{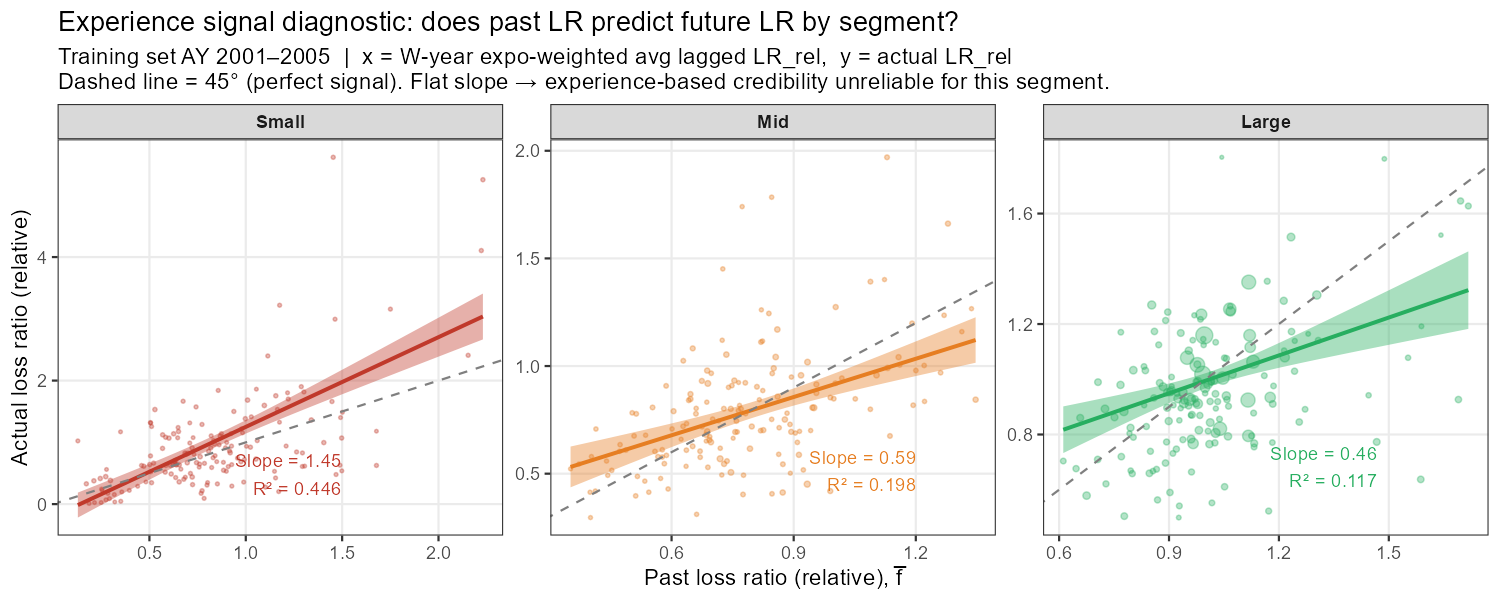} 

}

\caption{Pre-adoption signal check: actual next-year relative loss ratio versus the flat ($\lambda=1$) multi-year average $\bar{f}$ by size tercile (training data, AY 2001--2005). NEP-weighted OLS slope and $R^2$ annotated per panel. Signal is evident across all company size bands in the training data.}\label{fig:fig-eda-signal}
\end{figure}

\textbf{Past experience is predictive.}
Figure~\ref{fig:fig-eda-signal} plots the actual next-year relative loss ratio against \(\bar{f}\), the exposure-weighted average of past relative loss ratios across all available lag years (equal weight per year, i.e.~\(\lambda=1\)), by size tercile. This is a pre-modelling check of whether signal exists, complementary to the post-modelling calibration slope check (Section~\ref{sec:results-main}).
The NEP-weighted slope and \(R^2\) quantify predictive content per tercile; a flat relationship (slope \(\approx 0\)) in a given segment means the account's own history carries no genuine signal there, so a well-calibrated \(Z\) should assign it little weight.
Large's pooled \(R^2\) is the lowest of the three (0.12).
However, regressing on the single most recent lag alone instead gives \(R^2=0.43\) for Large, higher than the lag-1 figure for either Small (0.32) or Mid (0.15). Stale-year averaging, not a lack of signal, therefore drives Large's low pooled \(R^2\).

\begin{figure}[H]

{\centering \includegraphics[width=0.65\linewidth]{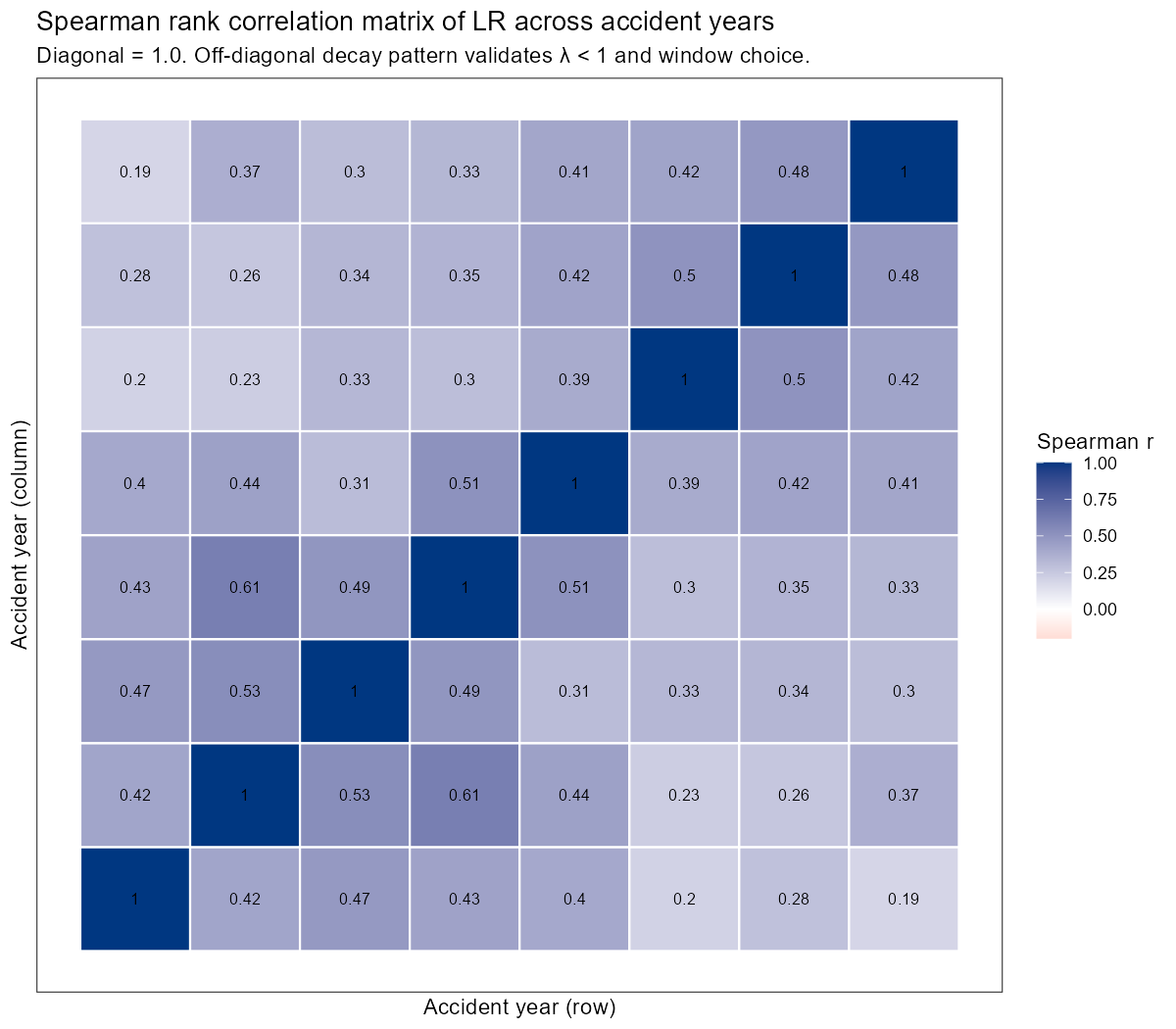} 

}

\caption{Spearman rank correlation of relative loss ratio between every pair of accident years (CAS Commercial Auto, 96 qualifying companies, AY 1998--2005).}\label{fig:fig-eda-corrmat}
\end{figure}

\begin{figure}[H]

{\centering \includegraphics[width=1.02\linewidth]{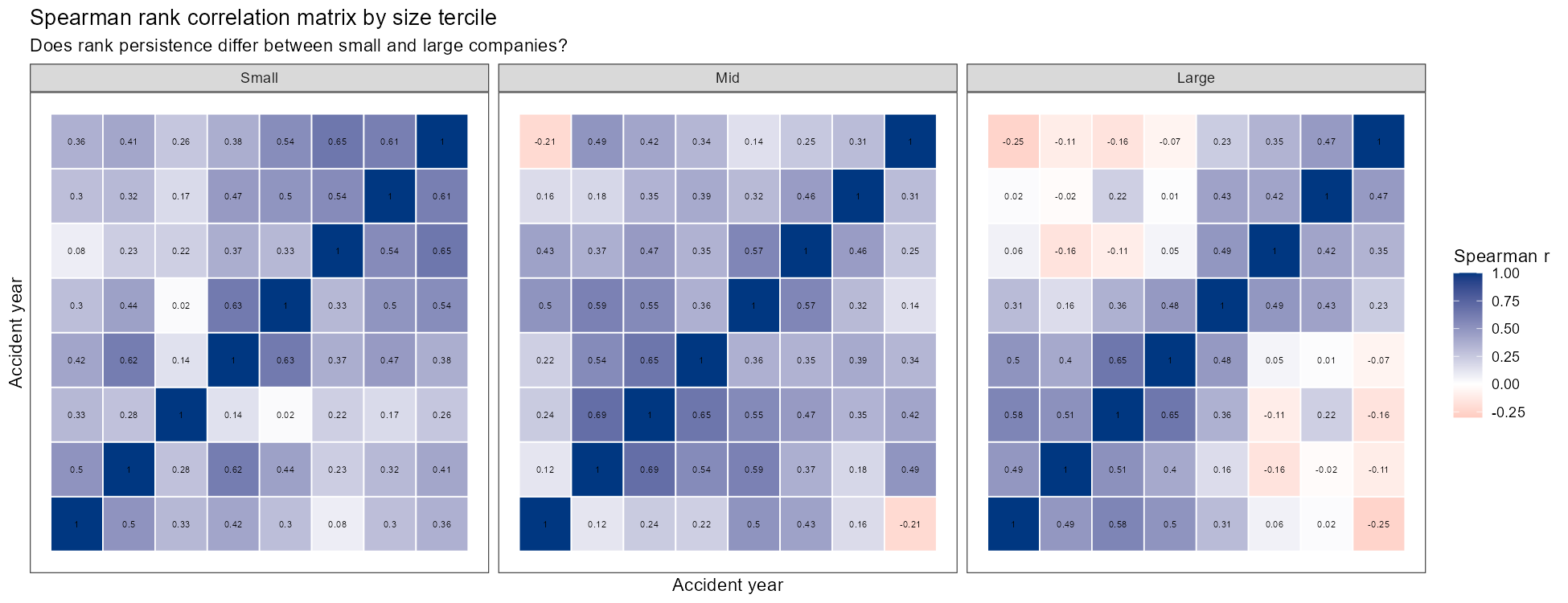} 

}

\caption{Spearman rank correlation matrices by account size tercile (same data as Figure~\ref{fig:fig-eda-corrmat}). Decay is fastest for Large accounts (right panel).}\label{fig:fig-eda-corrmat-bysize}
\end{figure}

\textbf{The signal decays with time, at a rate that differs by account size.}
Figure~\ref{fig:fig-eda-corrmat} shows the Spearman rank correlation matrix across the full training panel.
Correlation decays gradually away from the diagonal, from about 0.5 at lag 1 to about 0.2 at lag 7, confirming that older years carry less signal and arguing for EWMA (\(\lambda < 1\)) over a flat average.
Figure~\ref{fig:fig-eda-corrmat-bysize} splits the matrix by size tercile: off-diagonal colours fade faster for large accounts than small, motivating a size-varying \(\lambda\).

\begin{figure}[H]

{\centering \includegraphics[width=0.72\linewidth]{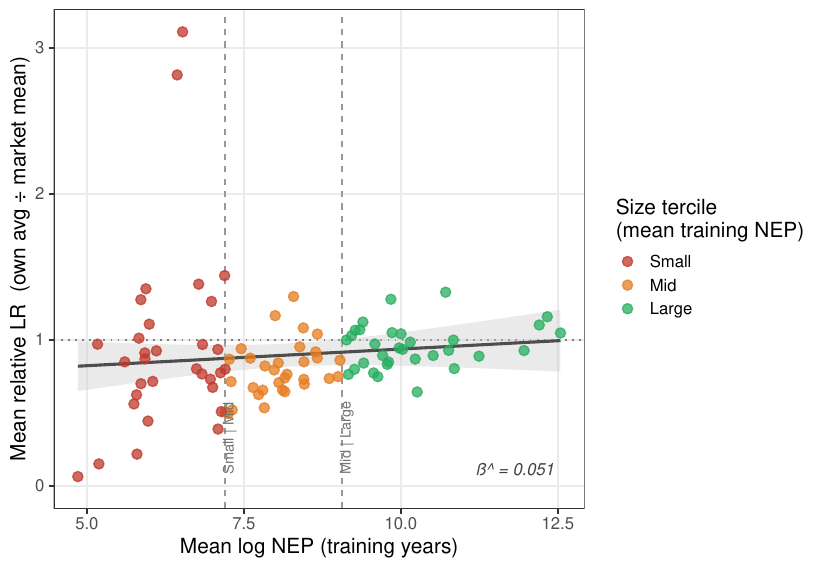} 

}

\caption{Each point is one company: mean relative loss ratio (own average over AY 2001--2005, divided by the market mean) against mean log NEP over the same period. Colour indicates size tercile (defined by mean training NEP). Dashed vertical lines mark the tercile breaks. The NEP-weighted OLS line (grey band = 95\% CI) rises from $\approx 0.74$ at the median Small company to $\approx 0.94$ at the median Large company, confirming a non-flat size gradient. This motivates $\exp(\alpha + \beta \cdot \log E_{it})$ as the complement functional form to test.}\label{fig:fig-eda-complement}
\end{figure}

\textbf{The complement is not flat across company sizes.}
Figure~\ref{fig:fig-eda-complement} plots each company's mean relative loss ratio against mean log NEP.
The NEP-weighted OLS line rises from \(\approx 0.74\) for the median Small company to \(\approx 0.94\) for the median Large company (slope 0.051 per unit of log NEP, 95\% CI 0.030--0.071). An unweighted fit is flatter and not distinguishable from zero (slope 0.023, 95\% CI \(-0.022\) to \(0.068\)).
This motivates the \(\exp(\alpha + \beta \cdot \log E_{it})\) complement: the non-zero \(\hat\beta\) justifies a size-varying prior rather than the flat grand mean that B-S uses by default.

\subsection{Results}\label{sec:results-main}

\noindent\textbf{Model selection (training data).}
Structural choices can be tested on the training data using likelihood-ratio tests and leave-one-out cross-validation (LOO-CV, a within-training proxy for out-of-sample fit).
The LOO-CV ratio is the gain in expected log predictive density divided by its standard error, so, as with a \(z\)-statistic, values above about 2 suggest a genuine improvement.
Freeing \(b_Z\) (vs the B-S-nested \(b_Z = 1\)) gives a MAP likelihood-ratio test \(\chi^2(1) = 39.4\) (\(p \approx 3.5 \times 10^{-10}\)) and a LOO-CV ratio of 1.8; estimating a decay rate at all (vs fixing \(\lambda = 1\)) gives \(\chi^2(1) = 75.6\) (\(p \approx 3.4 \times 10^{-18}\); conservative, as \(\lambda = 1\) lies on the boundary of \((0, 1]\)) and a ratio of 2.3.
Letting \(\lambda\) vary by size tercile (the proposed specification) gives a ratio of 2.1 over scalar \(\lambda\) (the first two tests use scalar \(\lambda\)). A two-group split (Small and Mid pooled against Large) fits equally well (ELPD difference \(-0.5\), SE 0.6).
The \(b_Z\) ratio is borderline even though its likelihood-ratio test is decisive and the held-out results point the same way: restricting \(b_Z = 1\) raises wMSE by 6\% (from 8.61 to 9.14 \(\times 10^{-3}\)).
Full diagnostics are in \hyperref[app:full-comparison]{Appendix~D}, and Section~\ref{sec:lambda-decay} examines how \(\lambda\) varies with size.

Model selection can instead be done directly on held-out performance, which some practitioners will prefer; there, sampling variation has to be assessed separately (for example by bootstrapping or cross-validation).
Training-only testing is a complementary strategy, particularly useful when a portfolio is too small to spare a held-out set without starving training.
The test-set results below (Table~\ref{tab:tbl-emp-results} onward) provide independent confirmation that these choices generalise (Section~\ref{sec:leakage-check}).

\medskip

\noindent\textbf{Evaluation metrics.}
wMSE is the NEP-weighted squared prediction error: it weights accounts by premium volume, giving larger accounts, where a given ratio error carries greater financial consequence, correspondingly more influence on the metric. This is a financial-importance weighting, not a precision weighting: the fitted variance falls more slowly than \(1/E_{it}\) (\hyperref[app:variance]{Appendix~C}).
It is the primary metric and the natural scale for comparison with the prior credibility literature. Rankings under log-wMSE (Table~\ref{tab:tbl-emp-results}) are broadly similar: the fully patched B-S edges ahead of the proposed model, and the row-9 diagnostic falls behind the scalar-\(\lambda\) model; the conclusions are unchanged.
Gini is the secondary metric (rank discrimination quality) and the held-out calibration slope (the exposure-weighted regression of actual on predicted; ideal 1.0) is a tertiary diagnostic.
All three size groups are right-skewed at 10-year development (skewness coefficient: Small = 2.3, Mid = 1.6, Large = 1.4), with Small accounts showing the longest right tail (maximum loss ratio 341\%, beyond the axis range plotted in Figure~\ref{fig:fig-intro-lr-dist}). We therefore report log-wMSE alongside wMSE throughout as a robustness check, particularly for Small accounts.
Deviance under the fitted likelihood is a legitimate alternative to wMSE.

\medskip

\begin{table}[!h]
\centering
\caption{\label{tab:tbl-emp-results}Key empirical results: US Commercial Auto, AY 2006--2007 test set ($n=192$).
    wMSE and log-wMSE: exposure-weighted MSE on the loss-ratio and log loss-ratio scales (lower is better).
    Gini$_{\text{pct}}$: Gini as \% of the maximum achievable (higher is better).
    Calibration slope: ideal = 1.00.
    Bold = best among rows 1--8; row 9 (diagnostic) and row 10$^\dag$ (optimised for interval calibration, \hyperref[app:variance]{Appendix~C}) are excluded.
    Fully patched B-S: tercile-stratified $K$, size complement and EWMA with tercile $\lambda$.
    wMSE uses the realised test-year portfolio mean as complement; the deployable gain is $\approx 25\%$.
    Full 21-model comparison in \hyperref[app:full-comparison]{Appendix~D}.}
\centering
\resizebox{\ifdim\width>\linewidth\linewidth\else\width\fi}{!}{
\begin{tabular}[t]{>{\raggedright\arraybackslash}p{6cm}cccc}
\toprule
Model & wMSE ($\times10^{-3}$) & log-wMSE ($\times10^{-3}$) & Gini$_{\text{pct}}$ & Calib. slope\\
\midrule
\addlinespace[0.3em]
\multicolumn{5}{l}{\textbf{Baselines}}\\
Market Mean (baseline) & 19.25 & 70.82 & 0.6\% & 1.00\\
Last Year LR (naive) & 13.14 & 49.59 & 75.1\% & 0.63\\
\addlinespace[0.3em]
\multicolumn{5}{l}{\textbf{B-S benchmark}}\\
B-S (standard, full history) & 10.97 & 48.78 & 69.6\% & 1.08\\
\addlinespace[0.3em]
\multicolumn{5}{l}{\textbf{Fully patched B-S}}\\
B-S (fully patched) & 8.16 & \textbf{32.15} & \textbf{79.0\%} & 0.85\\
\addlinespace[0.3em]
\multicolumn{5}{l}{\textbf{Best GLMM competitor}}\\
GLMM (random intercept + size) & 10.62 & 45.20 & 69.7\% & 1.06\\
\addlinespace[0.3em]
\multicolumn{5}{l}{\textbf{Logistic credibility (comparators)}}\\
LogCred (scalar $\lambda$) & 8.61 & 34.98 & 76.5\% & \textbf{1.00}\\
LogCred (continuous $\lambda$) & 8.38 & 34.32 & 77.2\% & 1.02\\
\addlinespace[0.3em]
\multicolumn{5}{l}{\textbf{Proposed model}}\\
LogCred (tercile $\lambda$) & \textbf{7.96} & 32.65 & 78.7\% & 1.03\\
\addlinespace[0.3em]
\multicolumn{5}{l}{\textbf{Diagnostic: B-S $Z$ variant}}\\
LogCred (B-S Z (tercile K) + tercile $\lambda$) & 8.36 & 35.60 & 78.9\% & 0.85\\
\addlinespace[0.3em]
\multicolumn{5}{l}{\textbf{Dispersion extension$^\dag$}}\\
LogCred (tercile $\lambda$ + disp.) & 8.11 & 33.17 & 80.1\% & 1.09\\
\bottomrule
\end{tabular}}
\end{table}

\medskip

\noindent\textbf{Overall results.}
The logistic credibility models substantially outperform standard B-S on wMSE and Gini (Table~\ref{tab:tbl-emp-results}): wMSE falls from 10.97 to between 7.96 and 8.61 \(\times 10^{-3}\) across rows 6--8, while the naive baselines and the GLMM sit well behind the logistic rows (the GLMM comparison is in Section~\ref{sec:glmm-comparison}).
LogCred (scalar \(\lambda\)) already achieves \(8.61 \times 10^{-3}\) with a single optimisation pass, a 22\% improvement over standard B-S.
Allowing \(\lambda\) to vary continuously with account size reduces wMSE to \(8.38 \times 10^{-3}\), and freeing \(\lambda\) to three independent size-tercile values (the proposed model) to \(7.96 \times 10^{-3}\), the lowest in the table.
Two further rows are not candidate models. Row 9 (B-S \(Z\) (tercile \(K\)) + tercile \(\lambda\)) is a diagnostic: it replaces the logistic \(Z\) with the B-S form \(Z = \tilde{E}/(\tilde{E} + K)\), with \(K\) estimated jointly for each size tercile, so comparing it with the proposed model isolates the contribution of a smooth logistic \(Z\) over stepwise \(K\). Row 10 (tercile \(\lambda\) + size-varying dispersion) is an extension of the proposed model for interval calibration (\hyperref[app:variance]{Appendix~C}). Its point-prediction accuracy is close to the proposed model's: Gini is 1.4 points higher and wMSE 1.8\% higher. We do not treat either difference as evidence of a point-prediction gain or loss.

The closest competitor in Table \ref{tab:tbl-emp-results} is the fully patched B-S (row 4): it comes within 2.5\% of the proposed model on wMSE and is marginally ahead on Gini (79.0\% vs 78.7\%) and log-wMSE, but is less well calibrated overall (slope 0.85 vs 1.03) and trails it on small accounts (wMSE 27.2\% higher; 90\% bootstrap interval 9.4\% to 44.2\%). The overall gap is not stable across the two test years: the patch is 9.7\% worse than the proposed model in AY 2006 and 2.7\% better in AY 2007.
The headline gain comes mainly from large accounts, which carry 91.4\% of NEP. Small accounts (1.2\% of NEP) are a different story: stratifying \(K\) alone cuts their wMSE by 41\% relative to standard B-S, slightly more than the proposed model's 38\%, and the logistic model's contribution is to avoid the reversal when EWMA is added (see below). Overall, stratifying \(K\) alone reduces wMSE by only 6\%.
Calibration by size is examined later in this section (Figure~\ref{fig:fig-slope-results}).

\medskip

\noindent\textbf{Proposed model.}
LogCred (tercile \(\lambda\)) achieves wMSE \(= 7.96 \times 10^{-3}\),
a 27\%
improvement over standard B-S
(90\% bootstrap interval: 9\%--44\%).\footnote{CIs use a company-level pairs bootstrap (1{,}000 resamples): companies are resampled with replacement, preserving each company's two test-year observations, and wMSE is recomputed on the resampled test set. The reported interval is the $[5\%, 95\%]$ percentile range of the bootstrap distribution of the percentage improvement relative to standard B-S.}
Excluding the highest-leverage company (GRCODE 26433, Harco Natl Ins Co, NEP \(\approx \$50\)M, LR \(\approx 101\%\) in AY 2007) leaves the wMSE improvement at 33\%, so the headline is not driven by a single account.
In a deployable setting (replacing the oracle test-year portfolio mean \(\mu_t\) with the prior-year mean \(\mu_{t-1}\)), the improvement for the tercile-\(\lambda\) model is 25\% (90\% CI: 7\%--42\%).

Test-set Gini (79\%) exceeds in-sample Gini (70\%), which is inconsistent with over-fitting (see the Data Leakage Check).
In-sample calibration slopes by tercile (proposed model on training set AY 2001--2005) are: Small 1.58, Mid 0.96, Large 1.05, overall 1.05. The overall slope is close to 1, consistent with convergence. The Small slope exceeds the 1.3 warning level suggested in Section \ref{sec:practical} and indicates in-sample under-crediting (actuals vary more than predictions), substantially attributable to the NEP-weighted fit favouring Large accounts; \hyperref[app:full-comparison]{Appendix~D} confirms this with an unweighted refit. The held-out Small slope is 0.99 (90\% CI: 0.73--1.14), so the pattern does not carry over to the test period.

Training on AY 2001--2004 and evaluating on AY 2005 alone, the wMSE improvement over the same standard B-S for the tercile-\(\lambda\) model is 30\%, and test-set Gini again exceeds in-sample Gini (75\% vs 70\%), confirming stability across test windows.

\medskip

\noindent\textbf{Estimation of $K$.}
Can B-S be repaired within its own framework? We first checked whether the method-of-moments estimator of \(K\) is the problem, by choosing \(K\) to minimise training error instead.
Minimising training wMSE, minimising training log-wMSE and Gamma maximum likelihood give the same predictions, because \(\hat{K}\) collapses to its lower bound (\(\hat{Z} \approx 1\), no shrinkage).
Held-out wMSE falls from 10.97 to 10.63 \(\times 10^{-3}\) (3\%) and Gini is unchanged (69.6\% vs 70.1\%), but the slope moves from 1.08 to 0.88 and wMSE stays 34\% above the best logistic model.
A constant \(K\) is equivalent to fixing \(b_Z = 1\) (Section~\ref{sec:bs-nesting}), whereas the estimated \(\hat{b}_Z < 1\) implies \(K_i \propto E_i^{1-b_Z}\), growing with account size, so no single \(K\) can recover it.

\medskip

\noindent\textbf{Limits of sequential patching.}
Sequential patching relaxes B-S's assumptions one at a time, keeping each earlier fix. Row 4 is the fully patched B-S, built from all the steps below (full sequence in \hyperref[app:full-comparison]{Appendix~D}).

\begin{enumerate}
\def\labelenumi{\arabic{enumi}.}
\tightlist
\item
  \textbf{Stratify \(K\) by size.} Overall wMSE falls 6\% (from \(10.97\) to \(10.34 \times 10^{-3}\)). Small accounts gain most (41\% lower wMSE), and their held-out calibration slope moves from 4 (90\% CI: 2.5--4.9) to 0.85.
\item
  \textbf{Add the size-varying complement.} Overall wMSE falls by a further 2.3 percentage points, but Mid wMSE rises 16\% (this step also switches the experience term from the training-period mean to the lookback window; 7\% with a direct least-squares complement, \hyperref[app:full-comparison]{Appendix~D}).
\item
  \textbf{Add EWMA with tercile \(\lambda\).} This is the largest step: overall wMSE falls by a further 17.5 percentage points (to \(8.16 \times 10^{-3}\)), driven by Large accounts (30\% lower wMSE). Small accounts give back about half of their step-1 gain (+34\% relative to step 2; 90\% bootstrap interval +3.5\% to +74\%). Re-estimating the complement or \(K\) for the EWMA experience does not remove this (\hyperref[app:full-comparison]{Appendix~D}); it appears to reflect the high stratified-\(K\) weight passing a recency weighting fitted in training almost fully into the prediction (Figure \ref{fig:fig-intro-patching}).
\end{enumerate}

Refitting the same patch under a log-wMSE objective comes closer still: 1.1\% above LogCred on wMSE and 1.6\% below on log-wMSE. Patching can therefore approach the joint model's accuracy, but only through several coordinated choices, which the joint fit makes in one step.

\medskip

\noindent\textbf{Complement calibration.}
The continuous complement underestimates the Small tercile's actual mean relative loss ratio (by
26\%, NEP-weighted; \hyperref[app:full-comparison]{Appendix~D}).
A stratified complement reveals a non-monotone pattern in mean relative loss ratios by size (Mid sits below both Small and Large) which the continuous size slope cannot reproduce.
The impact on performance is modest, indicating that the credibility mechanism can largely compensate for a mis-specified complement. Where pricing smoothness across size boundaries matters, or where the portfolio is too small to support three free intercepts, the continuous complement is the preferred default.

\subsubsection{Data Leakage Check}\label{sec:leakage-check}

Four leakage channels are verified not to affect the results.

\begin{itemize}
  \item \textbf{Oracle complement.} $\mu_t$ is the realised portfolio mean, unavailable at renewal; substituting the prior-year $\mu_{t-1}$ reduces the wMSE improvement from 27\% to 25\%, as reported above.
  \item \textbf{Covariate standardisation.} All scaling parameters are estimated from training years (AY 2001--2005) only.
  \item \textbf{In-sample / out-of-sample gap.} The naïve last-year predictor also achieves higher test Gini (75\%) than in-sample (65\%), confirming the gap reflects the intrinsic predictability of AY 2006--2007 vs the volatile 2001--2005 training window, not model overfitting.
  \item \textbf{Model selection.} The proposed model was designated on EDA motivation (Section \ref{sec:eda}) and LOO-CV comparison (\hyperref[app:full-comparison]{Appendix~D}); the test set is confirmation only, not the selection criterion.
\end{itemize}

\subsubsection{Calibration Diagnostics}\label{sec:empirical-diagnostics}

Two further diagnostics, both on the held-out test set (AY 2006--2007), check whether the improvement holds across the prediction range and by account size, in addition to the pooled average results above.

\begin{figure}[H]

{\centering \includegraphics[width=0.95\linewidth]{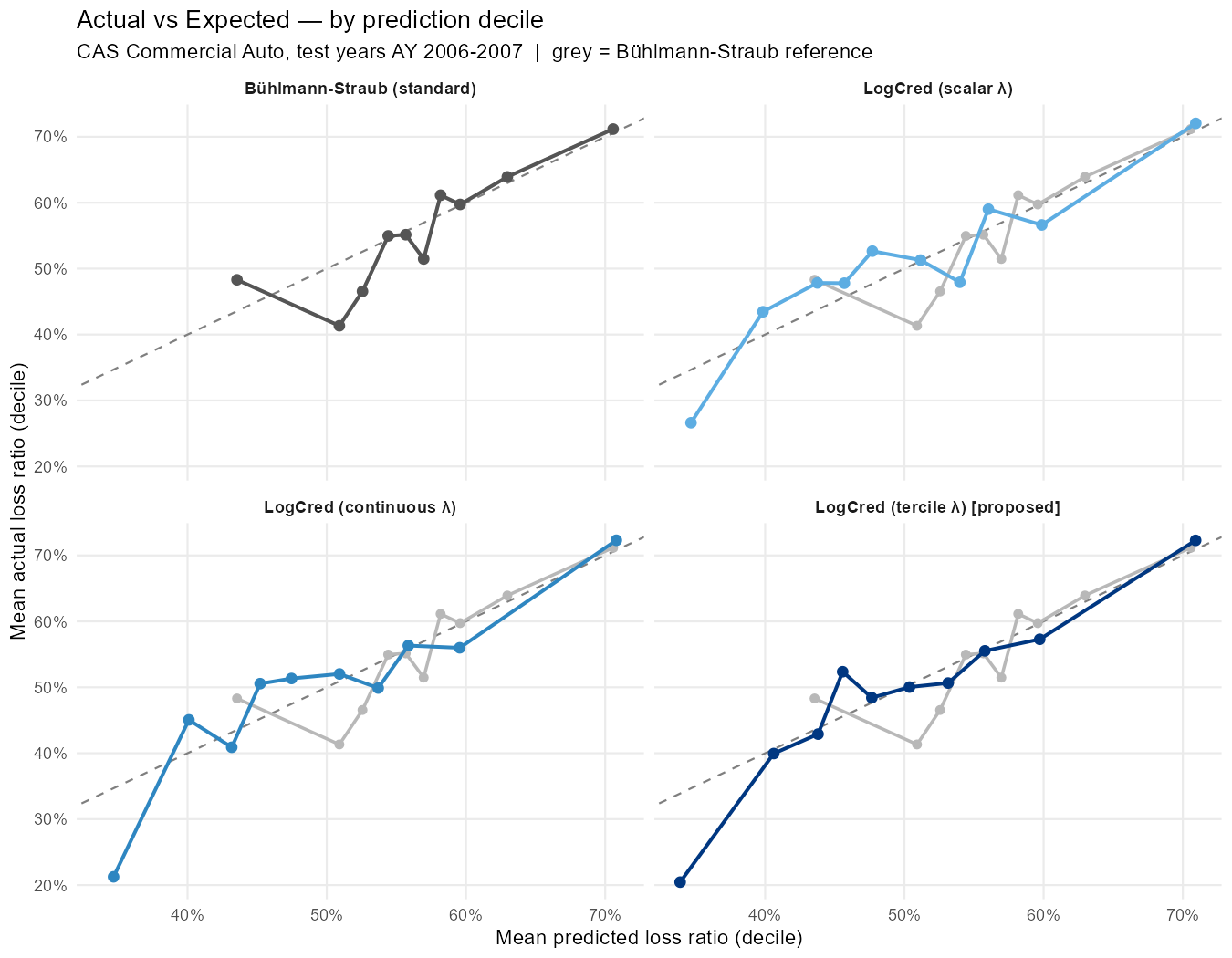} 

}

\caption{Actual vs expected loss ratio by prediction decile (NEP-weighted), held-out test set (AY 2006--2007). The top-left panel shows Bühlmann-Straub alone; the remaining three panels each show one logistic model (coloured) against Bühlmann-Straub (grey reference). Dashed diagonal = perfect calibration. Deciles are formed separately for each panel from that model's own predictions, so panels are not directly comparable point-for-point. The models are closer together at the top of the range; the logistic models' wider spread separates a low-risk tail that B-S does not (although they over-predict the lowest decile).}\label{fig:fig-emp-ave-decile}
\end{figure}

Figure~\ref{fig:fig-emp-ave-decile} shows where the improvement comes from. The models are close together at the top of the predicted range, but the logistic models identify a low-risk tail that B-S misses: the lowest-predicted decile has an actual loss ratio of about 20\% under the proposed model, versus 48\% under B-S.

\begin{figure}[H]

{\centering \includegraphics[width=0.9\linewidth]{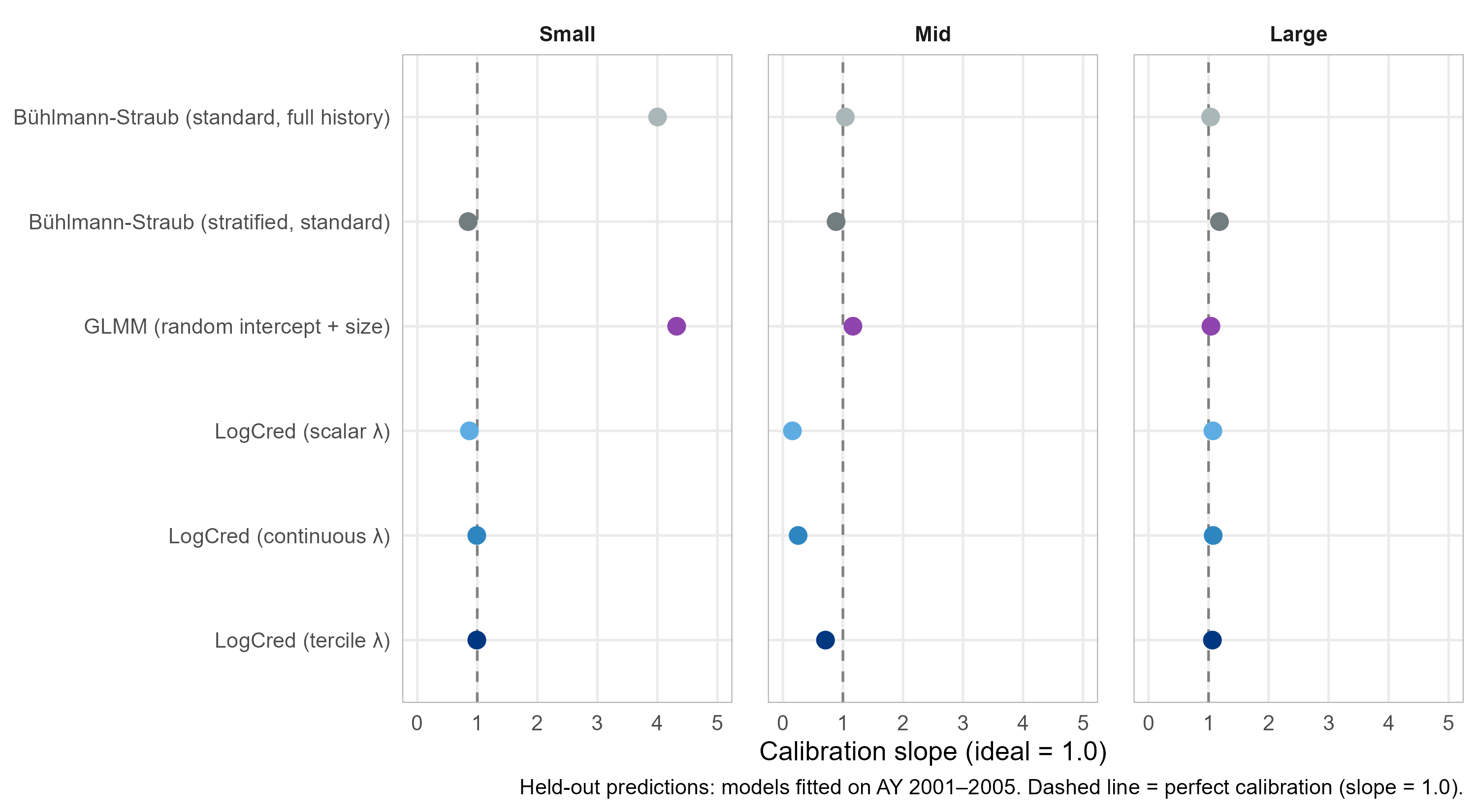} 

}

\caption{Calibration slope by size tercile on the held-out test set (AY 2006--2007). Ideal slope = 1.00. Bühlmann--Straub: Small slope 4 (90\% CI: 2.5--4.9), Large slope 1.03. Logistic Mid slopes: $\approx 0.16$ (scalar $\lambda$) to 0.71 (tercile $\lambda$; 90\% CI: 0.31--1.07). GLMM: Mid slope 1.17 (CI: 0.68--1.82), Small slope 4.3.}\label{fig:fig-slope-results}
\end{figure}

Figure~\ref{fig:fig-slope-results} shows the calibration slope by size tercile.
Standard B-S produces a held-out Small slope of \(\approx 4\) (90\% bootstrap CI: 2.5--4.9), indicating under-crediting, while the logistic framework brings the Small slope to near 1.0 (Large was already near 1.0 under B-S).
Pooled Mid slopes remain below 1.0 for all logistic variants (point estimates \(0.16\) to \(0.71\)), although the proposed model's 90\% bootstrap interval, 0.31 to 1.07, just includes 1.0. The GLMM's Mid advantage (Section~\ref{sec:glmm-comparison}) is likely sample-specific, reflecting a shift in predictability between training and test.

\subsection{The Temporal Decay Finding}\label{sec:lambda-decay}

\begin{figure}[H]

{\centering \includegraphics[width=0.85\linewidth]{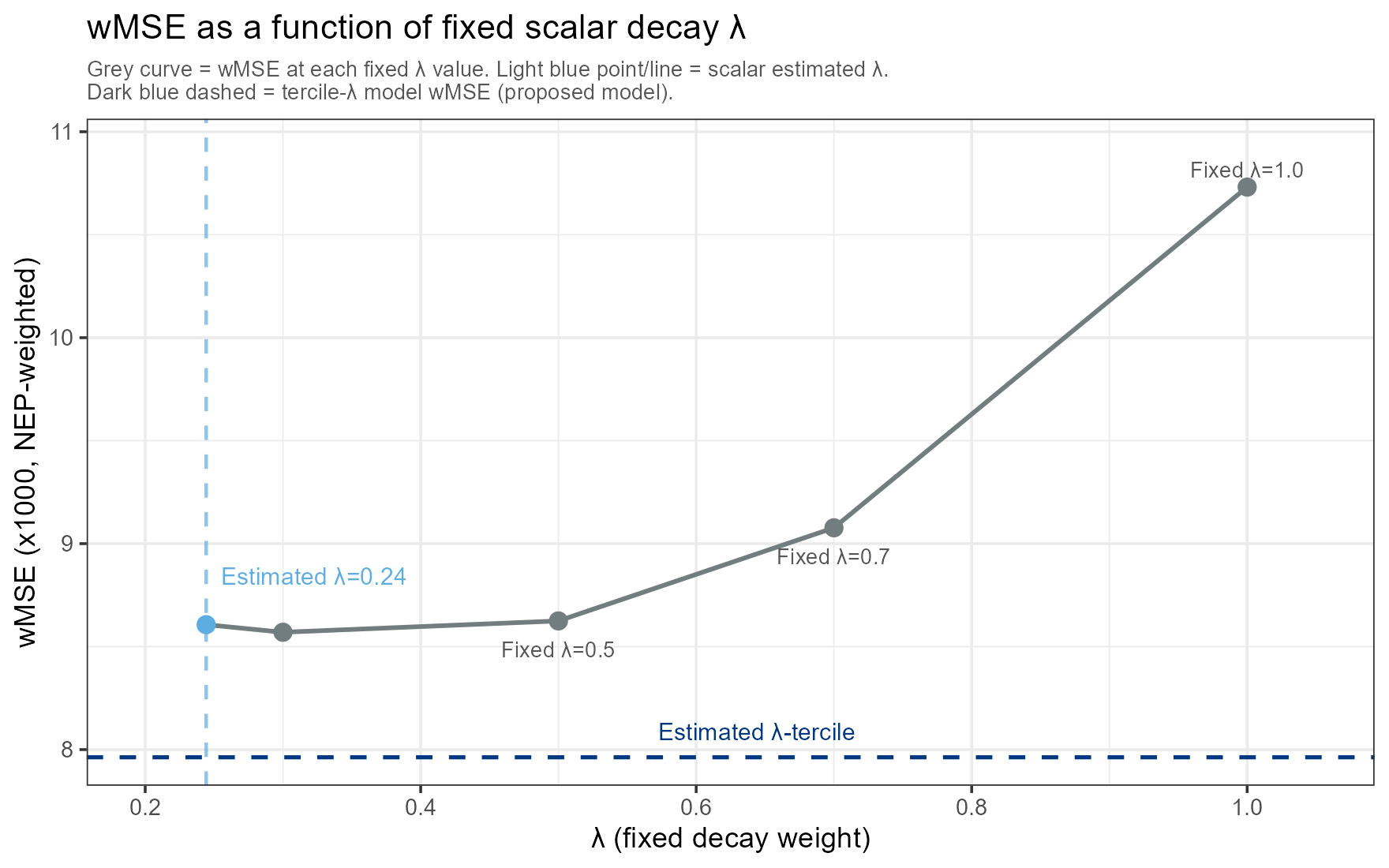} 

}

\caption{Test-set wMSE as a function of scalar decay $\lambda$, other parameters held at their scalar-$\lambda$ estimates (grey curve). The light blue point marks the estimated scalar $\hat\lambda$, close to the curve's minimum; $\lambda = 0.7$ or $1.0$ each carries a material wMSE penalty. The dark blue dashed line shows the tercile-$\lambda$ model wMSE, lower than any scalar $\lambda$, so no single decay rate recovers the gain from size-group estimation.}\label{fig:fig-emp-ewma-profile}
\end{figure}

The central finding is that the optimal decay rate depends on account size: large accounts are best served by recent experience, while small and mid-sized accounts benefit from several years of history.
Figure~\ref{fig:fig-emp-ewma-profile} plots held-out wMSE across a range of fixed EWMA decay rates.
The scalar \(\lambda\) estimated jointly on training data (\(\hat\lambda = 0.24\), light blue point) lands close to this out-of-sample profile's minimum.
Holding the other parameters at their scalar-\(\lambda\) estimates, setting \(\lambda\) to \(0.7\) or \(1.0\) each carries a material wMSE penalty.
The scalar estimate is nonetheless a compromise: the exposure-weighted likelihood is dominated by
large companies, and large companies prefer recency (low \(\lambda\)), so the
scalar estimate is pulled toward their optimum at the expense of small companies.

The principle extends beyond decay rates: \emph{any} pooled-parameter estimation in a
heterogeneous portfolio will be pulled toward the preferences of the
highest-exposure group.
Standard B-S and the GLMM pool in a similar way: B-S weights its between-account variance and collective mean by exposure, and the GLMM weights its likelihood by exposure. The same pull towards large accounts can therefore affect their effective credibility weights for small accounts.
Tercile or continuous \(\lambda\) avoids this by estimating decay rates within
size groups, preventing large-company dominance from distorting the small-account signal.

\begin{figure}[H]

{\centering \includegraphics[width=0.72\linewidth]{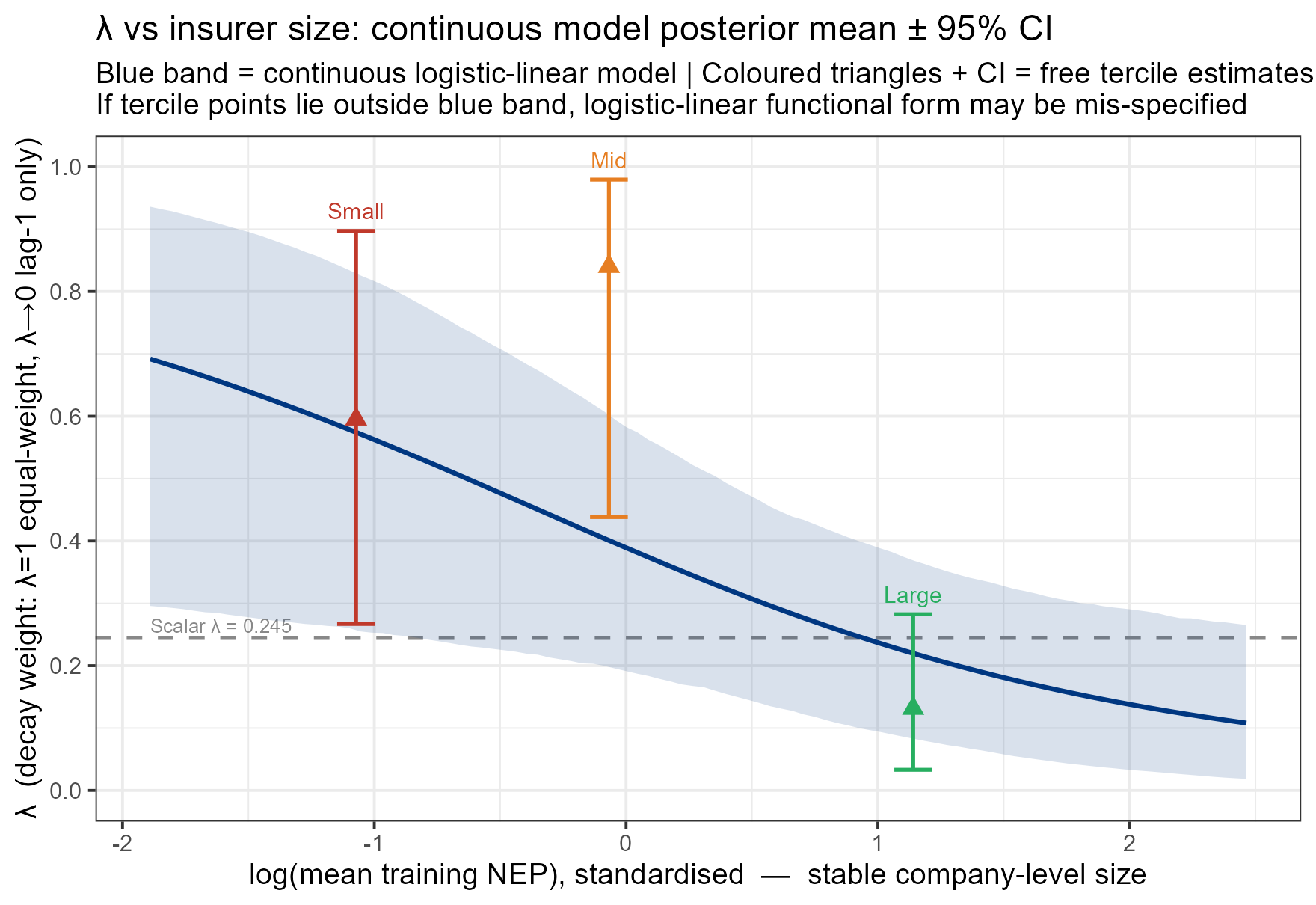} 

}

\caption{Estimated $\lambda$ as a function of insurer size (log mean training NEP $\bar{E}_i$). Dark blue band: continuous $\lambda$ model posterior mean and 95\% CI; coloured triangles with 95\% CIs: free tercile estimates (red = Small, orange = Mid, green = Large). All three CIs overlap the continuous band, but the Mid estimate lies above it. LOO-CV favours the tercile specification only weakly (ELPD ratio 1.6); held-out wMSE agrees (Table~\ref{tab:tbl-emp-results}).}\label{fig:fig-emp-lambda-cont}
\end{figure}

\begin{figure}[H]

{\centering \includegraphics[width=0.72\linewidth]{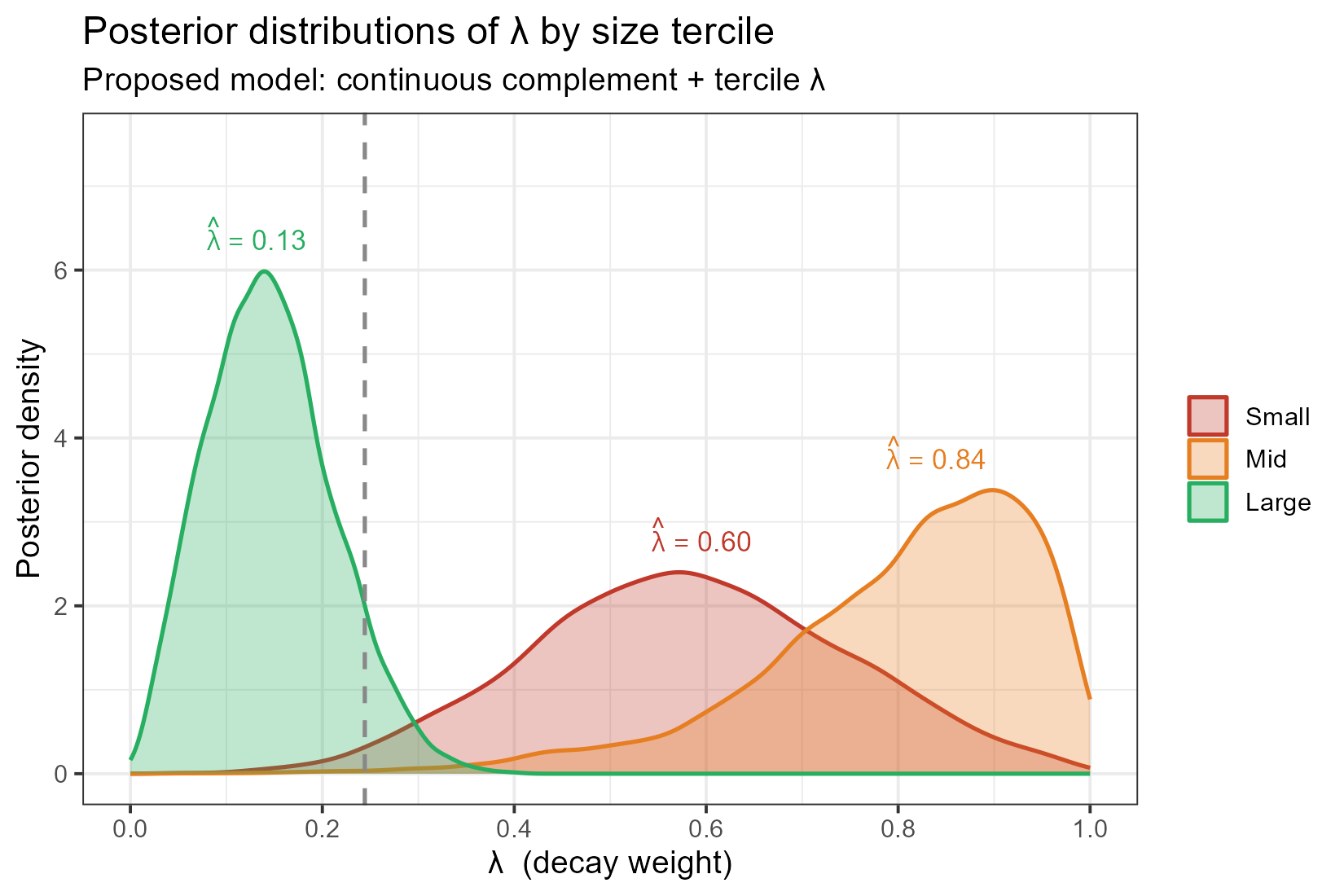} 

}

\caption{Posterior distributions of $\lambda$ by size tercile. Large is narrow and well-identified; Small and Mid are wide and largely overlapping (CI values in text above).}\label{fig:fig-emp-lambda-tercile}
\end{figure}

Figures \ref{fig:fig-emp-lambda-cont} and \ref{fig:fig-emp-lambda-tercile} show the size gradient in two complementary views: the first traces the continuous-\(\lambda\) model's posterior mean and 95\% posterior credible interval (CI) band across insurer size; the second shows the posterior distribution for each tercile separately.

The \emph{Large} posterior is narrow and well-identified
(\(\hat\lambda \approx 0.13\), 95\% CI: \([0.03,\, 0.28]\)),
confirming that large companies rely almost entirely on the most recent year.
The \emph{Small} and \emph{Mid} posteriors are both wide and largely overlap each other, but both are centred well above Large's:
\(\hat\lambda \approx 0.60\) (\([0.27,\, 0.9]\)) and
\(\hat\lambda \approx 0.84\) (\([0.44,\, 0.98]\)) respectively.

Only Large companies provide a strong, well-identified \(\lambda\) signal; the continuous-\(\lambda\) model's monotone constraint regularises Small and Mid by tying them to the size trend.
Widening the logit-scale prior on each tercile's \(\lambda\) from \(N(0, 1.5)\) to \(N(0, 2.5)\) raises the Mid estimate from 0.84 to 0.93 but changes held-out wMSE by only 0.6\%, so the headline does not depend on this prior.

The continuous-\(\lambda\) model lowers held-out wMSE from 8.61 (scalar \(\lambda\)) to 8.38 \(\times 10^{-3}\), and the tercile model lowers it further to 7.96 \(\times 10^{-3}\). A direct LOO-CV comparison favours the tercile specification only weakly (ELPD ratio 1.6\(\times\)SE), so this further 0.42 \(\times 10^{-3}\) is consistent with a discontinuity between Large and the other accounts but does not rule out a smooth gradient. A two-group specification (Small and Mid pooled against Large) performs the same (held-out wMSE 7.97 against 7.96 \(\times 10^{-3}\); LOO-CV difference \(-0.5\), SE 0.6), and its pooled Small-and-Mid \(\hat\lambda \approx 0.74\) (95\% CI 0.45 to 0.94) lies entirely above Large's (0.13; 0.03 to 0.28), so the tercile split is a convenient description of what is effectively a two-regime pattern.

\medskip

\noindent\textbf{Economic interpretation.}
A small company recording modest annual premium generates a claims record
with high year-to-year noise relative to the underlying signal.
To reliably separate signal from noise, the model must average across
multiple years: a single year's loss ratio could easily be driven by
one large claim.\footnote{This case study uses the CAS triangle database's company-year aggregated incurred losses; individual claim detail needed for per-occurrence capping is not available, and the aggregated figures themselves are not capped in any way. This noise therefore reflects low claim-count exposure and large-loss lumpiness conflated together; Section \ref{sec:applying}'s attritional-capping guidance would separate the two where per-occurrence data is available in a live deployment, and $\hat\lambda_\text{Sm}$ may run lower on capped data if lumpiness is the dominant driver.}
A large company generates many claims per year, which makes a single year's loss
ratio a more precise estimate of the true underlying rate and lowers the marginal value of an additional year.
Both forces push toward recency for large companies and longer memory for small ones.
The point estimates give Mid companies a slightly longer memory than Small (\(\hat\lambda_{\text{Md}} \approx 0.84\) against \(\hat\lambda_{\text{Sm}} \approx 0.60\)), but the intervals overlap almost entirely and the two groups' autocorrelations are alike (0.45 to 0.40 for Small and 0.46 to 0.37 for Mid, from lag 1 to lag 3, against 0.51 to 0.18 for Large), so we do not read the ordering as a real difference between Small and Mid.
\(Z\) and \(\lambda\) answer different questions: \(Z\) is \emph{how much to trust own experience}, and \(\lambda\) is \emph{how recent that experience should be}. They need not vary with size alike. In this dataset the proposed model's \(Z\) is nearly flat over the observed exposure range (fitted \(\hat{Z} \approx 0.61\)--\(0.67\); Section~\ref{sec:ffvalidation}), far flatter than the B-S weights imply (\hyperref[app:implied-z]{Appendix~E}), whereas \(\lambda\) falls sharply with size. The flatness depends on the exposure weighting of the likelihood: with weights closer to precision weights, the fitted slope turns negative (\hyperref[sec:weights-sens]{Appendix~D}). A weakly informative prior on \(b_Z\) (Section~\ref{sec:core-formula}) encodes \(Z\) as rising with account size, but unconstrained MLE on this data alone gives a slope indistinguishable from zero (\(\hat{b}_Z = 0.06\), 95\% CI \(-0.14\) to \(0.26\)).

\medskip

\noindent\textbf{Generalisability caveat.}
The specific values (\(\hat\lambda_\text{Lg} \approx 0.13\), \(\hat\lambda_\text{Sm} \approx 0.60\)) and the size gradient (including the non-monotone ordering of Small and Mid) are estimated on this dataset and should not be imported directly to other portfolios (in the case study, ``Large'' means mean training NEP above about \$8.7M): practitioners should re-estimate \(\lambda\) from their own data, using the LOO-CV profile as a guide to whether the size gradient is supported by the data in hand (Section~\ref{sec:cross-lob} shows the same direction on Other Liability, where \(\hat\lambda\) falls monotonically with size).
\hyperref[app:effective-memory]{Appendix~B} translates the estimated tercile
\(\hat\lambda\) values into effective lookback windows, providing a
practitioner-facing interpretation of the gradient.

\medskip

\noindent\textbf{Development uncertainty and $\lambda$.}
This case study uses losses at their most mature development, so the estimated \(\lambda\) values reflect genuine year-to-year drift and not development-estimation noise, and the model never sees an immature recent-year input of the kind recommended in Section~\ref{sec:applying} for live deployment. How \(\lambda\) changes once immature ultimates enter the training set is not established here, and the direction may depend on how the model is fitted and deployed. Testing it would need each training year's ultimate re-estimated using only the development available at that historical date, which we leave to future work. The effect should matter more for longer-tailed lines; the Other Liability validation in Section~\ref{sec:cross-lob} uses the same fully developed convention, so it does not test it either.

\subsection{Z Parameterisation: Form and Features}\label{z-parameterisation-form-and-features}

\subsubsection{Functional form validation}\label{sec:ffvalidation}

\begin{figure}[H]

{\centering \includegraphics[width=0.83\linewidth]{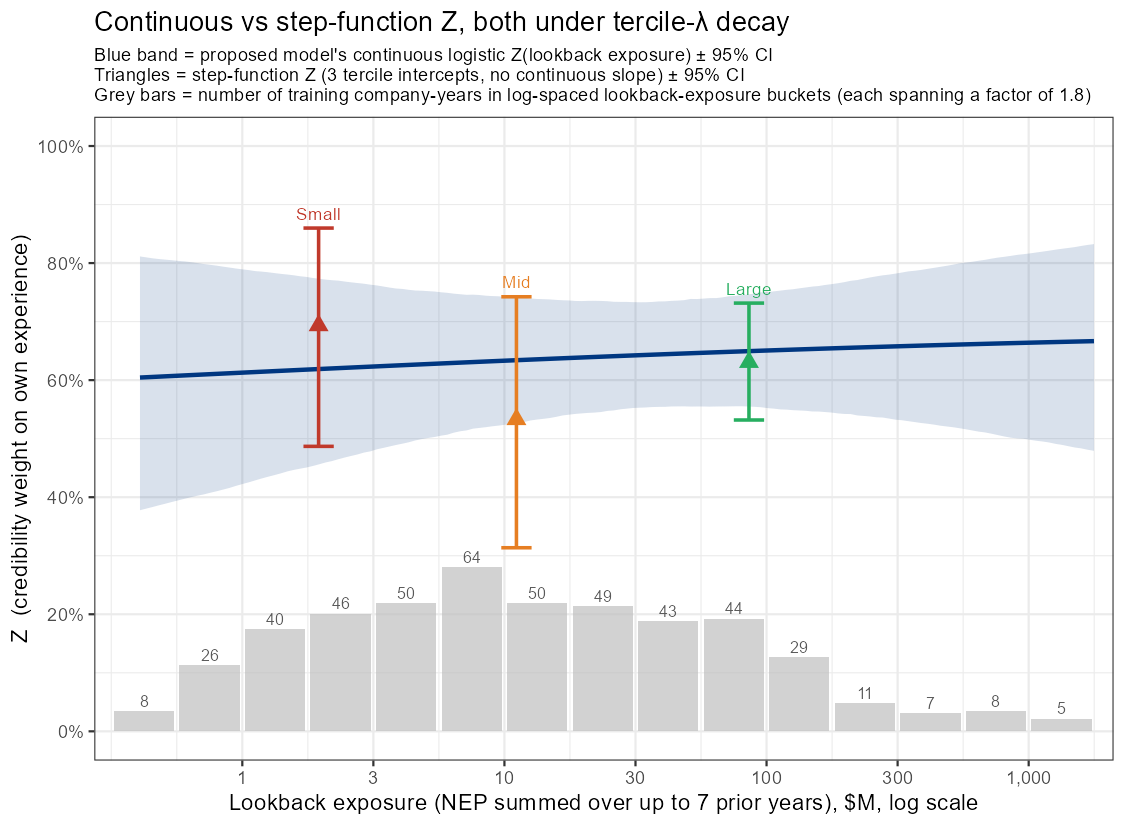} 

}

\caption{Continuous vs step-function $Z$, both under tercile-$\lambda$ decay. Blue band: proposed model's continuous logistic $Z(\tilde{E})$, posterior mean and 95\% CI. Coloured triangles: step-function $Z$ (three tercile intercepts, no continuous slope) with 95\% CIs. All three tercile point estimates lie within the logistic-linear confidence band, providing little statistical evidence against the assumed functional form. Grey bars: number of training company-years in log-spaced buckets of lookback exposure, each spanning a factor of 1.8 (\$M, log scale).}\label{fig:fig-z-shape}
\end{figure}

Is the logistic-linear functional form for \(Z\) itself adequate?
We answer this before asking which features matter, by stress-testing it against a step function with a separate \(Z\) intercept per size tercile and no continuous slope on exposure at all. Both specifications share the same tercile-\(\lambda\) decay structure (Section~\ref{sec:lambda-decay}); only the functional form of \(Z\) differs between them.
Figure~\ref{fig:fig-z-shape} plots the posterior mean and 95\% posterior credible interval (CI) of the
logistic-linear model's continuous \(Z(\tilde{E})\) curve (blue band) alongside the
step-function tercile estimates (coloured triangles with 95\% CIs).
If the logistic-linear form were badly misspecified, the tercile estimates would fall
outside the blue band.

The point estimates form a shallow U-shape rather than the smooth increase the logistic-linear form predicts:
Small \(\hat{Z} \approx 0.68\),
Mid \(\hat{Z} \approx 0.53\),
Large \(\hat{Z} \approx 0.63\).
This pattern is not statistically significant.
The Small and Mid tercile CIs (shown in Figure~\ref{fig:fig-z-shape}) are wide: each rests on just
32 training companies, which carry only 1.3\% and 7.0\% of training NEP respectively.
The CIs of all three groups overlap one another and the blue band substantially.
The held-out results agree: a pure step-function \(Z\) (three
tercile intercepts, no continuous slope, paired with tercile \(\lambda\))
achieves wMSE \(= 8.05 \times 10^{-3}\) on the
test set, marginally \emph{above} the proposed continuous-\(Z\),
tercile-\(\lambda\) model's
\(7.96 \times 10^{-3}\).

The elevated Small-company \(\hat{Z}\) is consistent with survivorship bias
(the balanced-panel filter retains only companies present throughout AY 1998--2007;
surviving small companies may genuinely be more stable).
However, the logistic-linear form remains the right deployment default: it matches
or improves on the unconstrained step function out-of-sample, and it
additionally encodes a basic prior any experience-rating framework should
respect: credibility rising monotonically with exposure. A non-monotone
step function would be difficult to defend to an underwriter (why should
a Mid-sized account receive less credibility than both smaller and larger
accounts?) even if its held-out performance were stronger than it is here.

\subsubsection{Feature selection}\label{feature-selection}

We tested three candidate features for \(Z\) beyond size: CoV of historical loss ratios, maximum absolute deviation from market mean and NEP growth. A fourth, years of history, cannot be separated from calendar year in a balanced panel, so it is not testable here.
Account size is the only feature that clears the training-based tests (LRT, and a LOO-CV ratio of about 2): CoV is insignificant once size is controlled; NEP growth adds nothing (LRT \(p = 1.0\)); the maximum deviation from the market mean improves LOO-CV only weakly (ratio 1.6) and does not improve held-out wMSE, so we do not adopt it.
Full test results are in \hyperref[app:full-comparison]{Appendix~D}.
Exposure-only \(Z\) is therefore the empirically supported choice. Candidate features in a deployment should be tested against this base using the same LRT and/or LOO-CV workflow.

Company-level random effects on \(a_Z\), \(b_Z\) or both are clearly positive in-sample (standard deviations of 1.3--1.5 on the logit scale), but LOO-CV shows no clear gain (ELPD differences of 2--5 against standard errors of 7--9) and held-out wMSE is no better (8.4--8.6 against 8.0 \(\times 10^{-3}\), evaluated the same way), so we keep \(Z\) a smooth function of size.

\subsection{Cross-Line Validation: Other Liability}\label{sec:cross-lob}

To test whether the findings from Commercial Auto (CA) generalise to a structurally different line, we apply the same framework to the Other Liability -- Occurrence (OL) line from the CAS triangle database (101 companies, 202 held-out company-year observations, AY 2001--2007).
Other Liability has more right-skewed small-company loss ratios than CA, stress-testing the framework's choice of blend form.

\begin{figure}[H]

{\centering \includegraphics[width=0.95\linewidth]{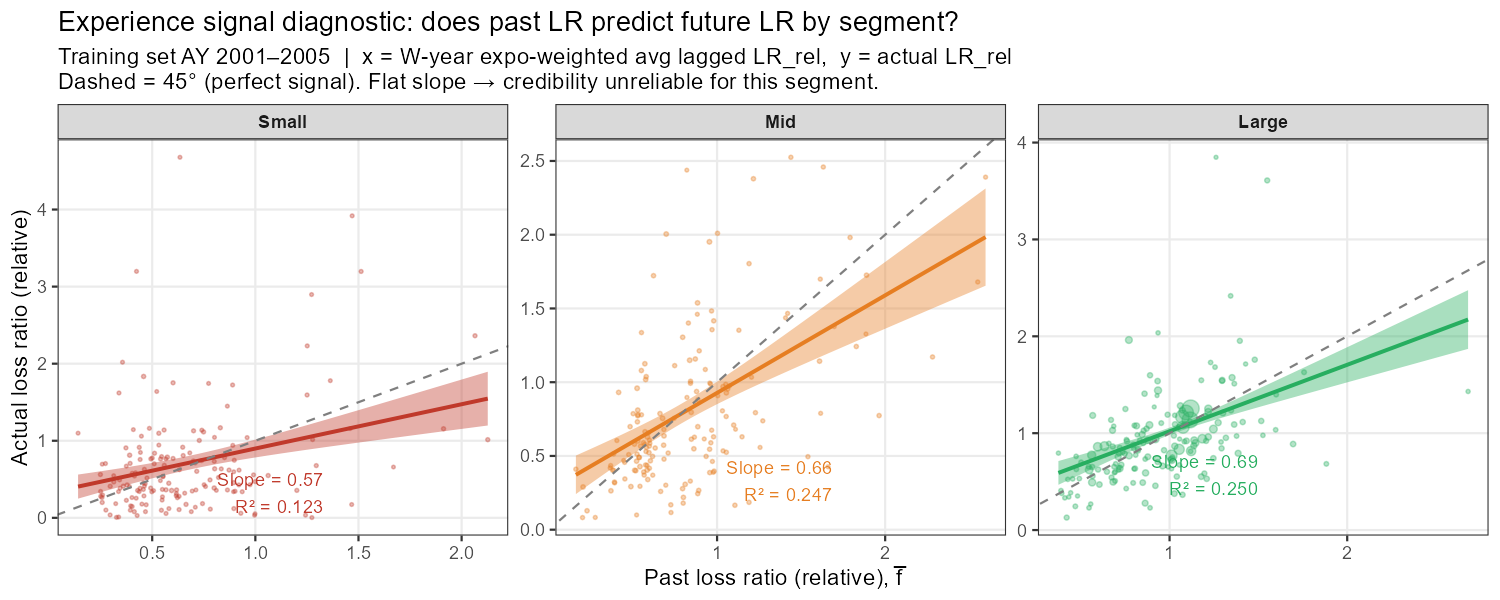} 

}

\caption{Pre-adoption signal check for Other Liability: actual next-year relative loss ratio versus the flat ($\lambda=1$) multi-year average $\bar{f}$ by size tercile (training data). NEP-weighted OLS slope and $R^2$ annotated per panel. Unlike Commercial Auto (Figure \ref{fig:fig-eda-signal}), where Large companies show the weakest predictive relationship under flat averaging ($R^2 = 0.12$, reflecting stale-year averaging rather than a lack of signal; Section \ref{sec:eda}) and Small the strongest ($R^2 = 0.45$), OL Small companies show the weakest signal ($R^2 = 0.12$), with Mid and Large nearly identical ($R^2 \approx 0.25$). All three slopes lie below the 45$^\circ$ line, indicating that the flat average does not fully capture the magnitude of year-to-year variation; the experience signal exists but is attenuated.}\label{fig:fig-ol-signal}
\end{figure}

\subsubsection{Temporal Decay and the Geometric Blend}\label{temporal-decay-and-the-geometric-blend}

In both lines, small companies favour multi-year averaging while large companies concentrate weight on the most recent year (Table~\ref{tab:lob-lambda}).

\begin{table}[!h]
\centering
\caption{\label{tab:lob-lambda}Estimated decay parameter $\hat{\lambda}$ by size group and line of business.
      CA estimates are posterior means from Bayesian (brms) model fits;
      OL estimates are maximum-likelihood point estimates, sufficient for the structural,
      cross-line comparison drawn here; Bayesian uncertainty quantification was not repeated for OL.}
\centering
\begin{tabular}[t]{lrr}
\toprule
Size group & CA $\hat{\lambda}$ (Bayesian) & OL $\hat{\lambda}$ (MLE)\\
\midrule
Small & 0.60 & 0.73\\
Mid & 0.84 & 0.47\\
Large & 0.13 & 0.24\\
Scalar & 0.24 & 0.32\\
\bottomrule
\end{tabular}
\end{table}

OL produces a monotone gradient (Small \(>\) Mid \(>\) Large), whereas in CA the Mid tercile sits above Small; the direction is consistent but the internal ordering differs.
The OL gradient is similar in size: \(\hat\lambda_{\text{Sm}} = 0.73\),
\(\hat\lambda_{\text{Md}} = 0.47\),
\(\hat\lambda_{\text{Lg}} = 0.24\), a spread of
\(0.49\) versus
\(0.46\) for CA.
A likelihood-ratio test confirms that estimating a single, shared decay
rate (along with complement flexibility) decisively improves on a
\(\lambda=1\), flat-complement restriction in both lines (the \(Z\)-slope
itself, held free on both sides of this test, is assessed separately in
Section~\ref{sec:zfeatures}):
\(\chi^2 = 126.4\), \(\text{df} = 3\),
\(p \approx 3.2 \times 10^{-27}\) for CA;
\(\chi^2 = 95.7\), \(\text{df} = 3\),
\(p \approx 1.3 \times 10^{-20}\) for OL ---
a decisive rejection of this restricted form in both lines.

Whether that shared rate should be split further by size tercile is a
separate question. Despite OL's wider gradient, the tercile-\(\lambda\)
model improves on the scalar for OL only negligibly (wMSE
16.56
versus 16.61 \(\times 10^{-3}\)):
an approximate (Hessian-based) 95\% CI on the tercile-\(\lambda\) estimates is wide for Small
(0.17--0.98) and Mid (0.23--0.73), too imprecise to add reliable predictive
value, while Large's is comparatively tight (0.11--0.45).

A further finding specific to Other Liability: geometric (log-space) blending of the EWMA reduces wMSE by 0.8--5.8\% across paired OL model variants (arithmetic is modestly better for CA), consistent with OL small-company loss ratios being more right-skewed.

With cross-line validation established, how does the framework compare to the natural statistical alternative, a mixed-effects model?

\subsection{Comparison with Mixed-Effects Models}\label{sec:glmm-comparison}

Table~\ref{tab:lob-results} compares the best logistic model, GLMM, and B-S standard across both lines on held-out wMSE.

\begin{table}[!h]
\centering
\caption{\label{tab:lob-results}Best-model wMSE comparison across both lines of business,
        held-out test set.
        Best logistic: LogCred (tercile $\lambda$) for CA;
        LogCred (two-rate $\lambda$, geometric blend)$^\dagger$ for OL.
        $^\dagger$Fitted $\hat\lambda_1=0.70$ (lag~1$\to$2 step), $\hat\lambda_2=0.03$ (lags~3+ decay); only lags~1--2 carry meaningful weight in the pooled estimate, consistent with large, short-memory accounts dominating the weighted fit (Section~\ref{sec:lambda-decay}), not necessarily an OL-specific property.
        Percentage improvements are relative to B{\"u}hlmann--Straub (standard, full history).
        wMSE uses the realised test-year portfolio mean $\mu_t$ (the same oracle convention as Table~\ref{tab:tbl-emp-results});
        the deployable-setting adjustment for CA is quantified in Section~\ref{sec:leakage-check}.}
\centering
\begin{tabular}[t]{lrlrl}
\toprule
Model & CA wMSE ($\times 10^{-3}$) & CA vs B-S & OL wMSE ($\times 10^{-3}$) & OL vs B-S\\
\midrule
Bühlmann--Straub (standard) & 10.97 & --- & 16.34 & ---\\
Best logistic & 7.96 & -27.4\% & 15.63 & -4.4\%\\
GLMM (random intercept + size) & 10.62 & -3.2\% & 16.18 & -1.0\%\\
\bottomrule
\end{tabular}
\end{table}

In CA, the logistic framework improves on standard B-S by
27.4\%, while the GLMM improves by only
3.2\%.
In OL the gap narrows markedly: the best logistic (two-rate \(\lambda\), geometric blend) improves on standard B-S by 4.4\% against the GLMM's 1.0\%.

The OL ``best logistic'' (two-rate \(\lambda\), pooled across sizes) is the best of many variants on the test set, so the margin is indicative only: the tercile-\(\lambda\) model used for CA is 1.3\% behind standard B-S and 2.3\% behind the GLMM. OL therefore shows the same gradient direction in-sample but no reliable held-out advantage.

The by-tercile breakdown below focuses on CA, where the logistic-GLMM gap is largest and bootstrap calibration diagnostics are available; it shows where the overall wMSE advantage arises and where it falls away. Held-out calibration slopes (90\% bootstrap CIs, company-level resample, 2,000 replications) diagnose why.

\medskip

\noindent\textbf{Small accounts:}
The logistic outperforms GLMM substantially on wMSE (\(67.88\) vs \(117.21\) \(\times 10^{-3}\)).
The slope diagnostic reveals why: the logistic achieves a slope of \(0.99\) (CI \([0.73,\,1.14]\)), tightly centred on ideal calibration.
GLMM's Small slope of \(4.3\) (CI \([2.73,\,5.54]\)) indicates under-crediting, as for standard B-S with slope 4 (the GLMM's own shrinkage gives \(\hat{Z} \approx 0.15\) for Small, as for standard B-S; \hyperref[app:implied-z]{Appendix~E}).

\noindent\textbf{Mid-sized accounts:}
The GLMM outperforms the logistic on wMSE (\(23.72\) vs \(25.67\) \(\times 10^{-3}\)), as does stratified-\(K\) B-S (\(23.71\)), while standard B-S is slightly behind the logistic (\(26.04\)). The GLMM's pooled Mid slope is also closer to 1.0 (\(1.17\), CI \([0.68,\,1.82]\), against \(0.71\), CI \([0.31,\,1.07]\), which just includes 1.0 at its upper end), as is standard B-S's (\(1.04\)), but this is weak evidence of better calibration, as the by-year slopes below show.

\begin{figure}[H]

{\centering \includegraphics[width=0.85\linewidth]{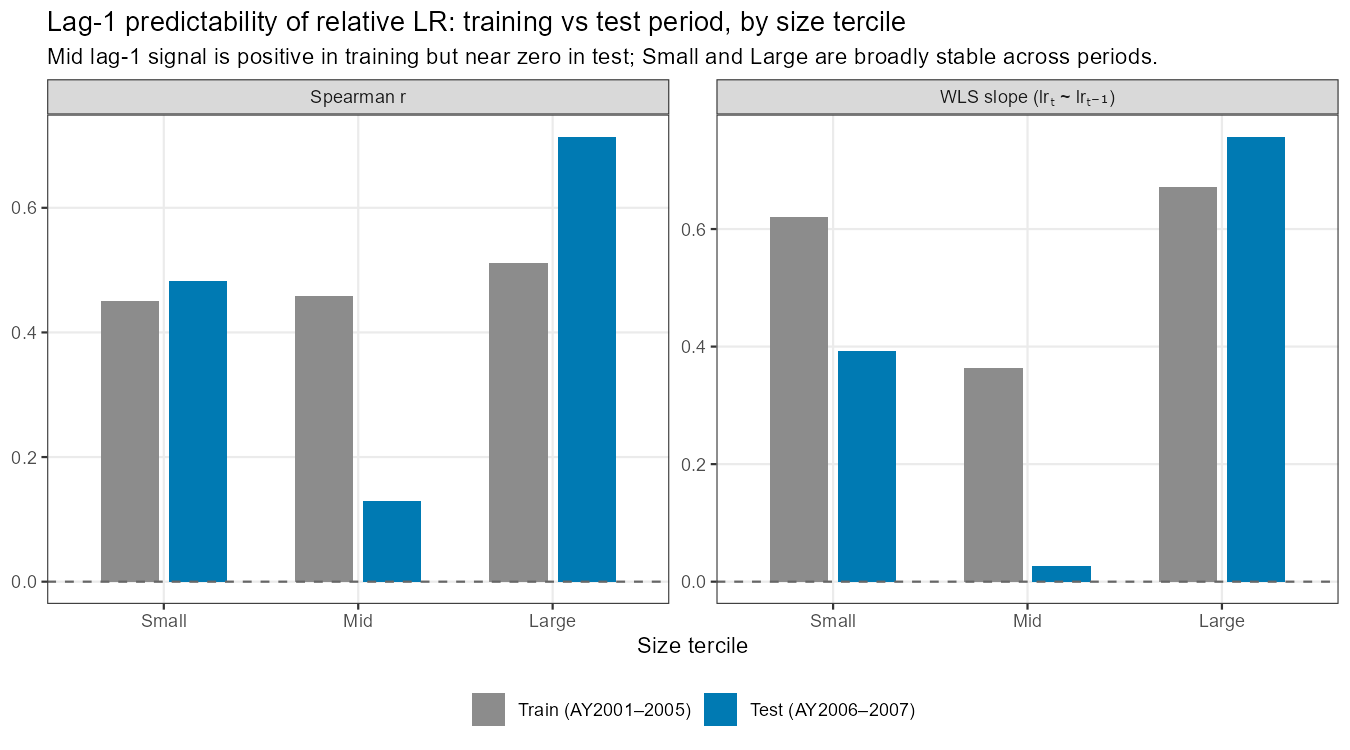} 

}

\caption{Lag-1 predictability of relative loss ratio (training AY 2001--2005 vs test AY 2006--2007) by size tercile. Spearman correlation (left) and exposure-weighted OLS slope of $\mathrm{LR}_t$ on $\mathrm{LR}_{t-1}$ (right). Mid's lag-1 signal is positive in training (slope $\approx 0.36$) but near zero in the test period (slope $\approx 0.03$); Small and Large are broadly stable.}\label{fig:fig-predictability-shift}
\end{figure}

The slope difference reflects a shift in predictability (Figure~\ref{fig:fig-predictability-shift}): Mid's lag-1 signal was positive in training but near zero in the test period, so the logistic's high \(\hat{Z}\) (\(\approx 0.64\)) over-weighted uninformative test experience, whereas the lower \(\hat{Z}\) of the GLMM and standard B-S (\(\approx 0.47\) and \(0.48\); \hyperref[app:implied-z]{Appendix~E}) was better matched to the collapsed signal. A bootstrap test (2000 company-level resamples) puts the GLMM\(-\)logistic wMSE difference at \(-1.91 \times 10^{-3}\) (90\% CI {[}\(-4.19\), \(0.35\){]}, mostly below zero): suggestive of a GLMM advantage in Mid, but not significant. Slope CIs also overlap substantially.

\medskip

Calibration slopes by test year show the instability directly: for Mid accounts, each of standard B-S, stratified B-S, the GLMM and the proposed model has a slope near zero in AY 2006 (0.03 to 0.20) and above 1 in AY 2007 (1.43 to 2.07). The pooled Mid slopes (0.71 to 1.17) therefore average two years that miscalibrate in opposite directions, and their distance from 1.0 should not be over-interpreted. The Mid wMSE gap is not statistically significant, and neither the credibility weight nor the complement explains it, so it may reflect this test period rather than a structural weakness.

\medskip

\noindent\textbf{Large accounts:}
The logistic clearly outperforms on both metrics: wMSE \(5.74\) vs \(8.14\) (GLMM) and \(8.45\) (B-S standard), all \(\times 10^{-3}\).
All logistic models are well-calibrated (slope \(\approx 1.06\), CI \([0.83,\,1.32]\)); GLMM is acceptable in slope terms (\(1.04\)) but with a wider CI and substantially worse wMSE.
Standard B-S (pooled \(K\)) assigns a typical Large account \(\hat{Z} \approx 0.93\), well above the logistic's near-flat weights (with no calibration cost: Large slope 1.03); its worse Large wMSE comes mainly from weighting every past year equally, not from \(Z\): B-S with EWMA decay reaches \(5.57\), close to the logistic's \(5.74\) (\(\times 10^{-3}\); \hyperref[app:full-comparison]{Appendix~D}).

\medskip

\noindent\textbf{GLMM summary:}
GLMM's mid-company performance comes at the cost of Small miscalibration (slope \(4.3\), CI entirely above 1.0) and worse Large predictions.\footnote{Two targeted logistic variants (free per-tercile $Z$ intercepts; per-company CoV as a $Z$-covariate) both fail to close the Mid gap, consistent with a data-period regime change rather than a modelling deficiency. A stratified GLMM (separate $\sigma^2_{u,s}$ per tercile) is unidentified at $\approx 32$ companies per tercile, with $\sigma^2_u$ and $\phi$ collapsing to boundary values whether $\phi$ is fixed at the pooled estimate or freely estimated, and its Mid wMSE of 26.86 is worse than the pooled GLMM's.}
The broader pattern reflects \textbf{variance-component pooling}: a single pooled \(K\) cannot simultaneously serve terciles whose stratified \(K\) values span almost two orders of magnitude (\(577\) to \(51,360\), Figure~\ref{fig:fig-intro-lr-dist}), producing a steeply rising GLMM credibility weight (\(\hat{Z} \approx 0.15\), \(0.47\), \(0.85\) on average for Small, Mid, Large accounts; implied \(K\) of \$18.3M against B-S's \$17.5M) versus the proposed logistic's near-flat \(\approx 0.61\)--\(0.67\). Figure~\ref{fig:fig-z-curve-comparison} (\hyperref[app:implied-z]{Appendix~E}) compares all three frameworks directly.

\subsection{Predictive Intervals}\label{sec:empirical-uq}

The Gamma family is well-suited to right-skewed loss ratios and exposes a dispersion parameter \(\phi\) that can be modelled as a function of account size, which is the key choice for interval width. Full derivation and coverage plots are in \hyperref[app:variance]{Appendix~C}.
Table~\ref{tab:uq-summary} summarises the key finding: constant dispersion produces severely under-wide intervals for small accounts (54.7\% Tier 3 coverage vs nominal 95\%); exposure-varying dispersion (\(\log\phi_{it} = \phi_0 + \phi_1\log E_{it}\)) raises coverage to about 90\% across all size groups (nominal 95\%) with little effect on point estimates in this case study.
Whether \(\phi\) is held constant or allowed to vary with account size determines coverage quality; exposure-weighting the likelihood has negligible effect.
The first two rows below isolate the effect of dispersion flexibility alone (scalar \(\lambda\) throughout, comparing constant vs varying \(\phi\)); the third row (proposed model for intervals) then adds the tercile-\(\lambda\) point-prediction gain on top.

\begin{table}[!h]
\centering
\caption{\label{tab:uq-summary}Tier 3 predictive interval coverage at nominal 95\%, held-out test set ($n=192$).
    Exposure-varying dispersion restores uniform coverage across size terciles;
    likelihood weighting has negligible effect. Bold = proposed model for intervals.}
\centering
\begin{tabular}[t]{lccccc}
\toprule
Model & wMSE ($\times10^{-3}$) & Gini (\%) & Overall & Small & Large\\
\midrule
Scalar $\lambda$, constant $\phi$ & 8.61 & 76.5 & 67.2 & 54.7 & 82.8\\
Scalar $\lambda$, $\phi{\sim}\log E$ & 8.92 & 77.9 & 90.6 & 89.1 & 89.1\\
\textbf{Tercile $\lambda$, $\phi{\sim}\log E$} & \textbf{8.11} & \textbf{80.1} & \textbf{90.6} & \textbf{89.1} & \textbf{89.1}\\
\bottomrule
\end{tabular}
\end{table}

Full seven-model comparison (including continuous-\(\lambda\) and weighted-likelihood variants) is in \hyperref[app:variance]{Appendix~C}.

\section{Simulation: Confirmation Under Controlled Conditions}\label{sec:simulation}

\begin{figure}[H]

{\centering \includegraphics[width=0.85\linewidth]{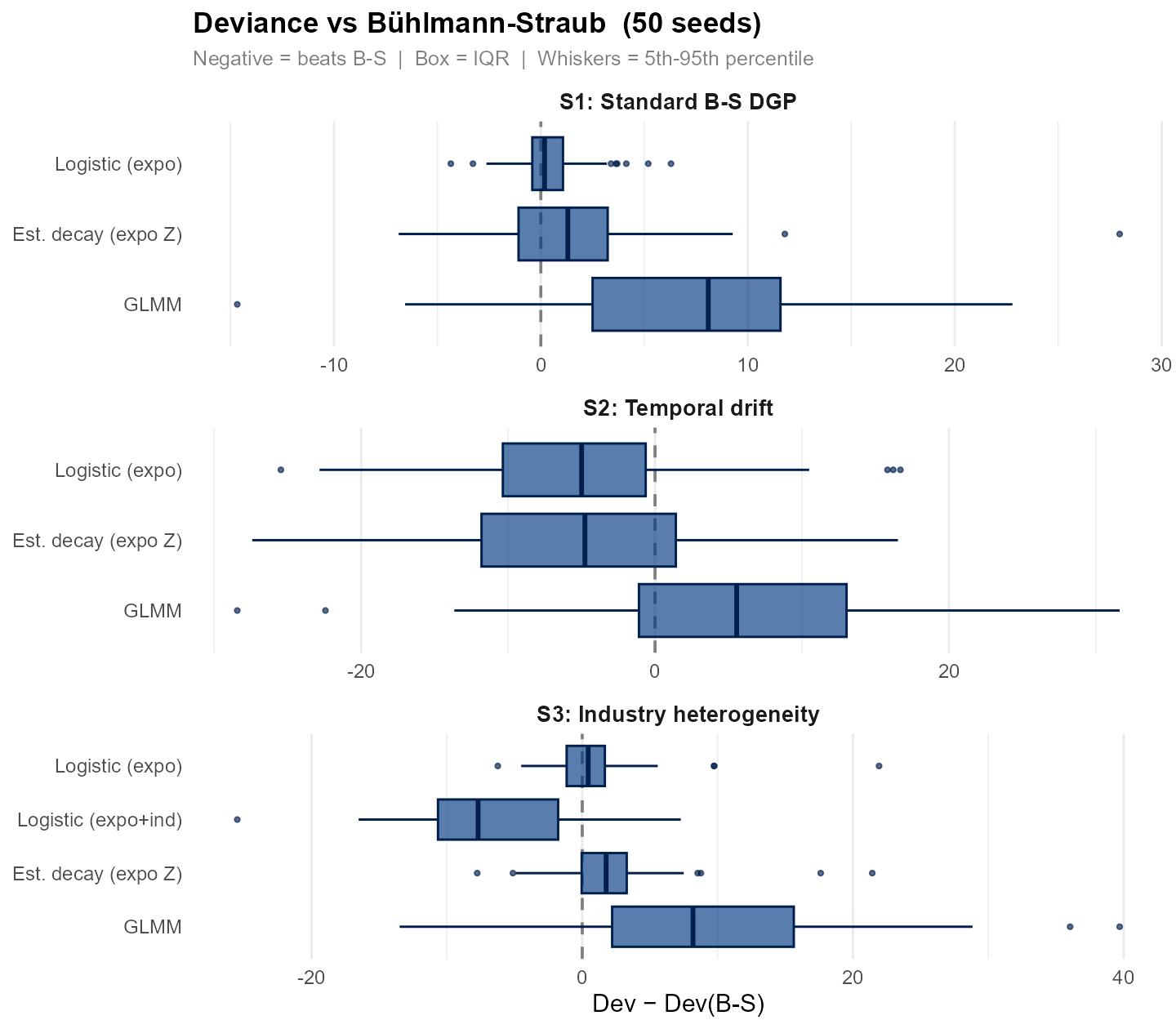} 

}

\caption{Deviance relative to Bühlmann--Straub across 50 seeds, four representative models (full model key in Appendix F). Negative = better than B-S; boxes = IQR; whiskers = 5th--95th percentile. S1 confirms no spurious gain when the null is true. In S2, estimated-decay (expo Z) leads while the GLMM, which averages over all training years equally, ranks near last. In S3, the logistic model with the industry covariate performs best; without the covariate it is close to B-S (slightly worse).}\label{fig:fig-sim-boxplot}
\end{figure}

Three simulation scenarios test the framework under controlled conditions using a Poisson data-generating process (DGP), distinct from the empirical Gamma likelihood and providing an out-of-family robustness check (Figure~\ref{fig:fig-sim-boxplot}). Each scenario simulates 1,000 accounts over 8 years, repeated for 50 seeds, with models ranked by Poisson deviance on the held-out final year.

S1 (standard B-S DGP) is a sanity check: when the null hypothesis is true (homogeneous \(K\), no temporal drift), the logistic model is indistinguishable from B-S, confirming the framework is conservative: it reduces to B-S when there is nothing additional to identify. S2 and S3 each test one structural property: S2 confirms that estimating \(\lambda\) from data improves predictions when account signals drift over time, and S3 confirms that the logistic \(Z\) recovers heterogeneous-\(K\) structure when a relevant account-level covariate is supplied. Full DGP specification is in \hyperref[app:dgp]{Appendix~F}.

\textbf{Scenario: temporal drift (S2).}
In S2, AR(1) persistence \(\psi\) is drawn from \(\mathcal{U}(0.1, 0.6)\) per seed. This scenario tests whether estimating \(\lambda\) from data recovers accuracy lost to a fixed-\(\lambda\) (or no-decay) B-S assumption when account-level signals genuinely drift over time.

\begin{table}[!h]
\centering
\caption{\label{tab:tbl-s2-sim}Multi-seed summary: temporal drift scenario, S2 (50 simulations). $\Delta$Dev = Dev(model) $-$ Dev(B-S); negative = better than B-S. Better = percentage of seeds in which the model has lower deviance than B-S. Mean Rank: 1 = best. Model definitions in \hyperref[app:dgp]{Appendix~F}. Ranks are over all models fitted, not only those shown. B-S ranks in the lower half on average (mean rank shown), so ignoring temporal structure is costly on average across drift speeds. Bold = best-performing model (lowest Mean Rank).}
\centering
\resizebox{\ifdim\width>\linewidth\linewidth\else\width\fi}{!}{
\begin{tabular}[t]{lrrrrr}
\toprule
Model & Mean Dev & SD Dev & \% Better & Mean Rank & \% Rank 1\\
\midrule
\textbf{Est. decay (expo Z)} & \textbf{-5.17} & \textbf{9.55} & \textbf{66} & \textbf{4.20} & \textbf{6}\\
Logistic (expo) & -4.85 & 9.23 & 76 & 4.34 & 18\\
EWMA (fixed λ) & -5.01 & 9.74 & 66 & 4.44 & 12\\
Logistic (full) & -2.94 & 10.45 & 68 & 5.00 & 2\\
Het-K B-S & -2.32 & 11.57 & 66 & 5.10 & 16\\
Est. decay (full Z) & -2.62 & 12.52 & 64 & 5.24 & 16\\
Bühlmann-Straub & 0.00 & 0.00 & 0 & 6.76 & 12\\
GLM naive & 0.69 & 11.48 & 50 & 6.96 & 6\\
GLMM (corr) & 5.68 & 12.82 & 34 & 8.14 & 6\\
GLMM & 6.42 & 13.12 & 34 & 8.46 & 2\\
Base rate & 14.14 & 17.85 & 22 & 10.66 & 2\\
\bottomrule
\end{tabular}}
\end{table}

The logistic model with estimated \(\lambda\) (estimated-decay, expo \(Z\)) achieves the lowest mean rank, adapting to whichever decay rate the data exhibit, while the GLMM, which averages over all training years equally, ranks near last. The finding holds across the full range of drift speeds: whether the
portfolio drifts slowly or rapidly, ignoring temporal decay is costly
and the framework adapts automatically.

\textbf{Scenario: industry heterogeneity (S3).}
Accounts belong to three industry sectors with permanent-risk SD \(\sigma_B \in \{0.10, 0.25, 0.40\}\) (larger \(\sigma_B\) gives smaller \(K\), which ranges 17-fold); the industry label can be supplied as a covariate to the logistic \(Z\) function.
This scenario tests whether the framework recovers heterogeneous-\(K\) structure when a relevant feature is supplied.

\begin{table}[!h]
\centering
\caption{\label{tab:tbl-s3b-sim}Multi-seed summary: industry heterogeneity scenario, S3 (50 simulations, heterogeneous $K$ by sector DGP). $\Delta$Dev = Dev(model) $-$ Dev(B-S); negative = better than B-S. Better = percentage of seeds in which the model has lower deviance than B-S. Mean Rank: 1 = best. Model definitions in \hyperref[app:dgp]{Appendix~F}. Ranks are over all models fitted, not only those shown. Industry label available as covariate in rows marked \textit{(expo+ind)}; absent in \textit{Logistic (expo)}. Bold = best-performing model (lowest Mean Rank).}
\centering
\resizebox{\ifdim\width>\linewidth\linewidth\else\width\fi}{!}{
\begin{tabular}[t]{lrrrrr}
\toprule
Model & Mean Dev & SD Dev & \% Better & Mean Rank & \% Rank 1\\
\midrule
\textbf{Logistic (expo+ind)} & \textbf{-6.85} & \textbf{6.67} & \textbf{86} & \textbf{2.26} & \textbf{52}\\
Stratified B-S & -7.08 & 5.91 & 88 & 2.28 & 30\\
Bühlmann-Straub & 0.00 & 0.00 & 0 & 5.06 & 4\\
Logistic (expo) & 0.96 & 4.29 & 36 & 5.78 & 0\\
Het-K B-S & 2.90 & 7.33 & 36 & 6.32 & 0\\
EWMA (fixed λ) & 6.94 & 7.09 & 16 & 9.20 & 2\\
\bottomrule
\end{tabular}}
\end{table}

The withheld-feature row (Logistic (expo), no industry label) confirms the framework is conservative: without the label it reduces to standard B-S, so gains require the relevant information to be present.
With the industry label, the logistic model outperforms standard B-S by
6.85 deviance
units on average and in 86\% of seeds, and ranks first in
52\% of seeds.

Stratified B-S also performs strongly (rank 2.28), confirming the industry information resolves B-S's \(K\)-pooling limitation. The logistic model's advantage is that it absorbs the industry effect as a covariate in the \(Z\) function, requiring no manual stratification.

Both simulation scenarios confirm the structural nature of the framework's advantage.

\section{Practical Implementation}\label{sec:practical}

The framework requires more initial setup than moment-estimated Bühlmann--Straub (a panel dataset, a chosen likelihood, and a numerical optimisation), but this is a one-time cost per portfolio, and \hyperref[app:implementation]{Appendix~G} provides a complete R implementation that converges in seconds on the case-study data.
Practitioners wanting the minimum viable checklist immediately can go directly to the Quick Start (Section~\ref{sec:mvd}); the phases below give the full guide.
The implementation has two distinct phases with different cadences.
\textbf{Phase 1} (fitting) is done once, or periodically when retraining,
on a panel of historical account-years.
It produces a set of estimated parameters: \((\hat{a}_Z, \hat{b}_Z, \ldots)\) governing
the credibility weight, the decay rate \(\hat{\lambda}\) (one rate, or varying with account characteristics such as size) governing how quickly older years lose weight, and the
complement parameters.
\textbf{Phase 2} (applying) is done whenever an account comes up for renewal
or is written as new business, using the fitted parameters to produce a
proposed rate.

The framework is agnostic about the level at which an ``account'' is
defined.
It can operate on a single policy,
a named insured (e.g.~a group of policies under one relationship), or a
portfolio segment (class, industry, territory).
The mathematical structure (EWMA, logistic \(Z\), credibility blend)
is identical at each level.
What changes is what the observables represent: the unit of exposure
\(E_{it}\) (e.g.~turnover, vehicle-years, TIV, premium, \ldots), the loss
statistic \(y_{it}\) (individual policy loss, account aggregate,
or segment aggregate), and the source and form of the complement \(\mu_i\).

\subsection{Phase 1: Fitting the Model}\label{sec:applying}

\subsubsection{Step 1: Prepare the training panel}\label{step-1-prepare-the-training-panel}

The training panel consists of account-year records containing
annual losses \(C_{it}\), exposure \(E_{it}\), and any account-level
covariates to be used in the \(Z\) equation.

\medskip

\noindent\textbf{Lookback window $W$.}
The lookback window \(W\) is set by the analyst before fitting and is not estimated.
All available history up to a practical maximum is preferred, except years excluded as set out under Excluded years below. Among the years retained, \(\lambda\) downweights distant years automatically, so a long window costs little; the effective lookback formula in \hyperref[app:effective-memory]{Appendix~B} converts \(\hat\lambda\) to an interpretable number of years.

\noindent\textbf{Excluded years.}
A year may be excluded where it bears no relevance to current risk following a material book change, where its experience is abnormal in ways that no covariate or benchmark adjustment (on-levelling, trend, calendar-year normalisation) captures, or where development uncertainty is too great to translate reliably to an ultimate basis. Exclusions should be justified on grounds set in advance, not on their effect on the results.

\noindent\textbf{Minimum history requirement.}
Accounts with only one lag year provide no temporal autocorrelation signal for estimating \(\lambda\); with two or three lag years the signal exists but remains weak. There is no hard minimum; rather, the uncertainty interval for \(\lambda\) widens as accounts have less history, and this is the operational diagnostic.
In practice, five or more years is strongly preferable: with a low \(\lambda\) (e.g.~\(\hat\lambda \approx 0.13\) for Large accounts in the CAS data), a three-year window places weight \(\lambda^2 \approx 0.02\) on the oldest year (effectively discarding it), so the effective history is shorter than the nominal count suggests.
Short-history accounts can be included in the fit, as they contribute to complement estimation regardless of history length, but their \(\lambda\) contribution is weak. Where short-history accounts dominate, \(\lambda\) may be identified by the few data-rich accounts alone: a wide interval flags weak identification, but a narrow one does not rule this out (Section~\ref{sec:mvd}).

\medskip

\noindent The following pre-processing steps apply to the entire panel before fitting.
Both the EWMA inputs \(C_{i,t-k}\), \(E_{i,t-k}\) in equation \eqref{eq:ewma} and the likelihood observations \(y_{it}\) are drawn from the same adjusted series, so adjustments should be made once, consistently, across the full panel.

\medskip

\noindent\textbf{Loss development.}
Losses should be translated to an estimated ultimate basis before computing \(y_{it}\). A Bornhuetter--Ferguson approach with a year-specific prior derived from \(\mu_i\) keeps the development and credibility steps consistent when the complement is supplied externally (Cape Cod is an alternative when no external complement is available, self-estimating the prior from used-up exposure). Origin years (accident, underwriting, or reporting year, as applicable) should generally be retained even where development is incomplete: excluding a year purely because it is not yet fully run off would mean \(\lambda\) is fit only on older, mature lags, and would never learn how recent experience should be weighted. A year may still be excluded where development is too uncertain to translate reliably (see Excluded years above).

\medskip

\noindent\textbf{Rate level and trend.}
Standard actuarial adjustments still apply. Experience should be on-levelled to a common rate level and trended to the renewal year before fitting, which matters most for loss ratios; the case study's calendar-year normalisation (Section~\ref{sec:data-setup}) is a substitute used only because the CAS data lacks rate-change history.

\medskip

\noindent\textbf{Layering adjustments.}
If deductibles or attachments shift across years, each account-year should be restated to the renewal attachment \(D_\text{ref}\) before fitting; formulas by target type can be found in standard actuarial increased-limits-factor (ILF) and excess-layer pricing references.

\medskip

\noindent\textbf{Attritional capping and large-loss separation.}
The credibility model is best applied to \emph{attritional} experience only: each year's per-occurrence losses should be capped at a consistent threshold (the reinsurance attachment or a per-occurrence large-loss cap) before computing \(y_{it}\).
The threshold should be consistent across years; if it has drifted with inflation, restating to a common real-terms attachment using standard severity trend factors is recommended before fitting.

The large-loss component can then be priced separately, for example via ILF/LAS (limited average severity) from portfolio-wide data (giving no weight to the account's own losses above the cap) and added back:
\[
  \hat{r}_i^{\text{total}} = \hat{r}_i^{\text{attritional}} + \hat{r}_i^{\text{large loss}}.
\]

\medskip

\noindent\textbf{Zero-loss years.}
A year with no claims above the attachment is an observed zero, not a missing observation; dropping it biases the EWMA upward and over-predicts future experience. A zero-loss lag year enters the EWMA numerator as zero and requires no special distributional treatment.

\medskip

\noindent\textbf{Complement scale.}
The complement \(\mu_i\) and the experience \(y_{it}\) must be on the same scale before fitting.
Where the complement comes from a GLM, the base rate should be applied before passing \(\mu_i\) to the credibility model: the GLM typically outputs a full prediction, but the base rate is often set in a separate step and may not yet be incorporated.
A useful check is to plot \(\mu_i\) against the first lag of \(y_{i,t-1}\): their distributions should overlap substantially.

\subsubsection{Step 2: Choose the model structure}\label{step-2-choose-the-model-structure}

\textbf{Target and likelihood.}
The target variable (\(y_{it}\)) is typically a loss ratio, pure premium,
claim frequency, or excess severity.
The likelihood should match the target: Gamma for severity or for loss ratios when zero-loss years are rare; Tweedie for loss ratios when zero-loss years are common; Poisson or negative binomial for claim frequency. Any distribution whose density can be reparameterised in terms of its mean is in principle compatible: the blend \((1-Z_i)\mu_i + Z_i\hat{f}_i\) is substituted as that mean in the log-likelihood.
For a Gamma loss-ratio target, let the shape vary with exposure, \(\log\phi_{it} = \phi_0 + \phi_1\log E_{it}\), if predictive intervals are needed. A constant shape gives every account the same coefficient of variation, which typically makes intervals too narrow for small accounts; \(\phi_1 = 1\) reproduces the B-S variance structure (\hyperref[app:variance]{Appendix~C}).

\medskip

\noindent\textbf{Blend scale.}
The arithmetic blend (\(\hat{r}_i = (1-Z_i)\mu_i + Z_i\hat{f}_i\)) and geometric blend (\(\hat{r}_i = \mu_i^{1-Z_i}\cdot\hat{f}_i^{Z_i}\)) should be treated as a separate modelling choice, fixed before fitting.
Geometric blending is the natural choice when the complement comes from a multiplicative GLM (as is typical in pricing), since both operate on the log scale; arithmetic blending is the simpler default when accounts can have zero losses in the lookback period (geometric requires flooring the experience to avoid \(\log 0\)), and may be preferred even with a multiplicative complement where interpretability to underwriters is a priority.
The empirical difference was modest on the two lines tested (Section~\ref{sec:cross-lob}): geometric reduced wMSE by 0.8--5.8\% across paired Other Liability variants, while arithmetic was modestly better for Commercial Auto; the gap may be larger for lines with more extreme skew, or for portfolios where the credibility target is defined at a less aggregated level than this case study's insurer-level accounts (e.g.~direct commercial accounts).

\medskip

\noindent\textbf{Evaluation metric.}
Use training-set likelihood ratio tests (LRT; for MLE fits) or leave-one-out cross-validation (LOO-CV; for Bayesian fits) for model structure decisions (which \(Z\) features, which \(\lambda\) specification). For predictive performance on a held-out period, use exposure-weighted MSE (wMSE, or log wMSE), with Gini (rank discrimination) and calibration slope alongside. wMSE is a reasonable default for ratio targets such as loss ratios, since exposure-weighting reflects financial materiality; for claim frequency or severity, deviance under the fitted likelihood is the natural alternative.

\textbf{$Z$ features.}
Log lookback exposure \(\log\tilde{E}_i\) (where
\(\tilde{E}_i = \sum_{k=1}^{W} E_{i,t-k}\)) is a structural input ---
it is the logistic analogue of the B-S exposure weight.
Additional features (industry, years of history,
claim volatility) can be added and tested via LRT
or LOO-CV.
All continuous features should be standardised to mean zero and unit variance across the training portfolio.
The assumed logistic-linear functional form can itself be checked by comparing against an unconstrained step function (e.g.~separate intercepts per size quantile group, no continuous slope): if the step-function estimates fall within the logistic curve's confidence band, the simpler continuous form is well supported (Section~\ref{sec:ffvalidation} demonstrates this on the CA data). If unconstrained MLE returns a counterintuitive sign for \(\hat{b}_Z\) (credibility falling with exposure), a weakly informative prior encoding the expected positive relationship (Section~\ref{sec:core-formula}) is the recommended remedy rather than discarding the exposure feature. Check also whether the sign is sensitive to the likelihood weights (\hyperref[sec:weights-sens]{Appendix~D}).

\medskip

\noindent\textbf{Complement.}
The complement \(\mu_i\) can be estimated jointly with the \(Z\) and
\(\lambda\) parameters (using a log-linear or richer GLM structure), or
supplied as fixed external values.
The fixed-complement path is the natural choice when the pricing team
already has a technical rate from an existing GLM: pass the GLM's
output as \(\mu_i\) and estimate only \((a_Z, b_Z, \ldots, \lambda)\).
This is fully valid: the underlying GLM is unchanged.

\subsubsection{Step 3: Estimate parameters}\label{step-3-estimate-parameters}

The log-likelihood is maximised jointly over the \(Z\) parameters \((a_Z, b_Z, \ldots)\), the decay rate(s) \(\lambda\) (one rate, or varying with account characteristics such as size), and any complement parameters. The blend prediction
\begin{equation}
\hat{r}_i = (1-Z_i)\,\mu_i + Z_i\,\hat{f}^{(\lambda_i)}_i
\end{equation}
is substituted as the mean for each account-year, with \(t\) suppressed as elsewhere (Section~\ref{sec:framework}). \hyperref[app:implementation]{Appendix~G} Listing 1 gives the full R implementation for loss ratio targets.
For Bayesian uncertainty quantification (posterior \(Z_i\) distribution and conservative rate \(R^*_i\)), fit via HMC once on the selected model (\hyperref[app:implementation]{Appendix~G}, Listing 2). MLE is a practical basis for exploration, feature selection and point predictions: in the case study its held-out wMSE was within 2\% of the Bayesian fit's (8.09 against 7.96 \(\times 10^{-3}\) for tercile \(\lambda\)), although the Mid \(\lambda\) ran to its upper bound of 1. Where informative priors are wanted (for example, a weakly identified \(b_Z\) or \(\lambda\), or actuarial judgement to be built in), use MAP (Listing 4) or the Bayesian fit.

\medskip

\noindent\textbf{Recommended starting point.}
A natural starting point is scalar \(\lambda\) with an existing complement held fixed, adding only \(b_Z\) and \(\lambda\) as free parameters over standard B-S.
A more flexible \(\lambda\) specification (varying by account characteristics such as exposure) and complement re-estimation can be introduced once the scalar model is validated. In the case study, two size bands (large accounts against the rest) captured the gain from a size-specific \(\lambda\) as well as three did.
The Quick Start in Section~\ref{sec:mvd} provides a decision tree for the \(\lambda\) specification.

\subsubsection{Step 4: Validate before deployment}\label{step-4-validate-before-deployment}

Run these checks before deploying to live renewals:

\textbf{$\hat{Z}_i$ profile.}
Plot the fitted \(\hat{Z}_i\) values against log exposure (and any other covariates in the logistic equation) across the portfolio.
A curve collapsed near 0 (the model ignores all account experience) or near 1 (the model ignores the complement entirely) indicates implausibly extreme parameters and warrants investigation.
A flat curve at an intermediate level is not itself a problem: it may simply mean that data richness is homogeneous across the portfolio and a scalar \(Z\) is appropriate.
Whether the size gradient \(b_Z\) is identified by the data or carried by the prior can be assessed from the width of the \(b_Z\) interval (Section~\ref{sec:zfeatures}).

\textbf{$\hat{\lambda}$ plausibility.}
Check each fitted \(\hat{\lambda}\) (one per group, if \(\lambda\) varies) for two warning signs: an interval spanning most of \((0, 1)\), or an estimate against a boundary (near 0, meaning last-year-only weighting, or near 1, meaning no discounting). Either can indicate weak identification or prior influence. A low estimate with a tight interval, such as the case study's Large estimate of 0.13, can be genuine and should stand.

\textbf{In-sample calibration by segment.}
The overall in-sample slope is usually close to 1.0 for a well-specified likelihood fit (1.05 in the case study), so a value far from 1 points to a convergence or specification problem, and the useful diagnostic is segment-level: regress actual on predicted separately by exposure band (and any other natural segmentation for your portfolio).
A segment with a persistently low slope is over-crediting that group; a high slope is under-crediting. Exposure band is the primary segmentation to check first, since \(Z_i\) varies with exposure. Choose the granularity (e.g.~terciles or quintiles) to suit the data volume in the portfolio.

\textbf{Held-out performance.}
If a held-out period is available, evaluate exposure-weighted MSE (and/or deviance), Gini and calibration slope, comparing the blend against the complement-only baseline \(\hat{r}_i = \hat{\mu}_i\) (Section~\ref{sec:empirical} illustrates this on AY 2006--2007).
If the blend does not improve on the complement, the account experience is not adding useful signal: either the portfolio is too small, the history too short, or the complement already captures the relevant variation.
Once performance is validated, refit on the full dataset (training + held-out) before deploying to live renewals.

\textbf{Comparing against Bühlmann--Straub.}
When benchmarking held-out performance, always compare against B-S. Report standard (full-history) B-S as the primary benchmark, as the reference point for the methodology, and rolling B-S with the same lookback window \(W\) as the like-for-like comparator for the nested likelihood-ratio test, since the nesting identity holds for it. In the case study, standard and rolling B-S gave almost identical held-out accuracy (\hyperref[app:nesting-proof]{Appendix~A}), but this need not hold on other portfolios.

\subsection{Phase 2: Applying at Renewal}\label{sec:applying-renewal}

Phase 2 applies the fitted \(Z\) parameters \((\hat{a}_Z, \hat{b}_Z, \ldots)\), decay rate(s) \(\hat\lambda\) and complement parameters to produce a renewal rate. Apply the same preprocessing as Phase 1, then compute:
\[
  \hat{Z}_i = \Lambda\!\bigl(\hat{a}_Z + \hat{b}_Z\log\tilde{E}_i + \cdots\bigr), \qquad
  \hat{f}_i = \frac{\sum_{k=1}^{W}\hat\lambda_i^{k-1} E_{i,t-k}\,y_{i,t-k}}{\sum_{k=1}^{W}\hat\lambda_i^{k-1} E_{i,t-k}}, \qquad
  \hat{r}_i = (1-\hat{Z}_i)\hat{\mu}_i + \hat{Z}_i\hat{f}_i.
\]
Here \(\hat\lambda_i\) is the estimated decay rate that applies to account \(i\): the common \(\hat\lambda\), or a group-specific rate (for example, by size band). Present the underwriter with \(\hat{Z}_i\) (how much the model trusts this account's own history), \(\hat{\mu}_i\) (what the portfolio benchmark says the risk should cost) and \(\hat{f}_i\) (what the account's own history says, weighted toward recent years by \(\hat\lambda_i\)), together with the renewal rate \(\hat{r}_i\) they produce. Here \(\hat{f}_i = \hat{f}_i^{(\hat\lambda_i)}\), that is, equation \eqref{eq:ewma} evaluated at the fitted decay rate. Showing the simple equal-weighted average alongside makes the effect of recency weighting visible when the two differ.

\medskip

\noindent\textbf{New accounts.}
For new accounts with no history, lookback exposure \(\tilde{E}_i = 0\), so \(\hat{Z}_i \to 0\) and the prediction reduces to the complement \(\hat{\mu}_i\). The EWMA accumulates from the second year onward, after which \(\hat{Z}_i\) is positive and grows with the account's accumulated exposure.

\medskip

\noindent\textbf{Refitting cadence.}
Refit annually or whenever the portfolio composition changes materially. As an indicative guide, investigate any parameter whose refitted value falls outside the previous fit's 95\% interval. Fixed absolute thresholds are unreliable because how much a parameter moves from sampling noise alone differs widely across parameters and portfolios.

\subsection{Worked Example}\label{sec:worked-example}

We apply the fitted model to two companies from the CAS dataset to illustrate the three numbers (\(\hat{Z}_i\), \(\hat{\mu}_i\), \(\hat{f}_i\)) and the renewal rate \(\hat{r}_i\) they produce, at the account level: one small company (Mountain West Farm Bureau) and one large company (Harco Natl Ins Co), chosen to isolate the key size-class differences.
All quantities are on an absolute loss-ratio scale.
For comparison, rolling Bühlmann--Straub (same lookback window, nesting exactly within the logistic model) is shown alongside; the full portfolio comparison uses standard B-S (Table~\ref{tab:tbl-emp-results}).

\begin{table}[!h]
\centering
\caption{\label{tab:worked-input}Input data for the AY 2007 prediction. NEP is the current accident year's net earned premium. Three lags are shown for readability; the fitted model uses all available history up to $W_{\max} = 7$ lags. Lags $t-3$, $t-2$, $t-1$ are the lag-10 ultimate loss ratios for AY 2004, 2005, 2006 respectively. Actual is the AY 2007 lag-10 ultimate outcome.}
\centering
\fontsize{10}{12}\selectfont
\begin{tabular}[t]{llllll}
\toprule
Account & NEP (\$k) & $t-3$ & $t-2$ & $t-1$ & Actual\\
\midrule
Mtn West Farm Bureau (Small) & 933 & 341\% & 327\% & 139\% & 218\%\\
Harco Natl Ins Co (Large) & 50,347 & 92\% & 81\% & 80\% & 101\%\\
\bottomrule
\end{tabular}
\end{table}

\textbf{Fitted parameters.}
This example uses the LogCred (tercile \(\lambda\)) variant and shows three input years for readability; the fitted model uses all available history up to \(W_{\max} = 7\) lags.
Key estimates: \(\hat\lambda_{\text{Sm}} = 0.6\) (Small), \(\hat\lambda_{\text{Lg}} = 0.13\) (Large); logistic \(Z\) parameters \(\hat{a}_Z = 0.561\), \(\hat{b}_Z = 0.058\) (on standardised \(\log\tilde{E}_i\)); size-varying complement with \(\hat\alpha = -0.222\), \(\hat\beta = 0.127\).
Rolling B-S uses equal-weight averaging (\(\lambda = 1\)), flat complement (market mean \(60\%\)), and pooled \(\hat{K} = 8{,}009\) (\$k).

\begin{table}[!h]
\centering
\caption{\label{tab:worked-output}Predicted and actual AY 2007 loss ratios for the two worked-example accounts. MWFB = Mountain West Farm Bureau; HN = Harco Natl Ins Co. All quantities expressed as percentages. Rolling B-S complement is the flat market mean ($60\%$); logistic complement is size-varying. All components use the full $W_{\max}=7$ lookback; the input table shows the three most recent years.}
\centering
\fontsize{9}{11}\selectfont
\begin{tabular}[t]{llllllllll}
\toprule
\multicolumn{1}{c}{ } & \multicolumn{2}{c}{Credibility Z} & \multicolumn{2}{c}{Complement} & \multicolumn{2}{c}{Experience blend} & \multicolumn{2}{c}{Prediction} & \multicolumn{1}{c}{ } \\
\cmidrule(l{3pt}r{3pt}){2-3} \cmidrule(l{3pt}r{3pt}){4-5} \cmidrule(l{3pt}r{3pt}){6-7} \cmidrule(l{3pt}r{3pt}){8-9}
Account & $\hat{Z}$ (B-S) & $\hat{Z}$ (lgt) & $\hat{\mu}$ (B-S) & $\hat{\mu}$ (lgt) & $\hat{f}$ (B-S) & $\hat{f}$ (lgt) & $\hat{r}$ (B-S) & $\hat{r}$ (lgt) & Actual\\
\midrule
MWFB & 37\% & 63\% & 60\% & 44\% & 181\% & 203\% & 105\% & 144\% & 218\%\\
HN & 97\% & 66\% & 60\% & 58\% & 80\% & 81\% & 79\% & 73\% & 101\%\\
\bottomrule
\end{tabular}
\end{table}

The two companies reveal how the model's components interact:

\begin{itemize}
\item
  \textbf{Mountain West Farm Bureau} (Small, bad history):
  The three years shown (341\%, 327\%, 139\%) are well above the 60\% market mean, and the earlier lags in the window are, on average, above it too.
  Rolling B-S assigns \(\hat{Z} = 37\%\) and blends its equal-weighted experience (181\%) with the market mean, giving a prediction of 105\%.
  The logistic model assigns \(\hat{Z} = 63\%\).
  With \(\hat\lambda_{\text{Sm}} = 0.60\) it places 47\% of the exposure weight on the most recent year (AY 2006, LR = 139\%) and discounts the older lags, including two years at or below the market mean (AY 2001 and 2003) that pull the equal-weighted average down. Its experience term is therefore higher (203\%) and, with a size-varying complement of 44\%, it predicts 144\%.
  The actual AY 2007 outcome was 218\%: both models missed the full
  magnitude (individual accident years are noisy), but the
  logistic prediction (144\%) was more informative than
  the B-S prediction (105\%), and the direction of the warning was correct.
  The Bayesian posterior additionally quantifies confidence in this prediction: \(R^*_i(0.95) \approx 159\%\), an uncertainty load of about 19 percentage points above the posterior-mean prediction of 141\%, reflecting genuine parameter uncertainty for a small, thinly-supported account (Tier 2 uncertainty, Section~\ref{sec:uq}).
\item
  \textbf{Harco Natl Ins Co} (Large, stable history):
  A large insurer with stable, above-market experience across the three years shown (92\%, 81\%, 80\%), against a market mean of 60\%.
  Rolling B-S gives it \(\hat{Z} = 97\%\) (standard B-S with the full training history: \(\hat{Z} \approx 93\%\)) and, with equal temporal weights, an experience term of \(\hat{f} = 80\%\), so it predicts 79\%.
  The logistic model gives a much lower \(\hat{Z} = 66\%\). Its decay \(\hat\lambda_{\text{Lg}} = 0.13\) concentrates 87\% of the exposure weight on the most recent year (weights \(1, \lambda, \lambda^2, \ldots\) before exposure scaling), but because the history is stable its experience term (81\%) is almost the same as B-S's. The logistic model shrinks that term towards the size-varying complement (58\%) and predicts 73\%.
  The actual AY 2007 outcome was 101\%. Both predictions fell below the outcome, and B-S was closer because it passes almost all of the above-market experience through, whereas the logistic model shrinks that experience.
  Temporal structure would add a further difference for an account whose trend is changing: the logistic model's low \(\hat\lambda_{\text{Lg}}\) would concentrate weight on the most recent year while B-S would average over the full window, yielding materially different predictions.
  The Bayesian posterior is far tighter here: \(R^*_i(0.95) \approx 76\%\), an uncertainty load of only about 3 percentage points above the posterior-mean prediction of 73\%, consistent with Harco's well-identified, high-exposure parameters.
\end{itemize}

\subsection{Quick Start: Minimum Viable Deployment}\label{sec:mvd}

For a pricing actuary deploying the framework for the first time, the following checklist sets out the essential steps from data checks to a first renewal rate, with optional extensions at the end.

\medskip

\noindent\textbf{Step 0: Go/no-go check before any fitting}

\begin{itemize}
  \item Sufficient distinct rated \emph{accounts}\footnote{At whatever level an account is defined: policy, named insured or portfolio segment (see the opening of Section \ref{sec:practical}).}, each with meaningful lag history (five or more lag years preferable; see Section \ref{sec:applying} for history guidance). Individual-insured data are likely to carry higher process variance per account than insurer-level (cedant) aggregates and may therefore require more accounts for equivalent identification of the decay rate $\lambda$ (how quickly older years lose weight).
  \item Account-level exposure by year is available (to construct log lookback exposure, the structural $Z$ input)
  \item The target loss statistic is comparable to the renewal being priced (restated for changes in rate level, attachment or trend) and, where loss skewness is high, attritional (large losses capped or excluded, depending on how much volume each account represents)
  \item Pre-adoption signal check: plot next-year loss statistic vs EWMA $\hat{f}_i$ by exposure band; a positive slope overall or in at least one band confirms the data support credibility weighting
  \item Temporal decay check: plot lag-by-lag rank correlation of the target loss statistic, net of calendar-year effects, split by account size (Figure \ref{fig:fig-eda-corrmat-bysize} demonstrates this on the CA data); faster decay for large accounts is a leading indicator that a size-varying $\lambda$ will be worth fitting, though the interval-width diagnostics in the decision tree below remain the deciding check.
  \item Complement--experience gap check: compare $\hat\mu_i$ and $\hat f_i$ across the portfolio; if they are already close for most accounts, the credibility weight matters little (Section \ref{sec:framework}) and a simpler fixed or judgemental $Z$ may suffice; the framework earns its value where the two diverge
  \item $\lambda$ go/no-go: if few accounts have sufficient lag history, $\lambda$ may be weakly identified. The width of the $\lambda$ interval after fitting will show this; if it remains diffuse, fix $\lambda$ at 1 or set it by actuarial judgement (see the decision tree below).
\end{itemize}

\medskip

\noindent\textbf{Decision tree for $\lambda$ specification.}
Always begin with scalar \(\lambda\). The \(\lambda\) intervals, not a fixed account count, are the diagnostic at every stage.
The tree lets the data decide. Where there are independent grounds for \(\lambda\) to differ by size or segment, build that in directly, either as a split set by judgement or as an informative prior in a Bayesian fit.

\begin{enumerate}
  \item \textbf{Fit scalar $\lambda$.}
  Inspect the $\lambda$ estimate and its 95\% interval (credible interval for Bayesian fits; profile-likelihood or bootstrap interval for MLE fits, since Hessian intervals can be unreliable for $\lambda$ near its bounds or with few accounts). A tight CI does not rule out size heterogeneity: exposure-weighting can make the pooled estimate converge tightly onto what the largest accounts prefer (Section~\ref{sec:lambda-decay}). In the case study, the scalar CI is tighter than the Small and Mid tercile intervals although its point estimate sits far from both. Proceed to step 2 regardless of CI width.

  \item \textbf{Fit an account-characteristic-varying $\lambda$ and check for separation between the group intervals directly} (not against the scalar CI).
  Account size is the natural first candidate (Section~\ref{sec:lambda-decay}), with terciles as a starting granularity.
  \begin{itemize}
    \item \emph{Little or no overlap between groups:} genuine heterogeneity confirmed; use the account-characteristic-varying specification.
    \item \emph{Substantial overlap:} try a coarser split (e.g. binary). If overlap persists, keep the scalar $\lambda$ provided its own CI was reasonably informative. If the scalar CI was itself too diffuse to trust, fix $\lambda$ externally (at 1 for standard B-S, or by judgement or observed autocorrelation) and estimate $Z$ alone.
  \end{itemize}

\end{enumerate}

\medskip

\noindent\textbf{Lookback window $W$.}
Use all available history up to a practical ceiling, subject to the exclusions in Section \ref{sec:applying} (the case study uses \(W_\text{max}=7\)). As a guide, set \(W\) large enough that less than 10\% of the EWMA weight lies beyond the window for the most persistent group, i.e.~\(\hat\lambda^W < 0.10\). The smallest such \(W\) is the effective lookback \(W^* = \lceil\log(0.1)/\log(\hat\lambda)\rceil\) (\hyperref[app:effective-memory]{Appendix~B}), which converts \(\hat\lambda\) to an interpretable number of years. At \(\hat\lambda \approx 0.84\) (Mid in the case study), \(W^*\) is 14, so \(W_\text{max}\) is the binding constraint there; with a lower \(\hat\lambda\) the formula would bind. Shorter-history accounts receive only their available lags but still contribute to the fit.

\medskip

\noindent\textbf{Fit the model (one function call, \hyperref[app:implementation]{Appendix~G} Listing 1):}

\begin{verbatim}
fit_mle <- nlminb(init, nll, df = df_train,
                  control = list(iter.max = 500, rel.tol = 1e-8))
\end{verbatim}

Estimates all parameters jointly, typically in seconds to a few minutes.

\medskip

\noindent\textbf{Three numbers and the rate they produce, per account at renewal (\hyperref[app:implementation]{Appendix~G} Listing 3):}

\begin{itemize}
  \item $\hat{Z}_i$: the credibility weight, how much to trust the account's own history (0 = full shrinkage to the complement; 1 = full own history)
  \item $\hat{\mu}_i$: the complement, what the portfolio says this risk should cost
  \item $\hat{f}_i$: what the account's own history says, with recent years counting more (weighted by $\hat\lambda_i$)
  \item $\hat{r}_i = (1-\hat{Z}_i)\hat{\mu}_i + \hat{Z}_i\hat{f}_i$: the recommended renewal rate
\end{itemize}

\medskip

\noindent\textbf{Two-line deployment validation (spreadsheet or R):}

\begin{verbatim}
slope <- coef(lm(actual ~ predicted, data = holdout, weights = exposure))[2]
slope   # compare with 1.0, and check its bootstrap interval (see below)
\end{verbatim}

Run this by your chosen segmentation dimension (size band, industry, territory, or whatever is natural for your book). Take the bootstrap interval by resampling accounts with replacement (the case study uses 2,000 draws) and report the 90\% interval.
A slope whose bootstrap interval excludes 1.0 in more than one consecutive cycle warrants investigation, especially if it also lies outside roughly \([0.7, 1.3]\): well below 1 indicates over-crediting, well above 1 indicates under-crediting. Requiring more than one cycle matters because band slopes can swing widely between years, particularly where a band holds few accounts or the predictability of experience shifts (Section \ref{sec:results-main} shows this for Mid accounts in the case study). A modest, persistent deviation specifically in the smallest exposure band may instead reflect the trade-off from NEP-weighted fitting, which favours larger accounts (Section \ref{sec:results-main}), rather than a mis-specified model.

\medskip

\noindent\textbf{When to add a Bayesian fit (optional):}
MLE (Listing 1) can produce the three renewal numbers (Listing 3). Run HMC once on the selected model (\hyperref[app:implementation]{Appendix~G} Listing 2) if you also want the conservative rate \(R^*_i(q)\) and an uncertainty dashboard, or where informative priors are wanted, for example where \(b_Z\) or \(\lambda\) is weakly identified or actuarial judgement should be built in. Under MLE alone, \(R^*_i(q)\) has no direct counterpart and intervals for \(\hat{Z}_i\) and \(\hat{r}_i\) need the delta method or a bootstrap. If predictive intervals are used to quantify uncertainty, also check, once a cycle's outcomes are known, that about 5\% exceed their 95\% predictive quantile, overall and by size band. Allow for noise of at least about \(\pm 1.96\sqrt{0.05\times 0.95/n}\) for \(n\) accounts (4 percentage points for 100, 1.4 for 1,000); a persistent departure warrants re-examining the variance specification.

\medskip

\noindent\textbf{Judgemental priors.}
Where an existing credibility-blend mechanism is in use with judgementally-set parameters, these can be translated into prior means for the logistic \(Z\) coefficients (\((a_Z, b_Z)\) for a size-only model, or the full coefficient vector if additional features are included). How direct the translation is depends on the form of the existing rater, and some work may be needed to make it. The Bayesian posterior then answers whether the data support, refute, or are too sparse to discriminate from the existing curve.

\section{Conclusion}\label{sec:conclusion}

Bühlmann--Straub credibility remains the standard, theoretically grounded way to blend an account's experience with a benchmark.
But three of its assumptions can break down in commercial lines: a single portfolio-wide \(K\), a flat complement (the same portfolio mean for every account) and equal weight on every year of history.
Where accounts differ in predictability, expected loss ratio or how quickly older experience loses relevance, the effects can be visible in the data and consequential in pricing.
In the case study, patching each limitation in turn (stratifying \(K\) by size, enriching the complement, adding EWMA weighting) recovered most of the improvement, but the elements interact: the EWMA step eroded about half of the small-account gain from stratification.

The logistic credibility framework addresses this with a simple generalisation: replace the fixed-\(K\) formula for \(Z\) with a logistic function of observable characteristics, fit by likelihood together with the complement and the decay rate on historical experience. In the case study this avoids the erosion, and the data themselves show which B-S assumptions to relax. The nesting result (Section~\ref{sec:bs-nesting}) means that, against rolling B-S, the question is not whether to replace it but how much of this additional flexibility the data actually support.
The answer from 96 real companies is favourable: calibration improves (chiefly for small accounts) and prediction error falls by 25\% relative to standard B-S with a deployable prior-year forecast complement (90\% bootstrap interval: 7\%--42\%). The oracle upper bound (using the realised portfolio mean) is 27\% (90\% bootstrap interval: 9\%--44\%). Both structural extensions, freeing the \(Z\) slope from 1 and estimating \(\lambda\) rather than fixing \(\lambda = 1\), are supported by likelihood-ratio tests and held-out wMSE. LOO-CV favours estimating \(\lambda\) (ratio 2.3) and, more narrowly, freeing \(b_Z\) (ratio 1.8, just below the usual threshold of 2).

The most commercially significant finding is the temporal decay gradient. In Commercial Auto the estimates are \(\hat\lambda_{\text{Sm}} \approx 0.60\), \(\hat\lambda_{\text{Md}} \approx 0.84\) and \(\hat\lambda_{\text{Lg}} \approx 0.13\), with the Small and Mid values not statistically distinguishable; in Other Liability they fall monotonically with size (0.73, 0.47, 0.24). The gradient is learned from the data rather than assumed in advance. The mechanism is consistent with the law of large numbers: small accounts generate few claims per year so each year's loss ratio is dominated by process noise, making multi-year averaging optimal. Large accounts generate many claims so each year is already a precise signal, making recent experience sufficient. A fixed-\(\lambda\) rule cannot recover this gradient. In Commercial Auto the proposed model outperforms a pooled GLMM in overall wMSE (the two are close in Other Liability). It can also price accounts that were not in the training data from the history they bring, which matters in portfolios with meaningful new business. GLMM's mid-account advantage in this case study is consistent with a dataset-specific regime change (Section~\ref{sec:glmm-comparison}). Whether it generalises is an empirical question.

The Bornhuetter--Ferguson method \citep{bornhuetter1972} occupies an analogous position in reserving: \(Z\) is set mechanically from the development pattern rather than jointly optimised, just as \(K\) is set mechanically in B-S. Whether estimating it jointly would help is an open question.

Other candidates for further work include:

\begin{itemize}
  \item \textbf{Cross-line validation}: extend beyond Commercial Auto and Other Liability to other lines, e.g.\ property, directors' and officers' liability (D\&O), or other specialty classes.
  \item \textbf{Direct individual-account panels}: higher process variance per account provides the most demanding test of robustness.
  \item \textbf{Separating development uncertainty from risk drift}: development must be estimated somewhere, either jointly within the model or as a pre-processing step that replicates deployment conditions (e.g.\ re-estimating training-year ultimates using only the development available at each historical point). Either way, residual development uncertainty is not yet separated from genuine risk drift, which a single $\lambda$ currently absorbs together (Section \ref{sec:lambda-decay}). Weighting each lag-year by its development maturity is one mechanism to test.
  \item \textbf{Richer covariates for $Z$}: industry, broker channel, geographic concentration.
  \item \textbf{Hierarchical extension}: a logistic $Z$ and $\lambda$ at each level of a hierarchical structure \citep[Chapter~9]{buhlmann2005}.
  \item \textbf{Shared estimation error between experience and complement}: model the correlation arising from shared estimation error in trend, development, and on-levelling factors, each of which can manifest as a shift affecting both quantities.
  \item \textbf{State-space equivalence}: formal treatment of the equivalence between the exposure-weighted EWMA and a Kalman-filtered AR(1) model with $\lambda$ as the persistence parameter. A similar comparison can be made to penalised-regression state-space models \citep{korn2021}.
\end{itemize}

Perhaps the most practically valuable feature of the framework is that it does not ask the analyst to decide where credibility meaningfully varies or how quickly historical data becomes irrelevant: both are estimated from the portfolio's own data in a single pass. Standard B-S requires the analyst to fix \(K\) (often from moment estimators calibrated elsewhere) and implicitly sets \(\lambda = 1\) (equal weight to all years). The joint framework instead asks: \emph{what does this portfolio say about credibility structure and data relevance?} Any portfolio with sufficient historical data can be interrogated in the same way, and the interval estimates make clear where the answer is well-identified and where judgement remains indispensable.

Credibility matters in any market, but the stakes are asymmetric. Systematic mis-weighting of account history is not a neutral modelling error: it can leave smaller accounts' prices too close to the portfolio average regardless of their true experience, or base prices on stale experience, costing margin or accounts depending on which way the market is moving. Are you currently fixing \(K\) by moment estimation and \(\lambda\) by eye, or not at all? In this framework both are estimated jointly from the portfolio's own data instead, while still delivering transparent numbers per account (experience, complement, credibility weight and the resulting rate) and optionally quantifying uncertainty over every weight assigned.

\newpage

\section*{Disclaimer}\label{disclaimer}
\addcontentsline{toc}{section}{Disclaimer}

The views expressed in this paper are those of the author in a personal
capacity.
The empirical case study was conducted using only the publicly available CAS loss triangle database. The simulation study used entirely synthetic data.

\section*{Acknowledgements}\label{acknowledgements}
\addcontentsline{toc}{section}{Acknowledgements}

I thank Dave Clark, Uri Korn, and Charlie Stone for their generous time and thoughtful comments on earlier drafts of this paper. I also thank Markus Gesmann, who independently applied the framework to a separate Lloyd's market dataset. Any remaining errors are my own.

\section*{Use of Generative AI}\label{use-of-generative-ai}
\addcontentsline{toc}{section}{Use of Generative AI}

During the preparation of this work the author used a generative AI assistant to review and edit the text for clarity and grammar, to write and review analysis code under the author's direction, and to cross-check the numerical results in the paper against the underlying code outputs. The research questions, model design, interpretation of results and conclusions are the author's own. The author reviewed and edited all AI-assisted content and takes full responsibility for the content of the paper.

\section*{Data and Code Availability}\label{data-and-code-availability}
\addcontentsline{toc}{section}{Data and Code Availability}

The empirical study uses the publicly available CAS loss reserve triangle
database \citep{meyershi2011}, which can be downloaded from the Casualty
Actuarial Society website.
The simulation study uses entirely synthetic data generated by the
data-generating process described in \hyperref[app:dgp]{Appendix~F}.

R code implementing the logistic credibility model (MLE via
\texttt{nlminb} and Bayesian HMC via \texttt{brms}/Stan; \hyperref[app:implementation]{Appendix~G})
and a Python reference implementation are available at:

\begin{center}
\url{https://github.com/jakem87/logistic-credibility}
\end{center}

\newpage
\renewcommand{\bibname}{References}
\bibliography{references-morris2026}
\newpage

\section*{§A: Nesting Proof and Rolling Bühlmann--Straub Exposition}\label{app:nesting-proof}
\addcontentsline{toc}{section}{§A: Nesting Proof and Rolling Bühlmann--Straub Exposition}

\subsection*{MLE Consistency Under the B-S Data-Generating Process}\label{mle-consistency-under-the-b-s-data-generating-process}
\addcontentsline{toc}{subsection}{MLE Consistency Under the B-S Data-Generating Process}

\textbf{Proposition (MLE recovery under a B-S data-generating process).}
Suppose the data are generated by the B-S mechanism with exact credibility (for example, Poisson claim counts with a Gamma-distributed risk parameter, so that the credibility estimator is the conditional mean) and true structural parameter \(K_0\): that is, the true credibility weight is \(Z_i = \tilde{E}_i/(\tilde{E}_i + K_0)\) (rolling B-S: computed over the same \(W\)-year window),
the complement is flat, and there is no temporal decay; the fit holds \(\lambda = 1\) and the flat complement at these true values. Suppose further that accounts are independent and lookback exposure \(\tilde{E}_i\) takes more than one value across the portfolio, so
that \(a_Z\) and \(b_Z\) are jointly identified.
Then the logistic MLE of \((a_Z, b_Z)\) is consistent for the corresponding
parameter values:
\begin{equation}
  \hat{a}_Z \xrightarrow{p} -\log K_0, \qquad \hat{b}_Z \xrightarrow{p} 1
  \label{eq:mle-consistency}
\end{equation}
(on the unstandardised \(\log \tilde{E}_i\) scale) as the number of accounts
\(N \to \infty\).

\textit{Proof sketch.}
For any observation with true expected rate \(r_0 = \mathbb{E}[\theta_i \mid \bar{f}_i, \tilde{E}_i]\) (\(\theta_i\) being the account's true rate), the Poisson negative log-likelihood \(\ell(r) = r - C\log r\) (with \(C\) the observed outcome per unit exposure) satisfies
\(\mathbb{E}[\ell(r)] = r - r_0\log r + \text{const}\), which is uniquely
minimised at \(r = r_0\) (a strictly consistent scoring function for the mean; exposure weights are functions of the covariates and do not change this pointwise minimiser).
Under the B-S data-generating process (DGP), the conditional mean is
\(r_0 = (1 - Z_i^*)\,\mu + Z_i^*\,\bar{f}_i\) with \(Z_i^* = \tilde{E}_i/(\tilde{E}_i + K_0)\).
For the logistic model to achieve \(\hat{r}_i = r_0\) for every account
simultaneously (i.e.~for every value of \(\tilde{E}_i\)) requires
\(\Lambda(a_Z + b_Z\log \tilde{E}_i) = \tilde{E}_i/(\tilde{E}_i + K_0)\) to hold identically in \(\tilde{E}_i\).
Using the identity \(\Lambda(\log(x/K)) = x/(x+K)\), this is satisfied if and
only if \(a_Z = -\log K_0\) and \(b_Z = 1\). Uniqueness requires at least two distinct exposure values (one value is consistent with infinitely many \((a_Z, b_Z)\) pairs).
Consistency of the sample estimator then follows from standard M-estimation
results \citep[Thm.\ 5.7]{van2000asymptotic} under mild regularity
conditions (a compact parameter space containing the true values, bounded log-exposure, a uniform law of large numbers, and the variation in \(\tilde{E}_i\) assumed above, which together ensure \((-\log K_0, 1)\) is a well-separated maximiser of the population log-likelihood).

\textit{Scope note.}
The sketch uses the Poisson negative log-likelihood because it gives a clean closed form. The same argument applies to the Gamma log-likelihood used in the empirical study, which is also a strictly consistent scoring function for the mean \citep{gneiting2011}: the Gamma negative log-likelihood \(\phi y/r + \phi\log r\) has expectation \(\phi r_0/r + \phi\log r\), uniquely minimised at \(r = r_0\) whatever the value of \(\phi\), so the consistency conclusion carries over unchanged. Consistency is for the mean, so it does not require \(\phi\) or the variance function to be correct; the finding in \hyperref[app:variance]{Appendix~C} that the variance falls more slowly than \(1/E_{it}\) affects efficiency and interval calibration, not this result. Without exact credibility the B-S estimator is only the best linear predictor, and the MLE converges instead to the best member of the logistic family under the chosen loss. Nesting (Section~\ref{sec:bs-nesting}) holds for any data; exact credibility makes the nested member the truth. Conversely, in a fit that holds \(\lambda = 1\) and a flat complement, \(\hat{b}_Z\) well away from 1 (on the unstandardised scale), or \(\hat{K} = \exp(-\hat{a}_Z)\) far from the moment estimate, indicates a departure from homogeneous \(K\) or from exact credibility.

\subsection*{Rolling vs.~Standard Bühlmann--Straub}\label{rolling-vs.-standard-buxfchlmannstraub}
\addcontentsline{toc}{subsection}{Rolling vs.~Standard Bühlmann--Straub}

As defined in Section \ref{sec:bs-nesting}, rolling B-S (same \(W\)-year lookback for exposure and experience, with \(K\) re-estimated on that scale) nests exactly in the logistic model. Standard B-S \citep{buhlmann1970} (full training history) does not nest in the model as fitted (different data window, and \(K\) estimated once by method of moments). Standard B-S is the primary benchmark in the tables and headline gains, because it is the reference point for the whole methodology: the textbook formulation, using all available history. Rolling B-S is the like-for-like comparator for the nested likelihood-ratio test and the worked example. In the case study, held-out accuracy is almost identical (rolling vs standard B-S: wMSE 10.90 vs 10.97 \(\times 10^{-3}\); Gini 69.9\% vs 69.6\%), and giving standard B-S the same lookback window as the logistic model gives 11.16, so the choice of benchmark does not drive the result here (this need not hold in general). In the case study the gain over B-S therefore comes mainly from relaxing constraints that define B-S (freeing the \(Z\) slope \(b_Z\) from 1 and letting the complement and decay rate vary by account size), not from the choice of data window.

\medskip

\newpage

\section*{§B: Effective Lookback Window}\label{app:effective-memory}
\addcontentsline{toc}{section}{§B: Effective Lookback Window}

The EWMA decay parameter \(\lambda\) is the factor by which each year further back is down-weighted,
so it determines how many years of history carry material weight in the
blended forecast. Two functions of \(\lambda\) express this as a number of years.

The \emph{mean memory} \(\bar{W} = 1/(1-\lambda)\) is the mean age of the experience in the blended forecast under the geometric weights, counting the most recent year as 1. It says where the weight is centred.

\begin{figure}[H]

{\centering \includegraphics[width=0.72\linewidth]{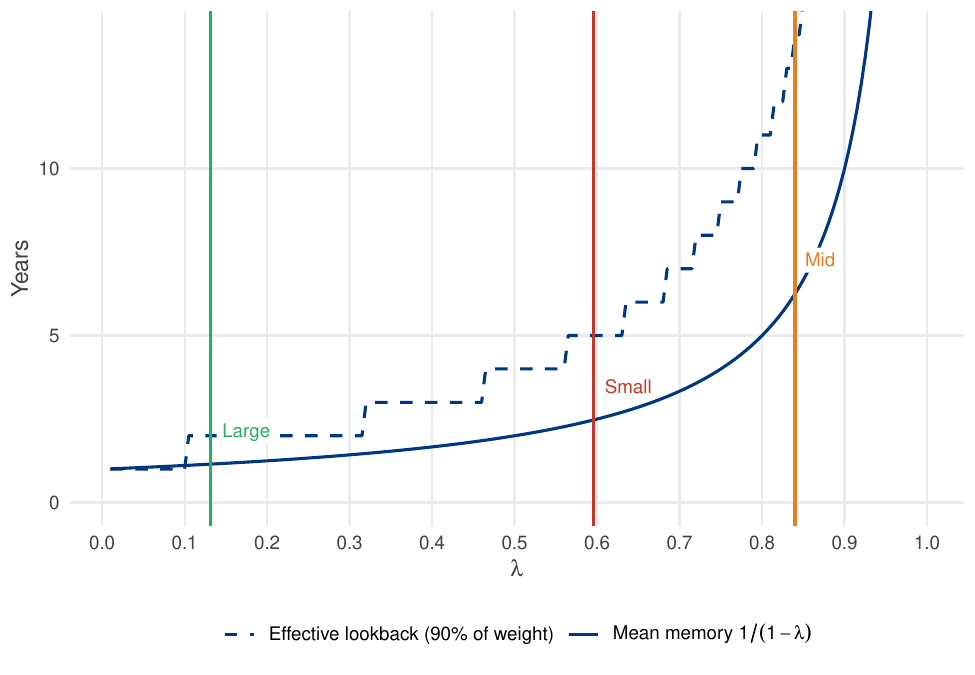} 

}

\caption{Mean memory $\bar{W} = 1/(1-\lambda)$ (solid) and effective lookback $W^*(\lambda, 0.90)$ (dashed) against the EWMA decay rate $\lambda$. Vertical lines: posterior mean $\hat\lambda$ by size tercile from the CAS Commercial Auto case study (Large 0.13, Mid 0.84, Small 0.60).}\label{fig:fig-app-effective-memory}
\end{figure}

The \emph{effective lookback} \(W^*(\lambda, \tau)\) says how far back the weight reaches. It is the smallest
integer \(w\) such that the cumulative normalised EWMA weight on the most
recent \(w\) years first reaches threshold \(\tau\):

\begin{equation}
  W^*(\lambda, \tau) = \min\left\{ w \in \mathbb{Z}^+ :
    (1 - \lambda^w) \geq \tau \right\},
  \label{eq:lookback}
\end{equation}

which simplifies to \(W^*(\lambda, \tau) = \lceil \log(1-\tau) / \log(\lambda) \rceil\).
Figure~\ref{fig:fig-app-effective-memory} plots \(\bar{W}\) (solid line) and \(W^*(\lambda, 0.90)\) (dashed line) for \(\lambda \in (0,1)\), with the three empirical tercile estimates marked. Both grow without bound as \(\lambda \to 1\), where the weights become equal.

In the CAS Commercial Auto case study, Large companies are rated almost entirely on the current year
(\(\hat\lambda_\text{Lg} \approx 0.13\), \(\bar{W} \approx 1.2\) years), while Small (2.5 years) and Mid (6.3 years) need a much longer window. Their intervals overlap, and a pooled Small-and-Mid estimate gives about 3.9 years (\(\hat\lambda \approx 0.74\)).
Both measures assume an unbounded history, so for Mid, where \(W^*(0.84, 0.90) = 14\), the 7-year window cuts off about 30\% of the weight (\(0.84^7\)).
These values are specific to this dataset and line of business and should not be assumed to transfer. Re-estimating \(\lambda\) from your own panel replaces the informal question ``how many years of data should I use?'' with a model-estimated answer testable against held-out data.

\newpage

\section*{§C: Variance Structure and Predictive Intervals}\label{app:variance}
\addcontentsline{toc}{section}{§C: Variance Structure and Predictive Intervals}

\subsection*{Variance proportionality check}\label{sec:var-prop-check}
\addcontentsline{toc}{subsection}{Variance proportionality check}

\textbf{Variance structure check.}
Bühlmann--Straub assumes that the process variance, given the risk parameter \(\theta_i\), satisfies \(\mathrm{Var}(\mathrm{LR}_{it} \mid \theta_i) \propto 1/E_{it}\).
We assess this by fitting a dispersion sub-model that lets the Gamma shape parameter vary with exposure: \(\log\phi_{it} = \phi_0 + \phi_1\log E_{it}\). The likelihood is unweighted, so that the exposure scaling is estimated rather than imposed.
Under a Gamma likelihood with mean \(\hat{r}_{it}\), the fitted variance given the prediction is \(\mathrm{Var}(\mathrm{LR}_{it} \mid \hat{r}_{it}) = \hat{r}_{it}^2/\phi_{it} = e^{-\phi_0}\,\hat{r}_{it}^2 E_{it}^{-\phi_1} \propto \hat{r}_{it}^2 E_{it}^{-\phi_1}\), and the coefficient of variation is \(1/\sqrt{\phi_{it}} \propto E_{it}^{-\phi_1/2}\). B-S assumes \(\mathrm{Var} \propto 1/E_{it}\), so it corresponds to \(\phi_1 = 1\); the \(\hat{r}_{it}^2\) factor is the Gamma's assumption, not B-S's.
The estimate is \(\hat\phi_1 = 0.57\) (95\% credible interval \([0.50, 0.63]\)) on raw \(\log E_{it}\); the shape is fitted on standardised log NEP, where the slope is 1.01 per standard deviation (SD 1.78).

At a given predicted loss ratio the fitted variance therefore falls with exposure, but more slowly than \(1/E_{it}\): roughly as \(E_{it}^{-0.57}\), so the coefficient of variation falls as \(E_{it}^{-0.28}\) rather than \(E_{it}^{-0.5}\).
It is the variance about the fitted prediction, so it includes anything the model leaves unexplained (year-to-year shocks, unmodelled differences between accounts) that need not shrink with exposure. The check shows that \(1/E_{it}\) does not describe it here, but cannot separate a slower decline in process variance, a variance floor that a power law cannot represent, and unexplained variation.

The primary models' likelihood weights (proportional to \(E_{it}\)) assume \(\mathrm{Var} \propto 1/E_{it}\), which over-weights large accounts relative to precision weights (\hyperref[sec:weights-sens]{Appendix~D} examines their effect on \(Z\)). They act in estimation only. The predictive distribution of a constant-\(\phi\) model has the same coefficient of variation for every account, which is why its intervals are too narrow for small accounts; the dispersion sub-model puts the estimated variance function into the predictive distribution. wMSE weights by NEP as a financial-importance choice, and the dispersion models change it by only about 1.8\% (Section~\ref{sec:dispersion}).

\begin{figure}[H]

{\centering \includegraphics[width=0.75\linewidth]{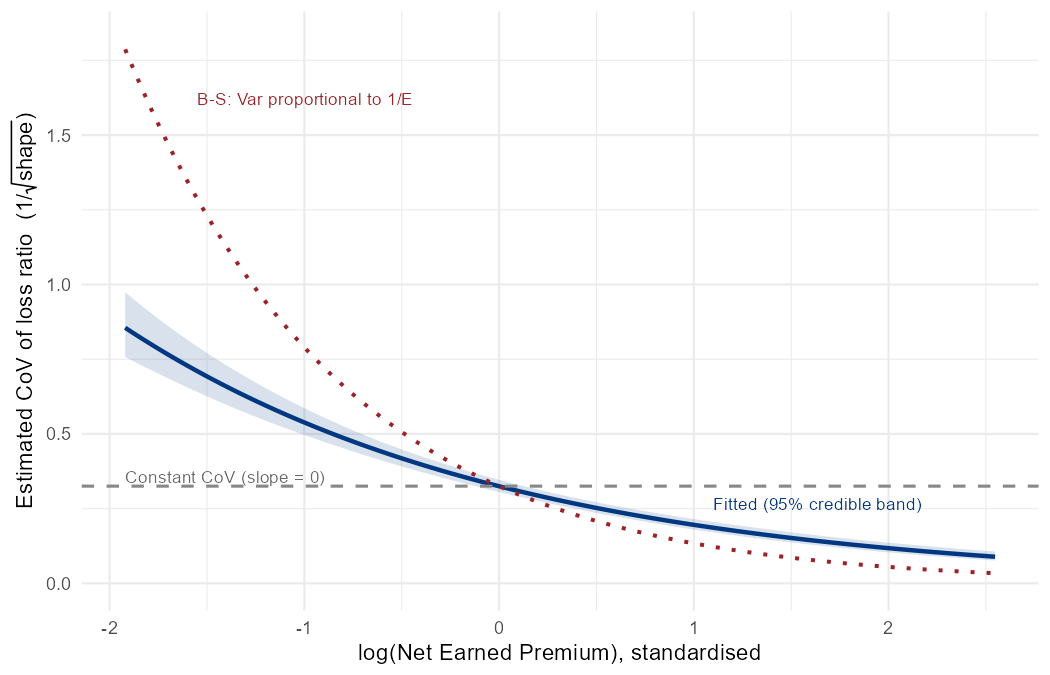} 

}

\caption{Estimated coefficient of variation of the loss ratio ($1/\sqrt{\phi}$) against standardised log NEP, with 95\% credible band. The dashed line is constant CoV; the dotted line is the CoV implied by B-S ($\mathrm{Var} \propto 1/E$), matched to the fit at the mean log NEP. The fitted slope on raw $\log E$ is $\hat\phi_1 = 0.57$ (95\% credible interval [0.50, 0.63]), against 1 under B-S.}\label{fig:fig-emp-shape-vs-size-app}
\end{figure}

\subsection*{Predictive Intervals}\label{sec:empirical-uq-app}
\addcontentsline{toc}{subsection}{Predictive Intervals}

Table~\ref{tab:coverage-comparison} compares seven models across point-prediction and coverage metrics, crossing three dimensions: \(\lambda\) specification (scalar vs.~tercile), dispersion specification (constant vs.~\(\phi{\sim}\log E_{it}\)), and NEP-proportional likelihood weighting (Section~\ref{sec:empirical-uq} summarises the results).

The dominant driver of coverage is the variance specification, not the weight.
Every model with \(\phi \sim \log E_{it}\) raises overall Tier 3 coverage to 88--91\% (against 67.2\% under constant dispersion), still a few points below the 95\% nominal level, with the improvement broadly uniform across size terciles.
By contrast, weighting the likelihood lowers overall coverage by only 1--3 points (88.0\% against 90.6\% for the scalar-\(\lambda\) pair).
The drop comes from the Mid tercile (82.8\% against 93.8\%); Small and Large change by under 2 points.
The scalar-\(\lambda\) model with \(\phi \sim \log E_{it}\) and an unweighted likelihood gives the best coverage (90.6\% overall, roughly uniform across terciles).
The combined tercile-\(\lambda\) specification matches it (90.6\%) while also improving point prediction (Gini 80.1\%, wMSE \(8.11 \times 10^{-3}\)).
The continuous-\(\lambda\) variant is slightly worse on both (89.6\% coverage), consistent with tercile \(\lambda\) as the preferred specification (Section~\ref{sec:empirical-diagnostics}).

\begin{table}[!h]
\centering
\caption{\label{tab:coverage-comparison}Point-prediction and coverage metrics for seven model variants.
        wMSE and Gini\% are exposure-weighted; T3 cov, Small, Mid, and Large are
        empirical Tier 3 (parameter + Gamma process noise) coverage rates (\%)
        of the nominal 95\% predictive interval on the held-out test set
        ($n=192$), overall and by size tercile. Bold = best in column.
        $^\ddagger$Tier 3 coverage for this variant is shown in Figures 25--26 (about 67\%, close to the scalar-$\lambda$ constant-$\phi$ row) and is not tabulated here.}
\centering
\resizebox{\ifdim\width>\linewidth\linewidth\else\width\fi}{!}{
\fontsize{9}{11}\selectfont
\begin{tabular}[t]{>{\raggedright\arraybackslash}p{4.5cm}lllllll}
\toprule
Model & $\text{wMSE}(\times 10^{-3})$ & Gini\% & Slope & T3 cov (\%) & Small (\%) & Mid (\%) & Large (\%)\\
\midrule
Scalar $\lambda$, const $\phi$, wtd & 8.61 & 76.5 & 1.00 & 67.2 & 54.7 & 64.1 & 82.8\\
Tercile $\lambda$, const $\phi$, wtd$^\ddagger$ & \textbf{7.96} & 78.7 & 1.03 & --- & --- & --- & ---\\
Scalar $\lambda$, $\phi{\sim}\log E$, no wt & 8.92 & 77.9 & 1.11 & \textbf{90.6} & 89.1 & \textbf{93.8} & 89.1\\
Scalar $\lambda$, $\phi{\sim}\log E$, wtd & 8.96 & 76.5 & 0.92 & 88.0 & \textbf{90.6} & 82.8 & \textbf{90.6}\\
Tercile $\lambda$, $\phi{\sim}\log E$, no wt & 8.11 & \textbf{80.1} & 1.09 & \textbf{90.6} & 89.1 & \textbf{93.8} & 89.1\\
\addlinespace
Tercile $\lambda$, $\phi{\sim}\log E$, wtd & 8.29 & 77.7 & 0.95 & 89.6 & 89.1 & 90.6 & 89.1\\
Continuous $\lambda$, $\phi{\sim}\log E$ & 8.48 & 78.6 & 1.09 & 89.6 & 89.1 & 89.1 & \textbf{90.6}\\
\bottomrule
\end{tabular}}
\end{table}

\begin{figure}[H]

{\centering \includegraphics[width=0.85\linewidth]{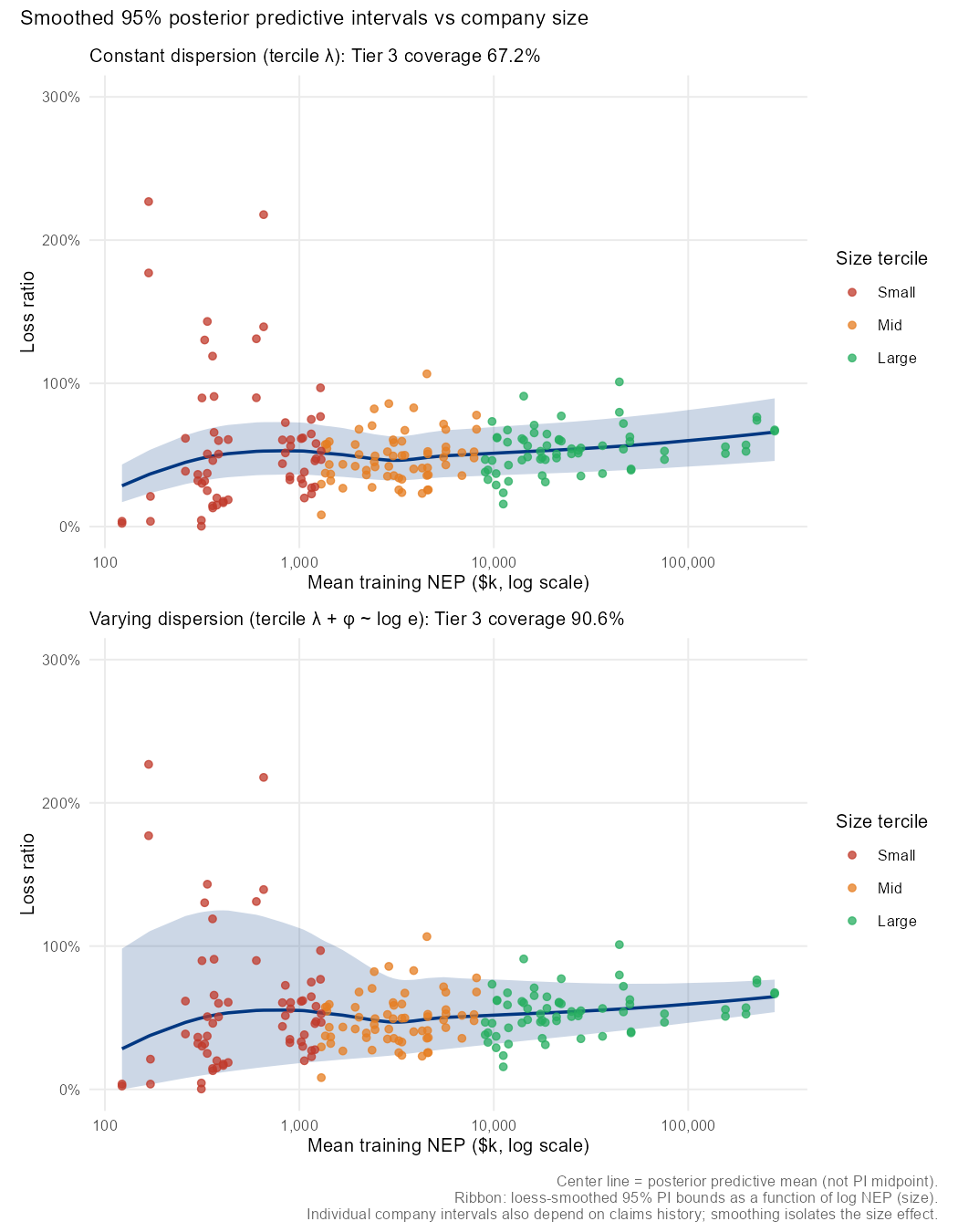} 

}

\caption{Held-out loss ratios (points, coloured by size tercile) with loess-smoothed 95\% posterior predictive ribbon. Top: best point-prediction model (tercile $\lambda$, constant $\phi$, exposure-weighted likelihood). Bottom: preferred combined model (tercile $\lambda$, $\phi \sim \log E_{it}$, unweighted likelihood). The ribbon represents the marginal effect of company size on interval width. Under constant dispersion the band is too narrow for small companies even with optimal $\lambda$; the varying-dispersion model corrects this.}\label{fig:fig-coverage-ribbon}
\end{figure}

\begin{figure}[H]

{\centering \includegraphics[width=0.65\linewidth]{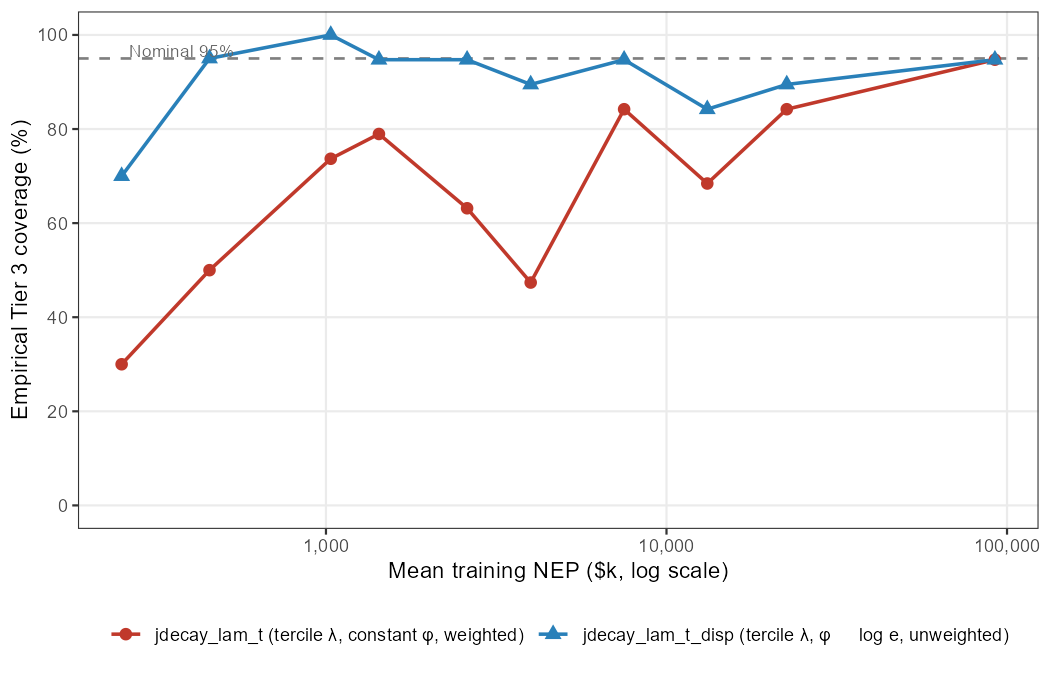} 

}

\caption{Empirical Tier 3 (parameter uncertainty + Gamma process noise) coverage by company size decile on the held-out test set ($n=192$). Red: best point-prediction model (tercile $\lambda$, constant $\phi$, weighted); size-graded coverage even with optimal $\lambda$. Blue: preferred combined model (tercile $\lambda$, $\phi \sim \log E_{it}$, unweighted); size gradient in coverage removed. Dashed line is the nominal 95\% target. The difference is driven by dispersion modelling (the weighting difference moves coverage by only 1--3 points): point-prediction optimality is not sufficient for calibrated predictive intervals.}\label{fig:fig-coverage-decile}
\end{figure}

Figures~\ref{fig:fig-coverage-ribbon} and~\ref{fig:fig-coverage-decile} show the mechanism.
With constant dispersion the band has the same relative width (constant CV) at every size, so it widens slightly with size as predicted loss ratios rise and small-company observations fall outside it; with exposure-varying dispersion the band fans out towards smaller companies and coverage is uniform across the size spectrum.

Most of the roughly 11\% wMSE gap between the weighted tercile-\(\lambda\) and unweighted scalar-\(\lambda\) dispersion models (7.96 against 8.92 \(\times 10^{-3}\)) is the tercile-\(\lambda\) gain itself (scalar \(\lambda\): 8.61), which the model-comparison tests support; the dispersion models cost only 1.8--3.6\% more, within what a single two-year holdout can resolve. The case for exposure-varying dispersion rests on coverage (67.2\% against 90.6\%).

In summary: use the weighted tercile-\(\lambda\) model for point predictions (wMSE \(= 7.96 \times 10^{-3}\), slope 1.03); for predictive intervals use the scalar-\(\lambda\) model with exposure-varying \(\phi\) and an unweighted likelihood (Tier 3 coverage 90.6\%, roughly uniform across size terciles), or the combined tercile-\(\lambda\) specification when one model must serve both (for example, simulating treaty structures across the portfolio).

\newpage

\section*{§D: Full Model Comparison}\label{app:full-comparison}
\addcontentsline{toc}{section}{§D: Full Model Comparison}

\begin{table}[H]
\centering
\caption{\label{tab:tbl-emp-results-full}Empirical results: US Commercial Auto, AY 2006--2007 test set.
    wMSE: exposure-weighted MSE (lower is better);
    Gini$_{\text{pct}}$: Gini as \% of the maximum achievable (higher is better);
    calibration slope: ideal = 1.00.
    Bold = best in column (lowest wMSE, highest Gini$_{\text{pct}}$, calibration slope closest to 1.00, excluding the market-mean baseline, whose slope is 1.00 by construction).
    $^\dag$Optimised for predictive interval calibration (Section \ref{sec:empirical-uq}). The row shown uses wide priors on $Z$ and $\lambda$; the version used there and in Appendix C, with the same priors as the primary model, has wMSE 8.92 and Gini 77.9\%.}
\centering
\resizebox{\ifdim\width>\linewidth\linewidth\else\width\fi}{!}{
\begin{tabular}[t]{>{\raggedright\arraybackslash}p{9cm}ccc}
\toprule
Model & wMSE ($\times10^{-3}$) & Gini$_{\text{pct}}$ & Calib. slope\\
\midrule
\addlinespace[0.3em]
\multicolumn{4}{l}{\textbf{Baselines}}\\
\hspace{1em}Market Mean (baseline) & 19.25 & 0.6\% & 1.00\\
\hspace{1em}Last Year LR (naive) & 13.14 & 75.1\% & 0.63\\
\addlinespace[0.3em]
\multicolumn{4}{l}{\textbf{Bühlmann--Straub sequential patching ladder}}\\
\hspace{1em}Bühlmann-Straub (standard, full history) & 10.97 & 69.6\% & 1.08\\
\hspace{1em}Buhlmann-Straub (min-wMSE K, full history) & 10.63 & 70.1\% & 0.88\\
\hspace{1em}Bühlmann-Straub (stratified, standard) & 10.34 & 69.8\% & 1.07\\
\hspace{1em}Bühlmann-Straub (pooled K + size comp) & 11.50 & 67.3\% & 1.01\\
\hspace{1em}Bühlmann-Straub (strat K + size comp) & 10.09 & 69.2\% & 1.02\\
\hspace{1em}Bühlmann-Straub (fully patched: strat K + size comp + EWMA tercile $\lambda$) & 8.16 & 79.0\% & 0.85\\
\addlinespace[0.3em]
\multicolumn{4}{l}{\textbf{GLMM competitors}}\\
\hspace{1em}GLMM (random intercept) & 10.85 & 69.9\% & 1.03\\
\hspace{1em}GLMM (random intercept + size) & 10.62 & 69.7\% & 1.06\\
\addlinespace[0.3em]
\multicolumn{4}{l}{\textbf{Logistic $Z$-structure hierarchy}}\\
\hspace{1em}B-S logistic (bz=1) & 9.14 & 77.2\% & 0.83\\
\hspace{1em}B-S strat logistic (bz=1, az tercile) & 9.03 & 77.1\% & 0.84\\
\hspace{1em}LogCred (scalar $\lambda$) & 8.61 & 76.5\% & \textbf{1.00}\\
\hspace{1em}LogCred (stratified Z intercept) & 8.57 & 77.1\% & 0.97\\
\addlinespace[0.3em]
\multicolumn{4}{l}{\textbf{Proposed: point prediction (varying $\lambda$)}}\\
\hspace{1em}LogCred (continuous $\lambda$) & 8.38 & 77.2\% & 1.02\\
\hspace{1em}LogCred (tercile $\lambda$) & 7.96 & 78.7\% & 1.03\\
\hspace{1em}LogCred (two-group $\lambda$: Small+Mid vs Large) & 7.97 & 78.5\% & 1.04\\
\hspace{1em}LogCred (strat comp + tercile $\lambda$) & \textbf{7.94} & 79.4\% & 0.98\\
\addlinespace[0.3em]
\multicolumn{4}{l}{\textbf{Fixed $\lambda$ comparison}}\\
\hspace{1em}Logistic (fixed $\lambda$=0.5) & 8.62 & 77.0\% & 1.08\\
\addlinespace[0.3em]
\multicolumn{4}{l}{\textbf{Dispersion extension$^\dag$ (predictive intervals)}}\\
\hspace{1em}LogCred (scalar $\lambda$, $\phi{\sim}\log E$, loose priors)$^\dag$ & 9.13 & 77.2\% & 1.14\\
\hspace{1em}LogCred (strat comp + tercile $\lambda$, $\phi{\sim}\log E$)$^\dag$ & 8.26 & \textbf{80.2\%} & 1.05\\
\bottomrule
\end{tabular}}
\end{table}

Table~\ref{tab:tbl-emp-results-full} gives the 21-model comparison across wMSE, Gini and calibration slope, covering all Bühlmann--Straub sequential patches, GLMM competitors, logistic hierarchy comparisons and dispersion extensions. Eight of the ten models in the main-body summary (Table~\ref{tab:tbl-emp-results}) are rows of the table; the other two (the B-S-\(Z\) diagnostic, discussed below, and the dispersion variant, in \hyperref[app:variance]{Appendix~C}) are not. The sections that follow cover the sequential-patching ladder, diagnostics of the logistic model's components, and the GLM variants with experience as a fixed predictor.

\subsection*{The sequential-patching ladder}\label{the-sequential-patching-ladder}
\addcontentsline{toc}{subsection}{The sequential-patching ladder}

\begin{figure}[h!]

{\centering \includegraphics[width=1\linewidth]{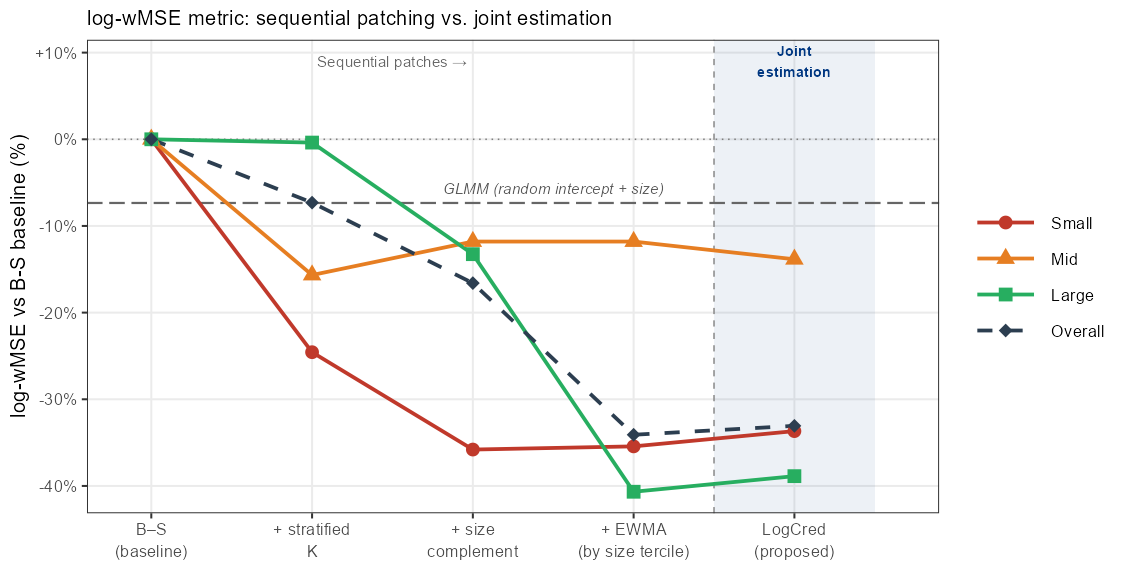} 

}

\caption{Prediction error (log-wMSE, \% change vs standard B-S) at each sequential patch step, by size tercile. Unlike raw wMSE (Figure~\ref{fig:fig-intro-patching}), the log scale shows no small-account reversal at the EWMA step: Small log-wMSE is 36\% below standard B-S after the size complement and 35\% below after EWMA with tercile $\lambda$, against 34\% below for the logistic credibility model with tercile $\lambda$ (right, shaded). The raw-scale reversal is consistent with a few large small-account losses, which dominate wMSE but are damped on the log scale.}\label{fig:fig-logwmse-patching}
\end{figure}

\textbf{Patching-ladder construction.} The ladder (Figure~\ref{fig:fig-intro-patching}, Section~\ref{sec:bs-failure}; log scale in Figure~\ref{fig:fig-logwmse-patching}) adds one patch at a time to standard B-S: stratifying \(K\) by size tercile, a size-varying complement, then EWMA with tercile \(\lambda\). At the stratified-\(K\) step the experience term is the account's exposure-weighted training-period mean, as in standard B-S. From the size-complement step onwards it is the lookback-window mean (equal-weighted at \(\lambda = 1\), then EWMA), while \(Z\) stays on the training-period exposure on which the stratified \(K\) values were estimated. A version using lookback exposure throughout, for both \(K\) and \(Z\), reduces wMSE by 24\% against standard B-S, compared with 26\% for the version shown, so the version shown is the stronger comparator and the small-account reversal appears in both.

\textbf{Size complement.} The size complement is fitted in two stages. For each training account-year we work out the complement that would have made the B-S prediction exactly right, given the account's credibility weight and experience, and we then regress that on log exposure. The regression is on the log scale, so it drops the 87 of 480 training rows (18\%) where that complement is negative, meaning even a zero complement would have over-predicted. Fitting by direct least squares on all rows cuts the Mid deterioration at this step, relative to the stratified-\(K\) step, from 16\% to 7\%, gives a fully patched B-S 24\% below standard B-S (26\% as shown), and leaves the Small reversal in place (+36\% against +34\%).

\textbf{Comparator caveats.} Rolling B-S estimates \(K\) from overlapping lookback windows of the same company, so its \(K\) is not a clean method-of-moments estimate. Standard B-S with its experience term updated to the same seven-year window as the logistic model, keeping the pooled \(K\) (Appendix~A), gives 11.16 \(\times 10^{-3}\), close to rolling B-S (10.90), so the conclusions do not depend on this. In the fully patched B-S the Mid decay rate sits at its upper bound (\(\hat\lambda = 1.00\)), so Mid is unchanged by the EWMA step and the ladder effectively has two free decay rates. Development maturity (Section~\ref{sec:lambda-decay}) and survivorship (\hyperref[app:survivorship]{Appendix~H}) apply to all models alike. An unshrunk seven-year average of each account's own year-normalised loss ratio, scaled by the realised \(\mu_t\), gives wMSE 10.81 \(\times 10^{-3}\) overall against 10.97 for standard B-S and 7.96 for the primary logistic model, and a Small-tercile calibration slope of 0.83 against 0.99.

\textbf{Why does adding EWMA erode the small-account gain?} Stratifying \(K\) takes Small wMSE (\(\times 10^{-3}\)) from 109.3 to 64.0 and the size complement leaves it at 64.2, but adding EWMA with tercile \(\lambda\) raises it to 86.4, a rise of 34.5\% (90\% company-bootstrap interval 3.5\% to 74\%). Refitting the complement or \(K\) to the EWMA experience does not remove the rise. Re-estimating the complement, iterated with \(\lambda\), gives 85.9, and dropping the size complement altogether gives 84.7. Refitting \(K\) makes it worse (95.6), because the training years favour a \(Z\) close to 1 for Small. A sweep over Small's decay rate and average \(Z\) shows why. Training error is lowest at a low \(\lambda_{\text{Sm}}\) (about 0.5) and a high \(Z\). In the test period, at \(Z \approx 0.8\), error rises as \(\lambda_{\text{Sm}}\) falls (64.2 at \(\lambda_{\text{Sm}} = 1\), 86.4 at \(\approx 0.51\)), and it is lower at \(Z \approx 0.6\) (78.5). The reversal therefore reflects a recency weighting that fits the training years but not the test years, combined with the high credibility the stratified \(K\) gives Small. The logistic model's lower \(Z\) for Small (about 0.6) limits the damage. This is a single panel (32 Small companies, two test years), so the evidence is indicative.

\subsection*{Diagnostics of the logistic model's components}\label{diagnostics-of-the-logistic-models-components}
\addcontentsline{toc}{subsection}{Diagnostics of the logistic model's components}

Fixing \(b_Z = 1\) gives held-out wMSE of 9.14 and 9.03 \(\times 10^{-3}\) for the pooled and stratified variants, against 8.61 with \(b_Z\) free, and overall calibration slopes of 0.83 and 0.84 respectively (mild over-crediting on balance). Stratifying the intercept by tercile barely changes either: it shifts the \(Z\)-curve vertically but leaves calibration little changed, because \(b_Z = 1\) is the wrong shape. Freeing \(b_Z\) recovers near-perfect calibration (slope \(= 1.00\)) and a lower wMSE. Adding tercile intercepts on top of free \(b_Z\) produces essentially the same result (slope \(= 0.97\), wMSE \(\approx\) identical) and the stratified intercept \emph{hurts} Small accounts (wMSE \(95.87\) vs \(86.32 \times 10^{-3}\)), where additional parameters overfit on 32 Small training companies. The performance gain is driven by the continuous free slope \(b_Z\), not by discretisation of the intercept.

\textbf{LOO-CV and likelihood-ratio tests.} These expand the training-data tests summarised in Section~\ref{sec:results-main} with their ELPD differences. Freeing \(b_Z\) against the B-S-nested \(b_Z = 1\) gains ELPD 16.5 (SE \(= 9.0\); ratio \(= 1.8\)), just below the \(2 \times \text{SE}\) threshold, although the MAP likelihood-ratio test is decisive (\(\chi^2(1) = 39.4\)); LOO standard errors are themselves uncertain with \(N\) this small \citep{vehtari2017practical}, so a ratio just below 2 is not a rejection. Estimating a scalar \(\lambda\) against \(\lambda = 1\) gains 36.9 (SE \(= 16.0\); ratio \(= 2.3\); \(\chi^2(1) = 75.6\)). Within the logistic family, tercile \(\lambda\) gains 7.3 over scalar \(\lambda\) (SE \(= 3.4\); ratio \(= 2.1\)) and continuous \(\lambda\) a borderline ratio of 1.5; head-to-head, tercile modestly beats continuous (ratio 1.6). A two-group split fits as well (wMSE 7.97 against 7.96 \(\times 10^{-3}\)), so the data support a Large-versus-rest distinction; the tercile specification is retained as primary because it was designated on the basis of the exploratory analysis (Section~\ref{sec:eda}) and costs nothing in fit. A stratified complement gives no LOO gain over the continuous complement at tercile \(\lambda\) (ratio 0.9), consistent with the negligible wMSE difference (Section~\ref{sec:results-main}).

Beyond this aggregate wMSE picture, the Small tercile's in-sample calibration slope specifically (1.58, Section~\ref{sec:results-main}) warrants a closer look. It is substantially attributable to NEP-weighting the fitting objective: shared parameters (\(a_Z\), \(b_Z\), the complement and \(\lambda\)) are estimated by minimising exposure-weighted error, so they are fit primarily to suit Large accounts, which dominate the weighted objective. In maximum-likelihood refits, moving from exposure weights to an unweighted likelihood takes the in-sample Small slope from 1.47 to 1.14 and the Large slope from 1.06 to 1.18, confirming the trade-off (table below, where the \(Z\)--size gradient moves with the weights in the same way).

\textbf{Likelihood weights and the $Z$--size gradient.}\phantomsection\label{sec:weights-sens}
The primary fits weight each observation's likelihood in proportion to its exposure \(E_{it}\), whereas the fitted variance falls roughly as \(E_{it}^{-0.57}\) (\hyperref[app:variance]{Appendix~C}), so precision weights would be proportional to \(E_{it}^{0.57}\). We refit the tercile-\(\lambda\) model by maximum likelihood with weights proportional to \(E_{it}^{\omega}\) for \(\omega = 1\) (primary), 0.57 and 0. The value \(\omega = 0.57\) is equivalent to an unweighted likelihood with \(\phi \propto E_{it}^{0.57}\), and \(\omega = 0\) (constant \(\phi\)) is an extreme.

\begin{table}[!h]
\centering
\caption{\label{tab:weights-sens}Sensitivity of the credibility weight to the likelihood weights (maximum likelihood, tercile $\lambda$, logistic $Z$, constant $\phi$). $\hat{b}_Z$ is on the standardised $\log\tilde{E}$ scale; $Z$ columns are mean fitted $Z$ by size tercile on the training years; wMSE is held-out and exposure-weighted; the two slope columns are the exposure-weighted in-sample calibration slopes (actual on predicted) on the training years. The $\omega = 1$ row differs slightly from the primary fit ($\hat{b}_Z = 0.06$) because the primary fit is Bayesian, with the priors in Appendix G (Listing 2).}
\centering
\resizebox{\ifdim\width>\linewidth\linewidth\else\width\fi}{!}{
\fontsize{9}{11}\selectfont
\begin{tabular}[t]{llllllll}
\toprule
Weight exponent $\omega$ & $\hat{b}_Z$ & $Z$ Small & $Z$ Mid & $Z$ Large & $\text{wMSE}(\times 10^{-3})$ & Slope Small & Slope Large\\
\midrule
1.00 & -0.08 & 0.67 & 0.65 & 0.63 & 8.09 & 1.47 & 1.06\\
0.57 & -0.29 & 0.75 & 0.69 & 0.61 & 8.44 & 1.35 & 1.10\\
0.00 & -0.98 & 0.88 & 0.74 & 0.49 & 12.59 & 1.14 & 1.18\\
\bottomrule
\end{tabular}}
\end{table}

Held-out accuracy changes little between \(\omega = 1\) and \(\omega = 0.57\), and the test of the B-S slope (\(b_Z = 1\)) is insensitive to the weighting (likelihood-ratio statistic 48, 31 and 31 under \(\omega = 1\), \(\omega = 0.57\) and a free dispersion exponent, all \(p < 10^{-7}\)). The shape of \(Z\) is what moves. It is nearly flat at \(\omega = 1\) (\(\hat{b}_Z = -0.08\)), but flatter weights (\(\omega = 0.57\), then \(\omega = 0\), equal weights) give small accounts a larger share, which raises their \(Z\) and turns the slope negative (-0.29 and -0.98); a free \(Z\) by tercile shifts the same way (Small/Mid/Large 0.77/0.58/0.62 at \(\omega = 1\), 0.90/0.73/0.56 at \(\omega = 0\)). These are point estimates from maximum-likelihood refits on one two-year holdout; the Bayesian fits without likelihood weights (with dispersion modelled) give negative slopes of \(-0.07\) to \(-0.19\) whose intervals include zero. Statements about the sign or flatness of the \(Z\)--size gradient are therefore conditional on exposure weighting.

The complement intercept and size slope are both clearly non-zero (\(\hat\alpha = -0.222\) {[}95\% CI: \(-0.321\), \(-0.121\){]}; \(\hat\beta = 0.127\) {[}95\% CI: \(0.072\), \(0.180\){]}), rejecting the standard grand-mean complement. A non-monotone tercile pattern (\(\exp(\hat\alpha_\text{Sm}) = 0.968\), \(\exp(\hat\alpha_\text{Md}) = 0.846\), \(\exp(\hat\alpha_\text{Lg}) = 0.986\)) is stable across \(\lambda\) specifications and reflects genuine complement structure rather than an identification artefact.

Table~\ref{tab:tbl-comp-calib} shows the complement calibration check: NEP-weighted mean fitted complement versus actual mean relative loss ratio by size tercile.

\begin{table}[!h]
\centering
\caption{\label{tab:tbl-comp-calib}Complement calibration check: NEP-weighted mean fitted complement vs
    actual mean relative loss ratio, by size tercile (training data AY 2001--2005).
    Continuous complement uses posterior mean of $\exp(\hat\alpha + \hat\beta \cdot \log E_{it})$;
    stratified complement uses posterior mean of $\exp(\hat\alpha_s)$ for each tercile $s$.
    Errors are fitted minus actual.}
\centering
\begin{tabular}[t]{lccccc}
\toprule
Tercile & Cont. comp. & Strat. comp. & Actual & Cont. error & Strat. error\\
\midrule
Small & 0.723 & 0.968 & 0.979 & -0.256 & -0.011\\
Mid & 0.813 & 0.846 & 0.819 & -0.006 & +0.027\\
Large & 1.016 & 0.986 & 1.014 & +0.002 & -0.028\\
\bottomrule
\end{tabular}
\end{table}

Mid-account calibration: the scalar-\(\lambda\) models give Mid slopes of \(\approx 0.12\)--\(0.16\), and the proposed model raises this to 0.71 (90\% bootstrap CI \([0.31,\, 1.07]\)). The GLMM's Mid slope near 1.0 (Section~\ref{sec:glmm-comparison}) reflects Mid's lag-1 signal, positive in training but near zero in the test period: the logistic \(Z\) of \(\approx 0.64\) over-weighted that uninformative test experience, whereas the GLMM's \(Z\) is \(\approx 0.47\) (\hyperref[app:implied-z]{Appendix~E}). Standard B-S and the GLMM, whose Mid weights are lower (\(\hat{Z} \approx 0.48\) and \(0.47\)), have pooled Mid slopes of 1.04 and 1.17, and those slopes are themselves a mixture of year-level miscalibrations.

Replacing the logistic \(Z\) with the B-S form \(Z_i = \tilde{E}_i/(\tilde{E}_i + K_\text{str})\), with \(K\) stratified by size tercile and estimated jointly, gives wMSE \(8.36 \times 10^{-3}\) (\(+5\%\) above the proposed model) and slope \(0.85\). \(K_\text{str}\) is only weakly identified, because the Gamma likelihood informs it through the conditional mean alone, so the estimates depend on the prior: \(N(3, 1.5)\) on \(\log K\) gives \(K_\text{str}\) of 107/46/45 (Small/Mid/Large), pushing \(Z\) toward 1, and wMSE \(9.01 \times 10^{-3}\), whereas \(N(8, 2)\), centred on the method-of-moments scale, gives 766/3,891/27,495 and the wMSE above (the version in Table \ref{tab:tbl-emp-results}). Under both priors the smooth logistic \(Z\) retains an edge, but its size depends on the prior and is not a property of the B-S form alone.

Freeing \(Z\) per tercile gives marginal Mid improvement but produces non-monotone estimates (\(\hat{Z}_\text{Sm} > \hat{Z}_\text{Lg} > \hat{Z}_\text{Md}\)), which credibility theory would not predict; the pattern also depends on the likelihood weights (see above). The continuous logistic \(Z\) is retained. Using lookback exposure instead of mean training NEP as the covariate for \(\lambda\) leaves the continuous-\(\lambda\) model essentially unchanged (wMSE 8.43 against 8.38 \(\times 10^{-3}\)).

\medskip

\noindent\textbf{GLM variants with experience as a fixed predictor.}
Three pricing-GLM variants that treat historical experience as a fixed covariate (Section~\ref{sec:framework}) were also tested on the held-out data. All are exposure-weighted Gamma GLMs with a log link and log exposure as a predictor: one with the log of the rolling average loss ratio (wMSE 4.9\% above standard B-S, slope 1.23), one adding an experience--exposure interaction (6.1\%, 1.25), and one with a tercile-stratified interaction (4.6\%, 1.22). All underperform both standard B-S and the proposed framework (27.4\% below B-S, slope 1.03).

\clearpage

\section*{§E: Credibility Weight (GLMM vs Logistic)}\label{app:implied-z}
\addcontentsline{toc}{section}{§E: Credibility Weight (GLMM vs Logistic)}

\noindent\textbf{GLMM credibility weight.} For the random-intercept GLMM (Gamma, log link, exposure weights) the fitted random intercept shrinks an account's own mean towards the fixed-effects prediction. The weight is approximately of the B-S form, \(Z_i \approx W_i/(W_i + K)\), where \(W_i\) is the account's total exposure weight in the training years and \(K = \phi/\sigma^2_u\), with \(\sigma^2_u\) the random-intercept variance (log scale) and \(\phi\) the Gamma dispersion. The form is not exact for a log link, because the fitted intercept also depends on the account's own level. The exact weight is the share of the gap between the account's own mean and the fixed-effects prediction that the fitted random intercept closes, \((e^{\hat{u}_i} - 1)/(\bar{Y}_i - 1)\), where \(\hat{u}_i\) is the fitted intercept and \(\bar{Y}_i\) the ratio of the account's own mean to that prediction. \(W_i/(W_i + K)\) differs from it by 0.007 on average and by at most 0.027 across the 82 accounts whose own mean differs from the fixed-effects prediction by more than 5\%. The estimated \(K\) is \(18{,}329\)k of NEP against \(17{,}547\)k for pooled B-S, so the GLMM shrinks almost exactly as B-S does: mean \(\hat{Z} \approx\) 0.15, 0.47 and 0.85 for Small, Mid and Large, against 0.15, 0.48 and 0.86.

\begin{figure}[H]

{\centering \includegraphics[width=0.9\linewidth]{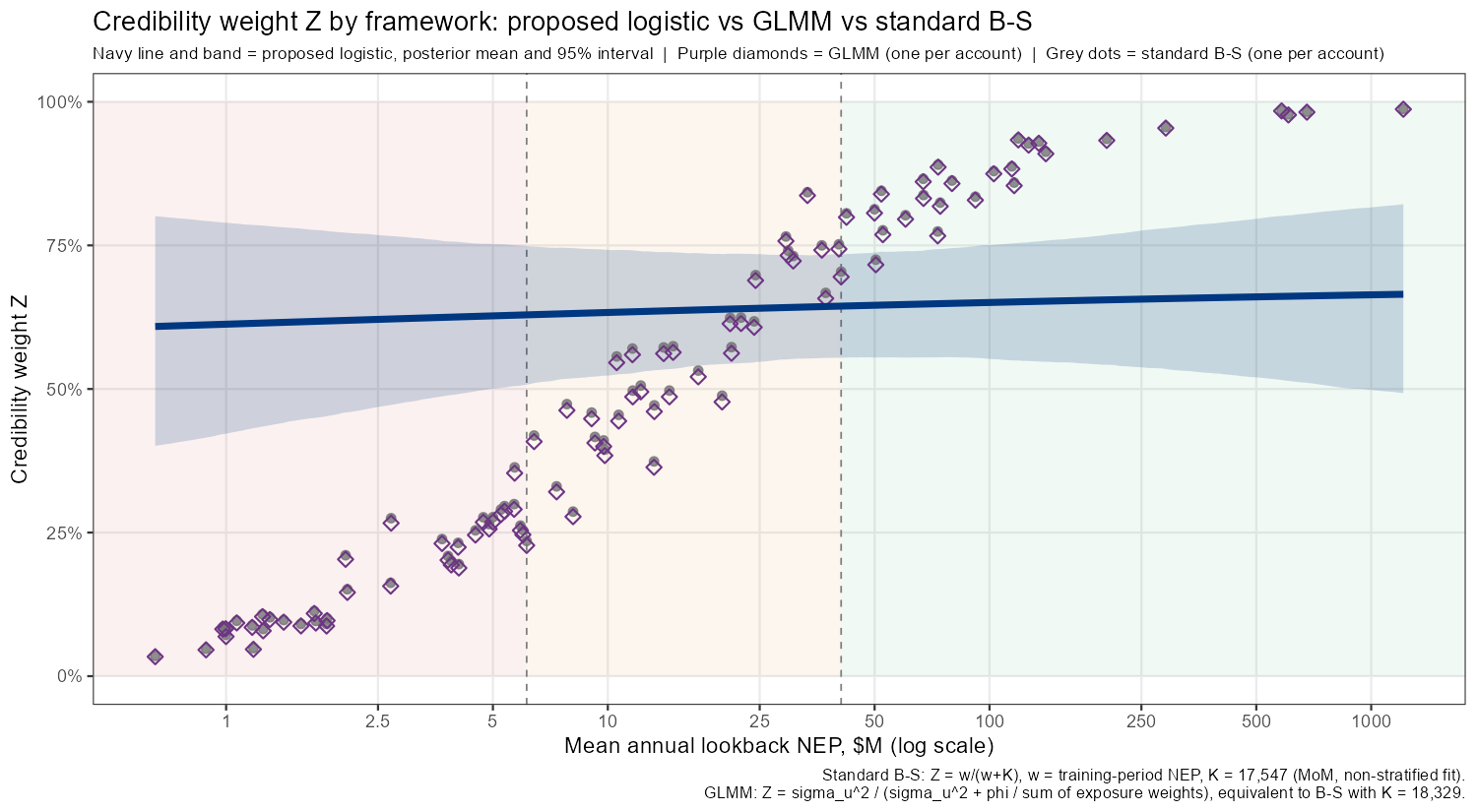} 

}

\caption{Credibility weight $\hat{Z}$ by framework against mean annual lookback NEP (\$M, log scale). Navy line and band = proposed logistic with 95\% posterior interval (nearly flat, $\hat{Z} \approx 0.61$--$0.67$ over the training exposure range). Purple diamonds = GLMM, $Z_i = \sigma^2_u/(\sigma^2_u + \phi/W_i)$, one per company. The two B-S-form frameworks nearly coincide. Grey dots = standard B-S, $Z = E_i/(E_i + K)$ with pooled method-of-moments (MoM) $K$, one dot per company; B-S weights use training-period NEP rather than lookback NEP, hence the scatter about a smooth curve. Vertical dashed lines = tercile boundaries on the log mean NEP scale. The logistic assigns substantially higher $\hat{Z}$ to Small accounts than either alternative; standard B-S rises steeply with size, assigning Large accounts a mean $\hat{Z} \approx 0.86$, well above the proposed logistic's $0.61$--$0.67$.}\label{fig:fig-z-curve-comparison}
\end{figure}

Figure~\ref{fig:fig-z-curve-comparison} compares the three frameworks. The proposed logistic is nearly flat (\(\hat{Z} \approx 0.61\)--\(0.67\) over the training exposure range), while the GLMM and standard B-S rise steeply together (standard B-S from \(\hat{Z} \approx 0.15\) on average for Small accounts to \(0.86\) for Large), which reflects the pooled-\(K\) constraint discussed in Section~\ref{sec:glmm-comparison}.

\newpage

\section*{§F: Simulation DGP}\label{app:dgp}
\addcontentsline{toc}{section}{§F: Simulation DGP}

\textbf{Common to all scenarios.} \(n = 1{,}000\) accounts observed for \(T = 8\) years with base frequency \(\mu_0 = 0.20\), over 50 seeds. Account exposure is log-normal (log-mean 1.5, log-SD 0.9) with annual log-normal noise (SD 0.15); accounts enter at random in years 3--7, so history at the test year (year 8) is 1--5 years. Claim counts are Poisson with mean \(\mu_0 E_{it}\theta_{it}\), where \(E_{it}\) is exposure and \(\theta_{it} = \exp(u_i + v_{it})\) is the account's relative risk: a permanent term \(u_i \sim \mathcal{N}(-\sigma_B^2/2, \sigma_B^2)\) and a stationary AR(1) transient \(v_{it}\) with persistence \(\psi\), mean \(-\sigma_W^2/2\) and SD \(\sigma_W\), so that \(E[\theta_{it}] = 1\). Every model uses the true base rate, \(\mu_0 E_{it}\), as its complement, and models are ranked by held-out Poisson deviance on the final test year.

\textbf{Standard B-S (S1).} Sanity-check scenario: the DGP satisfies the B-S variance-component assumptions (constant \(K\), no temporal drift); the log-normal risk parameter makes the B-S weight a close approximation to the conditional mean, not an exact one. Permanent risk \(\sigma_B = 0.25\) and no transient (\(\sigma_W = 0\)), so rates are stable: \(\theta_i = e^{u_i}\).
In B-S notation, the between-account variance of hypothetical mean frequencies is \(\sigma^2_u = \mu_0^2(e^{\sigma_B^2}-1)\), since a lognormal risk parameter with mean 1 has variance \(e^{\sigma_B^2}-1\), and the expected process variance per unit exposure is \(\sigma^2_\varepsilon = \mu_0\) (Poisson), so the true \(K = \sigma^2_\varepsilon/\sigma^2_u = 1/\{\mu_0(e^{\sigma_B^2}-1)\}\).
The logistic model should perform about as well as B-S here; any difference should be small.

\textbf{Temporal drift (S2).} Each seed draws persistence \(\psi \sim \mathcal{U}(0.10, 0.60)\); permanent risk \(\sigma_B = 0.20\) and transient SD \(\sigma_W = 0.35\) (log-normal scale).

\textbf{Industry heterogeneity (S3).} Three industry sectors with permanent risk \(\sigma_B \in \{0.10, 0.25, 0.40\}\) (stable to volatile), giving a wide spread of true \(K\) values, and modest temporal structure within sector (\(\psi = 0.15\), \(\sigma_W = 0.10\)). The industry label is available as a covariate in models marked \texttt{(expo+ind)} and withheld from the baseline \texttt{(expo)} model to isolate the information gain from supplying the segmentation label.

\medskip

\noindent\textbf{Simulation model key.}
All models use the same Poisson DGP observations; differences are in how \(Z_i\), \(\lambda\), and the complement are specified.

\begingroup\footnotesize
\begin{tabular}{p{3.5cm}p{12cm}}
\hline
\textbf{Label} & \textbf{Specification} \\
\hline
B\"{u}hlmann--Straub & Pooled method-of-moments (MoM) $K$ (process variance set to the known Poisson base rate), equal-weighted experience ($\lambda=1$). Baseline. \\
Het-K B-S & $Z_i = \Lambda(\log\tilde{E}_i - \log K_i)$, so $b_Z = 1$ as in B-S, with $\log K_i$ linear in $\text{n\_years}$ and CoV; $\lambda=1$. ML. \\
Stratified B-S & Separate MoM $K$ per industry sector; $\lambda=1$. S3 only. \\
Logistic (expo) & $Z_i = \Lambda(a_Z + b_Z\log\tilde{E}_i)$, free $b_Z$; $\lambda=1$ (cf.\ LogCred without decay). ML. \\
Logistic (expo+ind) & As Logistic (expo) with industry indicator shifts on $a_Z$. ML. S3 only. \\
Logistic (full) & $Z_i = \Lambda(a_Z + b_Z\log\tilde{E}_i + c_Z \cdot \text{n\_years} + d_Z \cdot \text{CoV})$; $\lambda=1$. Penalised ML (MAP). \\
EWMA (fixed $\lambda$) & $Z_i = \Lambda(a_Z + b_Z\log\tilde{E}_i)$; $\lambda$ set to 0.50 (uninformed practitioner default), with $a_Z, b_Z$ refitted at the fixed $\lambda$; the complement is the known base rate. \\
Est. decay (expo Z) & $Z_i = \Lambda(a_Z + b_Z\log\tilde{E}_i)$; $\lambda$ estimated jointly (cf.\ LogCred, scalar $\lambda$). Penalised ML (MAP). \\
Est. decay (full Z) & $Z_i = \Lambda(a_Z + b_Z\log\tilde{E}_i + c_Z\cdot\text{n\_years} + d_Z\cdot\text{CoV})$; $\lambda$ estimated jointly. Penalised ML (MAP). \\
GLM naive & Poisson GLM with the base rate as offset and the log ratio of the account's past claims to expected past claims (smoothed by 0.5) as a fitted regressor; no explicit credibility weight. \\
GLMM & Poisson GLMM with random intercept per account ($\lambda=1$ implicitly). Laplace-approximated ML. \\
GLMM (corr) & GLMM with each prediction multiplied by $e^{v_i/2}$, where $v_i$ is the posterior variance of the account's random intercept, correcting the bias from exponentiating a log-scale prediction. \\
Base rate & True base rate $\mu_0 E_{it}$; no account-level experience. \\
\hline
\end{tabular}

\smallskip

\noindent\textbf{Notation.} \(\tilde{E}_i\) is lookback exposure; \(\text{n\_years}\) is the number of years in the account's lookback window (1--5) and CoV the coefficient of variation of its past claim frequencies (0 with fewer than two years), both standardised; \(\Lambda\) is the logistic function. ML is maximum likelihood; MAP adds normal priors, \(\mathrm{N}(-1, 1.5)\) on the \(Z\) intercept and, where \(\lambda\) is estimated, \(\mathrm{N}(0, 1.5)\) on its logit; MoM is method of moments.

\endgroup

\newpage

\section*{§G: R Implementation}\label{app:implementation}
\addcontentsline{toc}{section}{§G: R Implementation}

All code is in R.
Four listings follow.
Listing 1 shows the fast MLE implementation: all parameters jointly in a single
\texttt{nlminb} call, converging in about a tenth of a second for the CAS case study on the machine used for this paper.
Listing 2 shows the Bayesian alternative via \texttt{brms}, which interfaces to Stan and delivers posterior uncertainty
and the conservative rate \(R^*_i(q)\).
Listing 3 shows how to compute the underwriter-facing numbers from
the MLE fit; the Bayesian fit works the same way with posterior means.
Listing 4 shows MAP estimation as a bridge between MLE and full Bayes,
useful when parameters are weakly identified in small portfolios.
All code assumes account-year panel data, one row per account-year, built by \texttt{R/01\_data.R} in the GitHub repository, with columns \texttt{lr\_rel} (loss ratio relative to the market mean), \texttt{expo\_wt} (exposure weight), \texttt{log\_expo\_sc} (standardised log current-year exposure), \texttt{log\_expo\_used\_sc} (standardised log cumulative lookback exposure, \(\log\tilde{E}_i\)), \texttt{log\_mean\_expo\_sc} (standardised log mean training exposure), \texttt{size\_tercile} (Small, Mid or Large, from tercile breaks of mean training exposure) and the lagged relative loss ratios and exposures \texttt{lr\_lag}\(k\)\texttt{\_rel} and \texttt{expo\_lag}\(k\).

\subsection*{Listing 1: MLE Quick-Start}\label{listing-1-mle-quick-start}
\addcontentsline{toc}{subsection}{Listing 1: MLE Quick-Start}

The model is a standard nonlinear optimisation: write down the negative
exposure-weighted Gamma log-likelihood and pass it to \texttt{nlminb}.
All parameters (credibility weight, decay rate, and base rate)
are estimated jointly.
Swap \texttt{dgamma} for \texttt{dpois} or a Tweedie density to change
the target type; the optimisation structure is unchanged.
The data preparation (\texttt{R/01\_data.R}, which builds \texttt{df\_train}) and the EWMA helper (\texttt{R/02\_features.R})
are in the GitHub repository.

\begin{Shaded}
\begin{Highlighting}[]
\CommentTok{\# par = c(az, bz, alpha, beta, lamB, lamS, log\_phi)}
\CommentTok{\# df\_train is built by R/01\_data.R; ewma\_fbar(df, lam) is in R/02\_features.R}
\NormalTok{nll }\OtherTok{\textless{}{-}} \ControlFlowTok{function}\NormalTok{(par, df) \{}
\NormalTok{  Z    }\OtherTok{\textless{}{-}} \FunctionTok{plogis}\NormalTok{(par[}\DecValTok{1}\NormalTok{] }\SpecialCharTok{+}\NormalTok{ par[}\DecValTok{2}\NormalTok{] }\SpecialCharTok{*}\NormalTok{ df}\SpecialCharTok{$}\NormalTok{log\_expo\_used\_sc)}
\NormalTok{  lam  }\OtherTok{\textless{}{-}} \FunctionTok{plogis}\NormalTok{(par[}\DecValTok{5}\NormalTok{] }\SpecialCharTok{+}\NormalTok{ par[}\DecValTok{6}\NormalTok{] }\SpecialCharTok{*}\NormalTok{ df}\SpecialCharTok{$}\NormalTok{log\_mean\_expo\_sc)}
\NormalTok{  mu0  }\OtherTok{\textless{}{-}} \FunctionTok{exp}\NormalTok{(   par[}\DecValTok{3}\NormalTok{] }\SpecialCharTok{+}\NormalTok{ par[}\DecValTok{4}\NormalTok{] }\SpecialCharTok{*}\NormalTok{ df}\SpecialCharTok{$}\NormalTok{log\_expo\_sc)}
\NormalTok{  fbar }\OtherTok{\textless{}{-}} \FunctionTok{ewma\_fbar}\NormalTok{(df, lam)   }\CommentTok{\# exposure{-}weighted EWMA over W lags}
\NormalTok{  mu   }\OtherTok{\textless{}{-}} \FunctionTok{pmax}\NormalTok{((}\DecValTok{1} \SpecialCharTok{{-}}\NormalTok{ Z) }\SpecialCharTok{*}\NormalTok{ mu0 }\SpecialCharTok{+}\NormalTok{ Z }\SpecialCharTok{*}\NormalTok{ fbar, }\FloatTok{1e{-}6}\NormalTok{)}
\NormalTok{  phi  }\OtherTok{\textless{}{-}} \FunctionTok{exp}\NormalTok{(par[}\DecValTok{7}\NormalTok{])}
  \SpecialCharTok{{-}}\FunctionTok{sum}\NormalTok{(df}\SpecialCharTok{$}\NormalTok{expo\_wt }\SpecialCharTok{*}
         \FunctionTok{dgamma}\NormalTok{(df}\SpecialCharTok{$}\NormalTok{lr\_rel, }\AttributeTok{shape =}\NormalTok{ phi, }\AttributeTok{rate =}\NormalTok{ phi }\SpecialCharTok{/}\NormalTok{ mu, }\AttributeTok{log =} \ConstantTok{TRUE}\NormalTok{))}
\NormalTok{\}}

\CommentTok{\# Converges in about 0.1 seconds for the 96{-}company case study and}
\CommentTok{\# 3{-}{-}10 seconds for a resampled 1,000{-}account panel (timings are from}
\CommentTok{\# the machine used for this paper and will vary with hardware)}
\NormalTok{init    }\OtherTok{\textless{}{-}} \FunctionTok{c}\NormalTok{(}\AttributeTok{az =} \SpecialCharTok{{-}}\FloatTok{0.5}\NormalTok{, }\AttributeTok{bz =} \FloatTok{0.5}\NormalTok{, }\AttributeTok{alpha =} \DecValTok{0}\NormalTok{, }\AttributeTok{beta =} \DecValTok{0}\NormalTok{,}
             \AttributeTok{lamB =} \DecValTok{0}\NormalTok{,  }\AttributeTok{lamS =} \DecValTok{0}\NormalTok{,  }\AttributeTok{log\_phi =} \DecValTok{2}\NormalTok{)}
\NormalTok{fit\_mle }\OtherTok{\textless{}{-}} \FunctionTok{nlminb}\NormalTok{(init, nll, }\AttributeTok{df =}\NormalTok{ df\_train,}
                  \AttributeTok{control =} \FunctionTok{list}\NormalTok{(}\AttributeTok{iter.max =} \DecValTok{500}\NormalTok{, }\AttributeTok{rel.tol =} \FloatTok{1e{-}8}\NormalTok{))}
\NormalTok{par\_hat }\OtherTok{\textless{}{-}} \FunctionTok{setNames}\NormalTok{(fit\_mle}\SpecialCharTok{$}\NormalTok{par, }\FunctionTok{names}\NormalTok{(init))}
\end{Highlighting}
\end{Shaded}

\subsection*{Listing 2: Bayesian Fit (brms)}\label{listing-2-bayesian-fit-brms}
\addcontentsline{toc}{subsection}{Listing 2: Bayesian Fit (brms)}

The \texttt{brms} formula implements the recommended LogCred tercile-\(\lambda\)
specification (Equations~\eqref{eq:pred}, \eqref{eq:logistic}, \eqref{eq:ewma} and~\eqref{eq:logit-lambda}).
Input data are as defined in the introduction to this appendix; three lags are shown for brevity (the case study and \texttt{R/03\_models\_ca.R} use seven).
Tercile indicators \texttt{is\_sm}, \texttt{is\_md}, \texttt{is\_lg}
are derived from size tercile breaks computed from the training data via \texttt{quantile(mean\_expo, c(1/3, 2/3))}.

\begin{Shaded}
\begin{Highlighting}[]
\FunctionTok{library}\NormalTok{(brms)}

\CommentTok{\# Add tercile indicators (breaks derived from training data via quantile())}
\NormalTok{df\_train}\SpecialCharTok{$}\NormalTok{is\_sm }\OtherTok{\textless{}{-}} \FunctionTok{as.numeric}\NormalTok{(df\_train}\SpecialCharTok{$}\NormalTok{size\_tercile }\SpecialCharTok{==} \StringTok{"Small"}\NormalTok{)}
\NormalTok{df\_train}\SpecialCharTok{$}\NormalTok{is\_md }\OtherTok{\textless{}{-}} \FunctionTok{as.numeric}\NormalTok{(df\_train}\SpecialCharTok{$}\NormalTok{size\_tercile }\SpecialCharTok{==} \StringTok{"Mid"}\NormalTok{)}
\NormalTok{df\_train}\SpecialCharTok{$}\NormalTok{is\_lg }\OtherTok{\textless{}{-}} \FunctionTok{as.numeric}\NormalTok{(df\_train}\SpecialCharTok{$}\NormalTok{size\_tercile }\SpecialCharTok{==} \StringTok{"Large"}\NormalTok{)}

\CommentTok{\# Non{-}linear formula: seven parameters (plus the Gamma shape) estimated jointly}
\NormalTok{form }\OtherTok{\textless{}{-}} \FunctionTok{bf}\NormalTok{(}
\NormalTok{  lr\_rel }\SpecialCharTok{|} \FunctionTok{weights}\NormalTok{(expo\_wt) }\SpecialCharTok{\textasciitilde{}}
    \CommentTok{\# Complement: (1 {-} Z\_i) * mu\_hat\_i}
\NormalTok{    (}\DecValTok{1} \SpecialCharTok{{-}} \FunctionTok{inv\_logit}\NormalTok{(az }\SpecialCharTok{+}\NormalTok{ bz }\SpecialCharTok{*}\NormalTok{ log\_expo\_used\_sc)) }\SpecialCharTok{*}
      \FunctionTok{exp}\NormalTok{(alpha }\SpecialCharTok{+}\NormalTok{ beta }\SpecialCharTok{*}\NormalTok{ log\_expo\_sc) }\SpecialCharTok{+}
    \CommentTok{\# Experience: Z\_i * EWMA{-}weighted past lr\_rel (tercile lambda)}
    \FunctionTok{inv\_logit}\NormalTok{(az }\SpecialCharTok{+}\NormalTok{ bz }\SpecialCharTok{*}\NormalTok{ log\_expo\_used\_sc) }\SpecialCharTok{*}
\NormalTok{      (lr\_lag1\_rel }\SpecialCharTok{*}\NormalTok{ expo\_lag1 }\SpecialCharTok{+}
\NormalTok{         lr\_lag2\_rel }\SpecialCharTok{*}\NormalTok{ expo\_lag2 }\SpecialCharTok{*}
           \FunctionTok{inv\_logit}\NormalTok{(lamSm}\SpecialCharTok{*}\NormalTok{is\_sm }\SpecialCharTok{+}\NormalTok{ lamMd}\SpecialCharTok{*}\NormalTok{is\_md }\SpecialCharTok{+}\NormalTok{ lamLg}\SpecialCharTok{*}\NormalTok{is\_lg) }\SpecialCharTok{+}
\NormalTok{         lr\_lag3\_rel }\SpecialCharTok{*}\NormalTok{ expo\_lag3 }\SpecialCharTok{*}
           \FunctionTok{inv\_logit}\NormalTok{(lamSm}\SpecialCharTok{*}\NormalTok{is\_sm }\SpecialCharTok{+}\NormalTok{ lamMd}\SpecialCharTok{*}\NormalTok{is\_md }\SpecialCharTok{+}\NormalTok{ lamLg}\SpecialCharTok{*}\NormalTok{is\_lg)}\SpecialCharTok{\^{}}\DecValTok{2}\NormalTok{) }\SpecialCharTok{/}
\NormalTok{      (expo\_lag1 }\SpecialCharTok{+}
\NormalTok{         expo\_lag2 }\SpecialCharTok{*}
           \FunctionTok{inv\_logit}\NormalTok{(lamSm}\SpecialCharTok{*}\NormalTok{is\_sm }\SpecialCharTok{+}\NormalTok{ lamMd}\SpecialCharTok{*}\NormalTok{is\_md }\SpecialCharTok{+}\NormalTok{ lamLg}\SpecialCharTok{*}\NormalTok{is\_lg) }\SpecialCharTok{+}
\NormalTok{         expo\_lag3 }\SpecialCharTok{*}
           \FunctionTok{inv\_logit}\NormalTok{(lamSm}\SpecialCharTok{*}\NormalTok{is\_sm }\SpecialCharTok{+}\NormalTok{ lamMd}\SpecialCharTok{*}\NormalTok{is\_md }\SpecialCharTok{+}\NormalTok{ lamLg}\SpecialCharTok{*}\NormalTok{is\_lg)}\SpecialCharTok{\^{}}\DecValTok{2}\NormalTok{),}
\NormalTok{  az }\SpecialCharTok{\textasciitilde{}} \DecValTok{1}\NormalTok{, bz }\SpecialCharTok{\textasciitilde{}} \DecValTok{1}\NormalTok{, alpha }\SpecialCharTok{\textasciitilde{}} \DecValTok{1}\NormalTok{, beta }\SpecialCharTok{\textasciitilde{}} \DecValTok{1}\NormalTok{,}
\NormalTok{  lamSm }\SpecialCharTok{\textasciitilde{}} \DecValTok{1}\NormalTok{, lamMd }\SpecialCharTok{\textasciitilde{}} \DecValTok{1}\NormalTok{, lamLg }\SpecialCharTok{\textasciitilde{}} \DecValTok{1}\NormalTok{,}
  \AttributeTok{nl =} \ConstantTok{TRUE}
\NormalTok{)}

\CommentTok{\# Weakly informative priors (see Section 2.1 for rationale)}
\NormalTok{pri }\OtherTok{\textless{}{-}} \FunctionTok{c}\NormalTok{(}
  \FunctionTok{prior}\NormalTok{(}\FunctionTok{normal}\NormalTok{( }\FloatTok{0.0}\NormalTok{, }\FloatTok{0.3}\NormalTok{), }\AttributeTok{nlpar =} \StringTok{"alpha"}\NormalTok{),}
  \FunctionTok{prior}\NormalTok{(}\FunctionTok{normal}\NormalTok{( }\FloatTok{0.0}\NormalTok{, }\FloatTok{0.3}\NormalTok{), }\AttributeTok{nlpar =} \StringTok{"beta"}\NormalTok{),}
  \FunctionTok{prior}\NormalTok{(}\FunctionTok{normal}\NormalTok{(}\SpecialCharTok{{-}}\FloatTok{0.5}\NormalTok{, }\FloatTok{1.0}\NormalTok{), }\AttributeTok{nlpar =} \StringTok{"az"}\NormalTok{),}
  \FunctionTok{prior}\NormalTok{(}\FunctionTok{normal}\NormalTok{( }\FloatTok{0.5}\NormalTok{, }\FloatTok{0.5}\NormalTok{), }\AttributeTok{nlpar =} \StringTok{"bz"}\NormalTok{),}
  \FunctionTok{prior}\NormalTok{(}\FunctionTok{normal}\NormalTok{( }\FloatTok{0.0}\NormalTok{, }\FloatTok{1.5}\NormalTok{), }\AttributeTok{nlpar =} \StringTok{"lamSm"}\NormalTok{),}
  \FunctionTok{prior}\NormalTok{(}\FunctionTok{normal}\NormalTok{( }\FloatTok{0.0}\NormalTok{, }\FloatTok{1.5}\NormalTok{), }\AttributeTok{nlpar =} \StringTok{"lamMd"}\NormalTok{),}
  \FunctionTok{prior}\NormalTok{(}\FunctionTok{normal}\NormalTok{( }\FloatTok{0.0}\NormalTok{, }\FloatTok{1.5}\NormalTok{), }\AttributeTok{nlpar =} \StringTok{"lamLg"}\NormalTok{)}
\NormalTok{)}

\NormalTok{fit }\OtherTok{\textless{}{-}} \FunctionTok{brm}\NormalTok{(}
\NormalTok{  form,}
  \AttributeTok{data    =}\NormalTok{ df\_train,}
  \AttributeTok{family  =} \FunctionTok{Gamma}\NormalTok{(}\AttributeTok{link =} \StringTok{"identity"}\NormalTok{),}
  \AttributeTok{prior   =}\NormalTok{ pri,}
  \AttributeTok{chains  =} \DecValTok{4}\NormalTok{, }\AttributeTok{cores =} \DecValTok{4}\NormalTok{, }\AttributeTok{iter =} \DecValTok{2000}\NormalTok{,}
  \AttributeTok{control =} \FunctionTok{list}\NormalTok{(}\AttributeTok{adapt\_delta =} \FloatTok{0.97}\NormalTok{),}
  \AttributeTok{seed    =} \DecValTok{48}
\NormalTok{)}

\CommentTok{\# Posterior lambda estimates (plogis() is R\textquotesingle{}s inverse logit)}
\NormalTok{p      }\OtherTok{\textless{}{-}} \FunctionTok{as\_draws\_df}\NormalTok{(fit)}
\NormalTok{lam\_sm }\OtherTok{\textless{}{-}} \FunctionTok{mean}\NormalTok{(}\FunctionTok{plogis}\NormalTok{(p}\SpecialCharTok{$}\NormalTok{b\_lamSm\_Intercept))  }\CommentTok{\# approx 0.69 (3 lags)}
\NormalTok{lam\_md }\OtherTok{\textless{}{-}} \FunctionTok{mean}\NormalTok{(}\FunctionTok{plogis}\NormalTok{(p}\SpecialCharTok{$}\NormalTok{b\_lamMd\_Intercept))  }\CommentTok{\# approx 0.66}
\NormalTok{lam\_lg }\OtherTok{\textless{}{-}} \FunctionTok{mean}\NormalTok{(}\FunctionTok{plogis}\NormalTok{(p}\SpecialCharTok{$}\NormalTok{b\_lamLg\_Intercept))  }\CommentTok{\# approx 0.16}
\end{Highlighting}
\end{Shaded}

\subsection*{Listing 3: Renewal Prediction}\label{listing-3-renewal-prediction}
\addcontentsline{toc}{subsection}{Listing 3: Renewal Prediction}

Given MLE parameter estimates (\texttt{par\_hat} from Listing 1), compute
the numbers delivered to the underwriter for each renewal account \(i\)
(three lags are shown for brevity):
the credibility weight \(\hat{Z}_i\), market-mean complement \(\hat\mu_i\)
and recency-weighted experience \(\hat f_i\), together with the renewal rate \(\hat r_i\) they produce.
(For Listing 2's tercile-\(\lambda\) specification, replace \texttt{lam\_hat} below with the fitted \(\lambda\) of each account's size tercile.)

\begin{Shaded}
\begin{Highlighting}[]
\CommentTok{\# Fitted parameters from Listing 1}
\NormalTok{az    }\OtherTok{\textless{}{-}}\NormalTok{ par\_hat[[}\StringTok{"az"}\NormalTok{]];    bz   }\OtherTok{\textless{}{-}}\NormalTok{ par\_hat[[}\StringTok{"bz"}\NormalTok{]]}
\NormalTok{alpha }\OtherTok{\textless{}{-}}\NormalTok{ par\_hat[[}\StringTok{"alpha"}\NormalTok{]]; beta }\OtherTok{\textless{}{-}}\NormalTok{ par\_hat[[}\StringTok{"beta"}\NormalTok{]]}
\NormalTok{lamB  }\OtherTok{\textless{}{-}}\NormalTok{ par\_hat[[}\StringTok{"lamB"}\NormalTok{]];  lamS }\OtherTok{\textless{}{-}}\NormalTok{ par\_hat[[}\StringTok{"lamS"}\NormalTok{]]}

\CommentTok{\# {-}{-}{-} Per{-}account renewal predictions (vectorised over df\_new) {-}{-}{-}{-}{-}{-}{-}{-}{-}{-}{-}{-}{-}{-}{-}}
\CommentTok{\# IMPORTANT: log\_expo\_used\_sc and log\_mean\_expo\_sc must be standardised using}
\CommentTok{\# the TRAINING{-}SET mean and SD (not the new{-}account\textquotesingle{}s own values):}
\CommentTok{\#   df\_new$log\_expo\_used\_sc \textless{}{-}}
\CommentTok{\#     (log(df\_new$expo\_used) {-} mu\_logE\_train) / sd\_logE\_train}
\CommentTok{\#   df\_new$log\_mean\_expo\_sc \textless{}{-}}
\CommentTok{\#     (log(df\_new$mean\_expo) {-} mu\_meanE\_train) / sd\_meanE\_train}
\CommentTok{\# where mu\_logE\_train, sd\_logE\_train etc. are saved from Phase 1 fitting.}

\CommentTok{\# Credibility weight (0 = full shrinkage to market mean, 1 = full experience)}
\NormalTok{Z\_hat  }\OtherTok{\textless{}{-}} \FunctionTok{plogis}\NormalTok{(az }\SpecialCharTok{+}\NormalTok{ bz }\SpecialCharTok{*}\NormalTok{ df\_new}\SpecialCharTok{$}\NormalTok{log\_expo\_used\_sc)}

\CommentTok{\# EWMA decay rate (varies by account size — uses stable mean training exposure)}
\NormalTok{lam\_hat }\OtherTok{\textless{}{-}} \FunctionTok{plogis}\NormalTok{(lamB }\SpecialCharTok{+}\NormalTok{ lamS }\SpecialCharTok{*}\NormalTok{ df\_new}\SpecialCharTok{$}\NormalTok{log\_mean\_expo\_sc)}

\CommentTok{\# Market{-}mean complement (size{-}adjusted prior)}
\NormalTok{mu\_hat  }\OtherTok{\textless{}{-}} \FunctionTok{exp}\NormalTok{(alpha }\SpecialCharTok{+}\NormalTok{ beta }\SpecialCharTok{*}\NormalTok{ df\_new}\SpecialCharTok{$}\NormalTok{log\_expo\_sc) }\SpecialCharTok{*}\NormalTok{ df\_new}\SpecialCharTok{$}\NormalTok{mu\_t}

\CommentTok{\# EWMA{-}weighted past loss ratio (exposure{-}weighted, relative lags)}
\NormalTok{wc }\OtherTok{\textless{}{-}}\NormalTok{ df\_new}\SpecialCharTok{$}\NormalTok{lr\_lag1\_rel }\SpecialCharTok{*}\NormalTok{ df\_new}\SpecialCharTok{$}\NormalTok{expo\_lag1 }\SpecialCharTok{+}
\NormalTok{      df\_new}\SpecialCharTok{$}\NormalTok{lr\_lag2\_rel }\SpecialCharTok{*}\NormalTok{ df\_new}\SpecialCharTok{$}\NormalTok{expo\_lag2 }\SpecialCharTok{*}\NormalTok{ lam\_hat }\SpecialCharTok{+}
\NormalTok{      df\_new}\SpecialCharTok{$}\NormalTok{lr\_lag3\_rel }\SpecialCharTok{*}\NormalTok{ df\_new}\SpecialCharTok{$}\NormalTok{expo\_lag3 }\SpecialCharTok{*}\NormalTok{ lam\_hat}\SpecialCharTok{\^{}}\DecValTok{2}
\NormalTok{we }\OtherTok{\textless{}{-}}\NormalTok{ df\_new}\SpecialCharTok{$}\NormalTok{expo\_lag1 }\SpecialCharTok{+}
\NormalTok{      df\_new}\SpecialCharTok{$}\NormalTok{expo\_lag2 }\SpecialCharTok{*}\NormalTok{ lam\_hat }\SpecialCharTok{+}
\NormalTok{      df\_new}\SpecialCharTok{$}\NormalTok{expo\_lag3 }\SpecialCharTok{*}\NormalTok{ lam\_hat}\SpecialCharTok{\^{}}\DecValTok{2}
\NormalTok{wf\_hat }\OtherTok{\textless{}{-}}\NormalTok{ wc }\SpecialCharTok{/} \FunctionTok{pmax}\NormalTok{(we, }\FloatTok{1e{-}8}\NormalTok{) }\SpecialCharTok{*}\NormalTok{ df\_new}\SpecialCharTok{$}\NormalTok{mu\_t   }\CommentTok{\# back on the absolute scale}

\CommentTok{\# Blended renewal rate}
\NormalTok{r\_hat }\OtherTok{\textless{}{-}}\NormalTok{ (}\DecValTok{1} \SpecialCharTok{{-}}\NormalTok{ Z\_hat) }\SpecialCharTok{*}\NormalTok{ mu\_hat }\SpecialCharTok{+}\NormalTok{ Z\_hat }\SpecialCharTok{*}\NormalTok{ wf\_hat}

\CommentTok{\# Return the underwriter{-}facing numbers per account}
\FunctionTok{data.frame}\NormalTok{(}
  \AttributeTok{GRCODE    =}\NormalTok{ df\_new}\SpecialCharTok{$}\NormalTok{GRCODE,}
  \AttributeTok{Z         =} \FunctionTok{round}\NormalTok{(Z\_hat,     }\DecValTok{2}\NormalTok{),   }\CommentTok{\# credibility weight}
  \AttributeTok{mu\_hat    =} \FunctionTok{round}\NormalTok{(mu\_hat,    }\DecValTok{4}\NormalTok{),   }\CommentTok{\# market{-}mean complement}
  \AttributeTok{f\_hat     =} \FunctionTok{round}\NormalTok{(wf\_hat,    }\DecValTok{4}\NormalTok{),   }\CommentTok{\# recency{-}weighted own experience}
  \AttributeTok{r\_hat     =} \FunctionTok{round}\NormalTok{(r\_hat,     }\DecValTok{4}\NormalTok{)    }\CommentTok{\# recommended renewal rate}
\NormalTok{)}
\end{Highlighting}
\end{Shaded}

\subsection*{Listing 4: MAP Estimation (Optional)}\label{listing-4-map-estimation-optional}
\addcontentsline{toc}{subsection}{Listing 4: MAP Estimation (Optional)}

MAP estimation adds log-prior penalty terms to the MLE objective (Equation \eqref{eq:map}).
This is useful when \(\lambda\) is weakly identified in small portfolios.
The priors below are illustrative and weakly informative: Normal(0, 1) on every
parameter except \(\log\phi\), which is Normal(2, 1). They differ from the
Listing 2 priors.

\begin{Shaded}
\begin{Highlighting}[]
\CommentTok{\# Negative log{-}posterior = negative log{-}likelihood + negative log{-}prior}
\NormalTok{nlp }\OtherTok{\textless{}{-}} \ControlFlowTok{function}\NormalTok{(par, df) \{}
\NormalTok{  nll\_val }\OtherTok{\textless{}{-}} \FunctionTok{nll}\NormalTok{(par, df)   }\CommentTok{\# reuse NLL from Listing 1}
  \CommentTok{\# Weakly informative priors:}
  \CommentTok{\#   Normal(0, 1) on every parameter except log\_phi; Normal(2, 1) on log\_phi}
\NormalTok{  prior }\OtherTok{\textless{}{-}} \SpecialCharTok{{-}}\FunctionTok{sum}\NormalTok{(}\FunctionTok{dnorm}\NormalTok{(par[}\SpecialCharTok{{-}}\FunctionTok{length}\NormalTok{(par)], }\AttributeTok{mean =} \DecValTok{0}\NormalTok{, }\AttributeTok{sd =} \DecValTok{1}\NormalTok{, }\AttributeTok{log =} \ConstantTok{TRUE}\NormalTok{)) }\SpecialCharTok{{-}}
            \FunctionTok{dnorm}\NormalTok{(par[}\FunctionTok{length}\NormalTok{(par)], }\AttributeTok{mean =} \DecValTok{2}\NormalTok{, }\AttributeTok{sd =} \DecValTok{1}\NormalTok{, }\AttributeTok{log =} \ConstantTok{TRUE}\NormalTok{)}
\NormalTok{  nll\_val }\SpecialCharTok{+}\NormalTok{ prior}
\NormalTok{\}}

\CommentTok{\# MAP fit: same call structure as Listing 1}
\NormalTok{fit\_map }\OtherTok{\textless{}{-}} \FunctionTok{nlminb}\NormalTok{(init, nlp, }\AttributeTok{df =}\NormalTok{ df\_train,}
                  \AttributeTok{control =} \FunctionTok{list}\NormalTok{(}\AttributeTok{iter.max =} \DecValTok{500}\NormalTok{, }\AttributeTok{rel.tol =} \FloatTok{1e{-}8}\NormalTok{))}

\CommentTok{\# MAP estimates are the posterior mode in this parameterisation.}
\CommentTok{\# Well{-}identified parameters stay close to the MLE; weakly identified}
\CommentTok{\# ones (here the lambda parameters) move toward their prior means.}
\NormalTok{fit\_map}\SpecialCharTok{$}\NormalTok{par}
\end{Highlighting}
\end{Shaded}

\newpage

\section*{§H: Supplementary Data Diagnostics}\label{app:deployment-diagnostics}
\addcontentsline{toc}{section}{§H: Supplementary Data Diagnostics}

\subsection*{Calendar-Year Normalisation}\label{app:trajectories}
\addcontentsline{toc}{subsection}{Calendar-Year Normalisation}

\begin{figure}[H]

{\centering \includegraphics[width=0.95\linewidth]{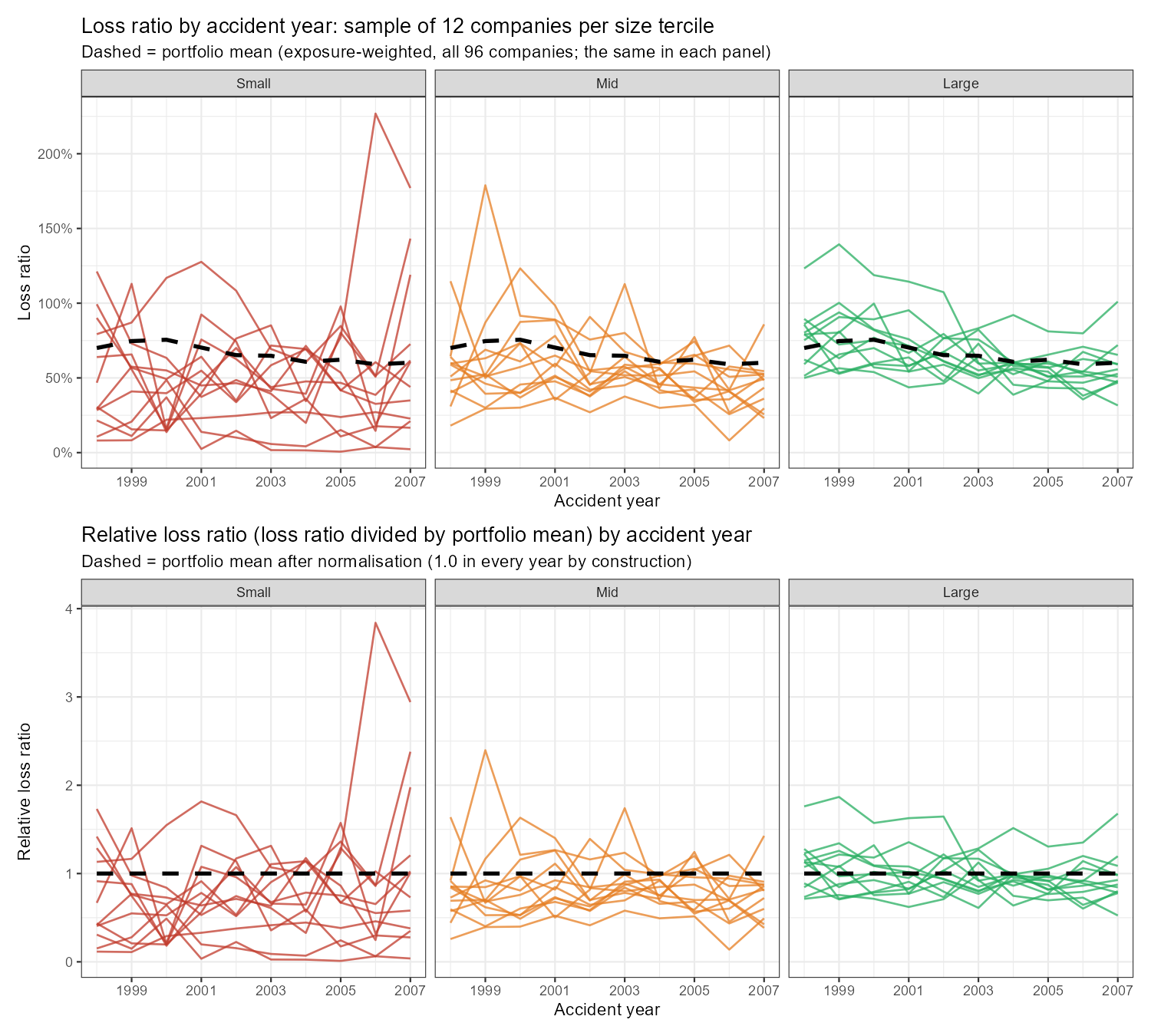} 

}

\caption{Loss ratio trajectories for a random sample of 12 companies per size tercile (same seed), before (upper panels) and after (lower panels) calendar-year normalisation. The dashed line is the exposure-weighted portfolio mean of all 96 companies, shown in every panel; after normalisation it is 1.0 by construction. The portfolio loss ratio falls from about 75\% (AY 1999--2000) to about 60\% (AY 2006--2007). Large companies follow it, and normalisation removes this common movement (accident year explains about 16\% of the variation in their loss ratios before and essentially none after). Small and Mid companies show little common movement to begin with.}\label{fig:fig-eda-trajectories}
\end{figure}

\subsection*{Survivorship Bias Check}\label{app:survivorship}
\addcontentsline{toc}{subsection}{Survivorship Bias Check}

As noted in Section~\ref{sec:empirical}, approximately 25 of the 41 incomplete-panel companies would classify as Small, about twice the Small share in the retained set, so the balanced panel skews toward stable long-tenured Small accounts. The likely direction of bias is toward slightly overestimating Small predictability; \(\hat{Z} \approx 0.62\) for Small may be modestly optimistic for a full population including entrants and exiters. The balanced filter was adopted for testing, to give every account a common held-out period, and is not a requirement of the framework, since short-history accounts can be fitted. Training only on accounts that stayed on the book, or that were written, gives a selected sample and may bias the estimates, so the training data should include lapsed accounts up to the point of exit and, where captured, the histories of submissions that were not written.

We ran a full sensitivity analysis on the CAS data using the proposed model (LogCred tercile \(\lambda\), continuous complement): the Bayesian fit for the accuracy, calibration, \(b_Z\) and LOO diagnostics, and maximum-likelihood refits for the \(\lambda\) comparison. Relaxing the balanced-panel filter to all companies with \(\geq 2\) training years and at least one test year expands the panel from 96 to 106 companies, adding 10 predominantly small entrants and exiters. Six diagnostics were evaluated:

\textbf{$\lambda$ gradient.} Tercile \(\lambda\) estimates (maximum-likelihood refits, so not identical to the Bayesian posterior means in the main text) are stable across panels: \(\hat\lambda_{\text{Sm}}\) 0.38 vs 0.35, \(\hat\lambda_{\text{Md}}\) 0.94 vs 0.98, \(\hat\lambda_{\text{Lg}}\) 0.16 vs 0.11. The gradient direction and magnitude replicate without change.

\textbf{$b_Z$ identification.} On the balanced panel \(b_Z\) is weakly identified (\(P(b_Z > 0) = 0.63\)) because all 96 companies share a full 10-year history, leaving little variation in data richness to identify the gradient. On the expanded panel, where companies range from 2 to 10 training years, \(b_Z\) is clearly positive (\(P(b_Z > 0) = 1\), 90\% CI \([0.32,\, 0.79]\)), confirming that the size gradient is a genuine data finding when sufficient variation is present. Adding log history length as a further \(Z\) covariate shows both size and history length are independently positive (\(\hat{b}_Z = 0.37\), \(\hat{w}_Z = 0.34\), each with \(P > 0.98\)), with a marginal LOO improvement of \(2.7 \pm 3.0\) ELPD, within one standard error.

\textbf{Influential observations.} One large short-history company (Federal Ins Co Grp, 5 training years, mean NEP \(\approx \$226\)M) is influential in the expanded-panel fit, distorting the size complement and the Large-tercile estimates (\(\hat\lambda_{\text{Lg}}\) = 0.11 on the expanded panel). Even so, the proposed model's wMSE improvement over standard B-S is 27.8\% on the expanded panel, against 28.3\% on the balanced panel (the 96 companies with complete histories; portfolio means here are computed over all companies, hence slightly different from the 27.4\% headline), so the headline result does not depend on the filter. This is precisely the class of account (large, short-history, recently performing well) that a balanced-panel filter protects against.

\textbf{Small calibration.} Bootstrap slope CI for Small accounts: balanced \(0.99\) \([0.73,\, 1.14]\) vs expanded \(1.26\) \([1.05,\, 1.69]\). The expanded panel Small CI lies entirely above 1.0 (statistically significant under-crediting), driven by short-history small accounts whose thin EWMA signal cannot track their outcome volatility. The balanced panel's headline Small calibration result (\(\approx 1.0\)) requires the filter.

Taken together, these diagnostics support the balanced panel as the primary case study: it identifies \(\lambda\) and \(b_Z\) cleanly, avoids influential short-history outliers, and delivers the Small calibration result. The framework still deploys on unbalanced portfolios via Phase 2 (Section \ref{sec:applying-renewal}); practitioners should monitor the \(\lambda\) interval width and refit without any account whose removal materially shifts the complement parameters.

\begin{figure}[H]

{\centering \includegraphics[width=0.72\linewidth]{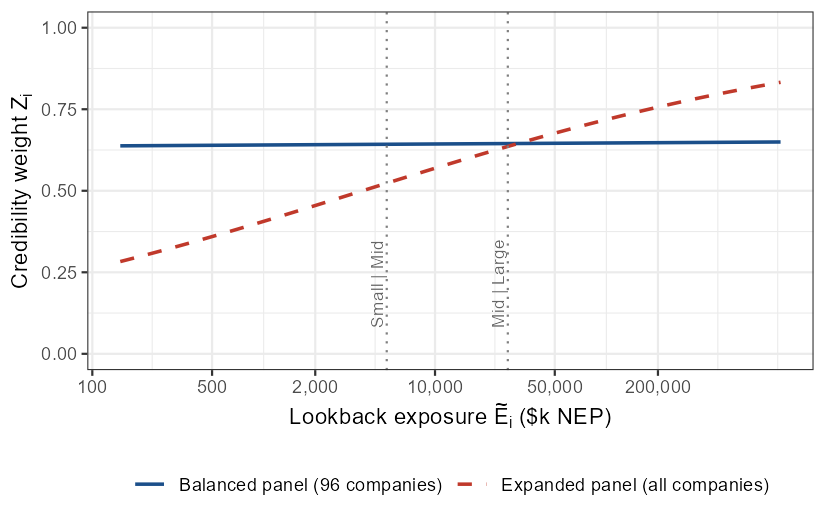} 

}

\caption{Fitted logistic $Z$ curves for the balanced panel (96 companies, solid) and the expanded panel (all companies with $\geq 2$ training years, dashed), plotted against log lookback exposure $\log\tilde{E}_i$. Vertical lines mark the tercile breaks. The two curves are nearly identical in the mid-to-large range; the main difference is at small exposures, where the expanded panel includes less predictable entrants and exiters, pulling Small $Z$ down slightly and making the size gradient ($\hat{b}_Z$) more clearly positive.}\label{fig:fig-surv-z-curves}
\end{figure}

\end{document}